%% file: draft_v2.tex
\documentclass[aps,showpacs,pra,superscriptaddress,preprintnumbers,amsmath,amssymb,twocolumn,longbibliography]{revtex4-2}
\usepackage[utf8]{inputenc}
\usepackage[T1]{fontenc}
\usepackage{graphicx}% Include figure files
\usepackage{dcolumn}% Align table columns on decimal point
\usepackage{appendix}
\usepackage{amsmath}
\usepackage{bm}% bold math
\usepackage[dvipsnames]{xcolor}
\usepackage{multirow}
\usepackage{notes2bib}
\usepackage{titletoc}
\usepackage{lipsum}

\usepackage{subfigure}
\usepackage{bbm}
\usepackage{enumerate}

\usepackage[
bookmarks=true,
colorlinks,
linkcolor=black,
urlcolor=black,
citecolor=black,
plainpages=false,
pdfpagelabels,
final,
breaklinks=true
]{hyperref}

\usepackage{titlesec}

\usepackage{array}

\usepackage{physics}

\begin{document}
\allowdisplaybreaks

\title{Structured squeezed light allows for high-harmonic generation in classical forbidden geometries}

\author{J.~Rivera-Dean}
\email{javier.rivera@icfo.eu}
\affiliation{ICFO -- Institut de Ciencies Fotoniques, The Barcelona Institute of Science and Technology, 08860 Castelldefels (Barcelona)}

\author{P.~Stammer}
\affiliation{ICFO -- Institut de Ciencies Fotoniques, The Barcelona Institute of Science and Technology, 08860 Castelldefels (Barcelona)}
\affiliation{Atominstitut, Technische Universität Wien, 1020 Vienna, Austria}

\author{M.~F.~Ciappina}
\affiliation{Physics Program, Guangdong Technion--Israel Institute of Technology, Shantou, Guangdong 515063, China}
\affiliation{Guangdong Provincial Key Laboratory of Materials and Technologies for Energy Conversion, Guangdong Technion – Israel Institute of Technology, Shantou, Guangdong 515063, China}
\affiliation{Technion -- Israel Institute of Technology, Haifa, 32000, Israel}

\author{M. Lewenstein}
\affiliation{ICFO -- Institut de Ciencies Fotoniques, The Barcelona Institute of Science and Technology, 08860 Castelldefels (Barcelona)}
\affiliation{ICREA, Pg. Llu\'{\i}s Companys 23, 08010 Barcelona, Spain}

%TC:ignore
\begin{abstract}
   High-harmonic generation (HHG) is a nonlinear process in which a strong driving field interacts with a material, resulting in the frequency up-conversion of the driver into its high-order harmonics. This process is highly sensitive to the field's polarization: circular polarization, for instance, inhibits HHG. In this work, we demonstrate that the use of non-classical structured light enables HHG in this otherwise prohibitive configuration for classical drivers.~We consider circularly polarized light with non-classical fluctuations, introduced via squeezing along one polarization direction, and show that these non-classical features prompt the HHG process. We find that the spectral properties of the emitted harmonics depend on the type of squeezing applied and, by analyzing the inner electron dynamics, we relate the observed differences to modifications of the HHG three-step mechanism induced by the specific squeezing type.~This approach opens new pathways for integrating quantum optics in HHG, providing novel means of controlling the light-matter interaction dynamics.
\end{abstract}
%TC:endignore

\maketitle
\emph{Introducion.---}High-harmonic generation (HHG) has emerged as a fundamental application for probing matter on its intrinsic ultrafast timescales~\cite{corkum_attosecond_2007,krausz_attosecond_2009,goulielmakis_high_2022}, enabled by the high-intensity driving conditions and sub-femtosecond dynamics involved in the process.~HHG begins with the interaction of a strong-laser field with a given matter system~\cite{burnett_harmonic_1977,mcpherson_studies_1987,ferray_multiple-harmonic_1988,ghimire_observation_2011,luu_extremeultraviolet_2018}, prompting a sequence of three steps~\cite{krause_high-order_1992,corkum_plasma_1993,lewenstein_theory_1994,vampa_theoretical_2014} in which electrons undergo tunneling ionization, propagate in the continuum, and recombine with their parent system.~This recombination results in the release of attosecond-duration bursts of radiation~\cite{antoine_attosecond_1996,drescher_x-ray_2001,paul_observation_2001}, corresponding to the energy accumulated by the electron in the laser field~\cite{krause_high-order_1992,vampa_theoretical_2014}.~By probing these three steps in different materials~\cite{marangos_development_2016,goulielmakis_high_2022,zong_emerging_2023}, or by using this attosecond radiation to investigate other systems~\cite{corkum_attosecond_2007,krausz_attosecond_2009,ciappina_attosecond_2017}, HHG enables insights into the electronic structure and ultrafast dynamics of matter with sub-femtosecond resolution.

\begin{figure}
	\centering
	\includegraphics[width=1\columnwidth]{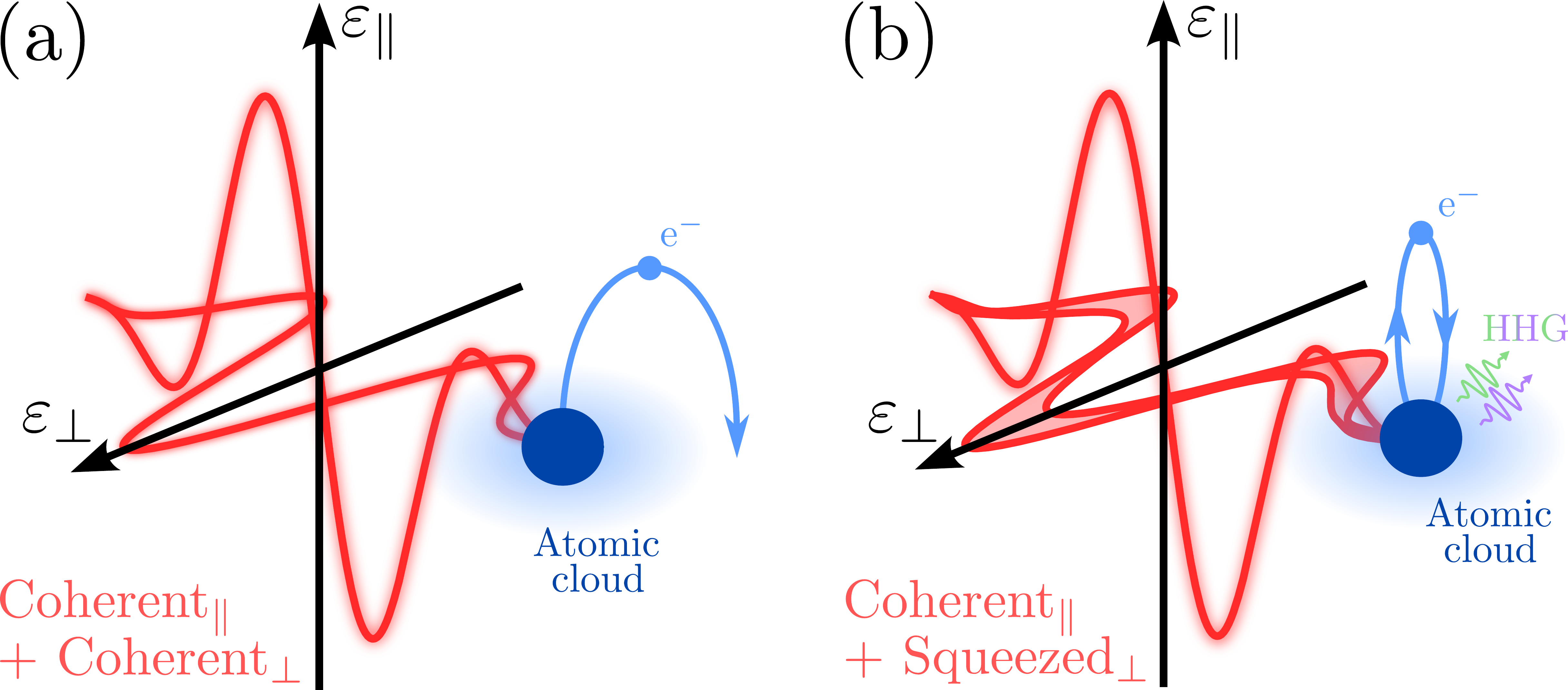}
	\caption{(a) Strong, circularly polarized classical field interacting with an atom. Circular polarization prevents the electron from returning to the parent ion, disabling HHG. (b) Configuration investigated in this work, with one polarization component in a squeezed state.~The additional uncertainty facilitates the electron’s return to the parent ion, enabling HHG despite the circular-polarization conditions.}
	\label{Fig:Scheme}
\end{figure}

Besides the material system, HHG is also sensitive to the driving field parameters.~By tuning the driver's intensity and frequency, one can control the spectral range of the emitted radiation and the HHG efficiency~\cite{krause_high-order_1992,lewenstein_theory_1994,lhuillier_high-order_1993,macklin_high-order_1993}.~The polarization of the driving field, in particular, has a profound impact: for increasing driving field ellipticity, HHG efficiency diminishes~\cite{budil_influence_1993,dietrich_high-harmonic_1994,corkum_subfemtosecond_1994,liang_experimental_1995,burnett_ellipticity_1995}, to the extent that circularly-polarized light can suppress HHG altogether~\cite{pisanty_spin_2014}. This effect can be understood within the three-step model: in a circular-polarized field, electrons are driven away from their atom during the propagation step, preventing recombination, and thus inhibiting radiation emission~\cite{corkum_plasma_1993,antoine_theory_1996,dudovich_attosecond_2006} [Fig.~\ref{Fig:Scheme}~(a)].

Recent theoretical~\cite{gorlach_high-harmonic_2023,even_tzur_photon-statistics_2023,stammer_absence_2024,tzur_generation_2023} and experimental~\cite{rasputnyi_high_2024,lemieux_photon_2024} studies have added an extra degree of freedom by employing non-classical states of light as HHG drivers.~Beyond having the potential to induce non-Poissonian photon statistics in the emitted light~\cite{lemieux_photon_2024,tzur_generation_2023}, non-classical drivers significantly impact the HHG dynamics and spectral characteristics.~In atoms driven by linearly polarized fields, non-classical light sources can generate radiation with extended spectral bandwidths~\cite{gorlach_high-harmonic_2023}.~This effect is particularly pronounced when using squeezed states of light where reduced uncertainty in one property, such as phase, is balanced by increased uncertainty in the conjugate property, such as amplitude.~Ref.~\cite{even_tzur_photon-statistics_2023} further observed that the photon statistics of these states can be interpreted as generating an effective force on the electron during its excursion in the continuum.

\begin{figure}
	\centering
	\includegraphics[width=1\columnwidth]{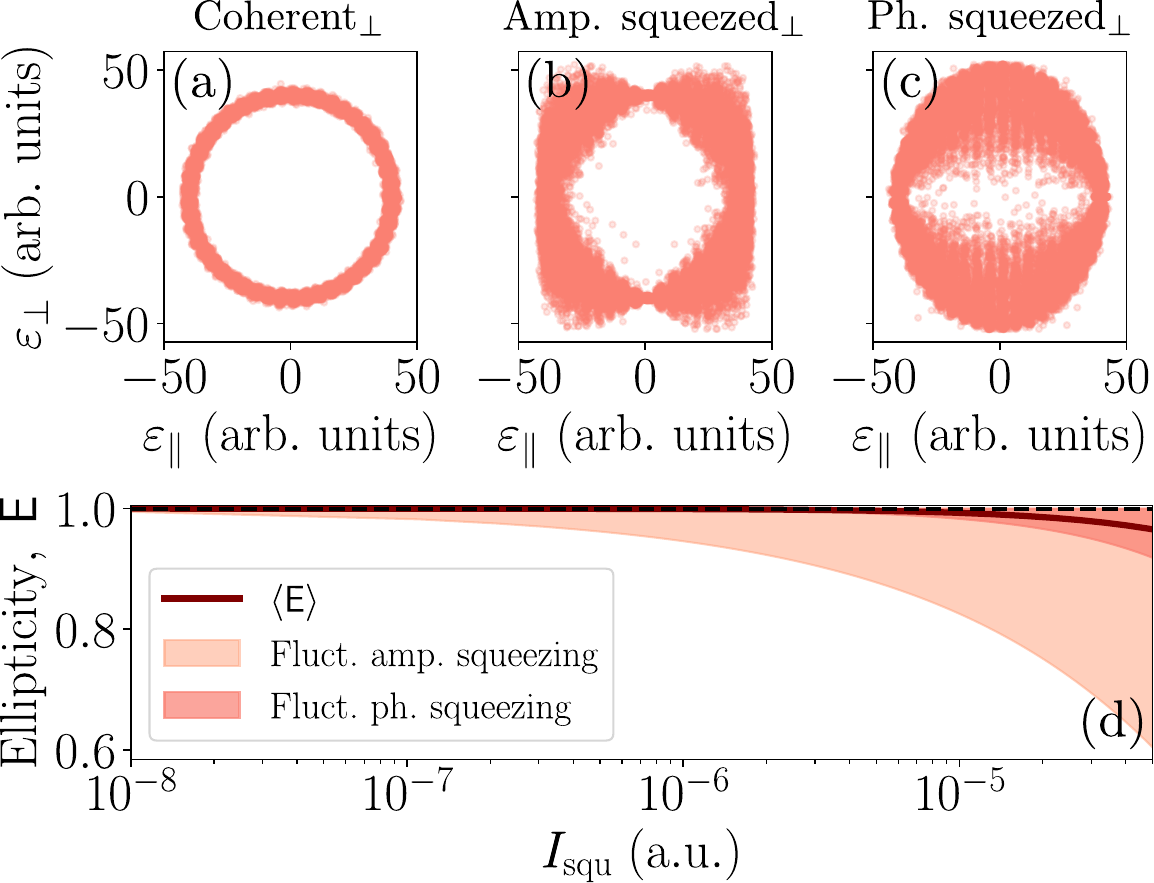}
	\caption{(a)-(c) Lissajous figures illustrating the considered states for different squeezing conditions: (a) no squeezing, (b) amplitude squeezing, and (c) phase squeezing.~(d) Mean ellipticity and its fluctuations as a function of the squeezing contribution to the intensity, $I_{\text{squ}}$.~The dashed horizontal black line indicates $\mathsf{E} = 1$ (circularly polarized light).~In the classical scenario, $\Delta \mathsf{E} \to 0$ as observed when $I_{\text{squ}}\to 0$.}
	\label{Fig:Liss:and:Ellip}
\end{figure}

In this Letter, we demonstrate that non-classical structured light---specifically, non-classical circularly polarized light---enables HHG, a phenomenon otherwise inhibited in the fully classical scenario.~We investigate configurations where along one driving field polarization component we have a coherent state, while along the orthogonal component a squeezed state [Fig.~\ref{Fig:Scheme}~(b)].~These configurations effectively reduce the field's ellipticity, with characteristics depending on the squeezed property; and enable the generation of harmonic radiation, with the emission becoming more pronounced as the level of squeezing increases.~We observe that the spectral features of the outgoing radiation depend on the type of squeezing.~Finally, we analyze how the photon statistics associated with these squeezed states influence the electronic trajectories, providing insights into the origin of the spectral characteristics. %In a broader picture, this work situates within the rapidly growing intersection of quantum optics and strong-field physics~\cite{bhattacharya_stronglaserfield_2023,cruz-rodriguez_quantum_2024}, demonstrating that non-classical light allows for effects otherwise forbidden in classical scenarios.

\emph{Effective ellipticity induced by squeezed light.---}We focus on scenarios where the initial quantum state of the field is
$
	\ket{\Phi} 
	= \ket{0,\alpha_1}_{\parallel} 
	\otimes \ket{r,\alpha_{2}}_{\perp}
$ [Fig.~\ref{Fig:Scheme}~(b)], with $\ket{r,\alpha}_{\mu} = \hat{D}_\mu(\alpha)\hat{S}_\mu(r)\ket{0}$ a displaced squeezed vacuum (DSV) state in the polarization direction $\mu$, where $\hat{D}_\mu(\alpha) = \text{exp}[\alpha \hat{a}^\dagger_{\vb{k}_0,\mu} - \alpha^*\hat{a}_{\vb{k}_0,\mu}]$ is the displacement operator and $\hat{S}_\mu(r) = \text{exp}[r^*\hat{a}^2_{\vb{k}_0,\mu} - r\hat{a}^{\dagger 2}_{\vb{k}_0,\mu}]$ the squeeze operator with $r$ the squeezing parameter.~Here, $\hat{a}_{\vb{k}_0,\mu}$ ($\hat{a}^\dagger_{\vb{k}_0,\mu}$) is the annihilation (creation) operator acting on the driving field mode $(\vb{k}_0,\mu)$.~We choose $\alpha_2 = i \alpha_1$, ensuring that $r=0$ leads to circular-polarized fields.~Crucially, we consider DSV states exhibiting substantial squeezing---sufficient to significantly perturb the HHG process~\cite{even_tzur_photon-statistics_2023}.~Squeezed states with these characteristics have recently been employed to  drive~\cite{rasputnyi_high_2024} or perturb~\cite{lemieux_photon_2024} HHG in semiconductors.~Notably, Ref.~\cite{rasputnyi_high_2024} observed HHG using squeezing intensities of $I_{\text{squ}}\sim10^{12}$ W/cm$^2$, two orders of magnitude below the $I_{\text{coh}}\sim10^{14}$ W/cm$^2$ required for atomic HHG. Due to their experimental feasibility, these values set the intensity scales for the squeezed component used in this work, i.e., $I_{\text{squ}}/I_{\text{coh}}\sim10^{-2}$.~Nonetheless, we highlight that, while these squeezed states were initially produced via nonlinear optical processes~\cite{spasibko_multiphoton_2017,manceau_indefinite-mean_2019}, recent theoretical~\cite{stammer_squeezing_2023,yi_generation_2024,rivera-dean_squeezed_2024,lange_electron-correlation-induced_2024,lange_electron-correlation-induced_2024} and experimental~\cite{theidel_evidence_2024,theidel_observation_2024} studies indicate that HHG itself could serve as source of DSV states.

In cases where circular-like fields drive HHG, harmonic emission is enabled by appropriately tuning the frequencies of the $\perp$- and $\parallel$-field components to direct the electron towards recombination~\cite{pisanty_spin_2014,fleischer_spin_2014}, resulting in Lissajous figures with field components vanishing simultaneously for both polarizations.~To determine whether non-classical features can induce similar effects by modifying the field's ellipticity $\mathsf{E}$, we analyze its Lissajous figure using statistical sampling to capture the intrinsic field fluctuations.~This involves identifying the outcomes $\{o\}$ and associated probabilities $\{p(o)\}$ for a given operator $\hat{O}$, then numerically sampling from $\{o\}$ according to $\{p(o)\}$.~We apply this approach to both polarization components of the electric field operator [Fig.~\ref{Fig:Liss:and:Ellip}~(a)-(c)], $\hat{E}_{\parallel,\vb{k}_0}(t)$ and $\hat{E}_{\perp,\vb{k}_0}(t)$, with $\hat{E}_{\mu,\vb{k}_0}(t) = i\epsilon_{\vb{k}_0}[\hat{a}_{\vb{k}_0,\mu} e^{i\omega_{\vb{k}_0}t}- a_{\vb{k}_0,\mu}^\dagger e^{-i\omega_{\vb{k}_0}t}]$, where $\epsilon_{\vb{k}} = \sqrt{\hbar \omega_{\vb{k}_0}/(2\epsilon_0 V)}$ and $V$ the quantization volume.

When squeezing is absent (panel~(a)), the Lissajous figure forms a near-perfect circle with finite width, reflecting the coherent state's inherent uncertainty.~Introducing squeezing along the $\perp$-component, the circular shape deforms based on the squeezing type.~For amplitude squeezing (panel~(b)), fluctuations create a fourfold structure along the distribution's diagonal axes.~This arises because uncertainties in phase not only induce ellipticity but also cause rotations of the polarization ellipse, as phase determines the ellipse's spatial orientation relative to a predefined axis. Phase squeezing (panel~(c)) instead yields an elliptical structure without rotations since phase is more precisely defined.~Conversely, fluctuations are enhanced along the $\varepsilon_{\perp}$-axis, corresponding to intensity variations.

We quantitatively assess these modifications by computing the ellipticity and its fluctuations~\cite{Supplementary}.~As we consider fields propagating in free space where the quantization volume $V\to \infty$ ($\epsilon_{\vb{k}}\to 0$)~\cite{tannoudji_classical_1997}, we evaluate physical observables in the classical limit.~This implies setting $\alpha \to \infty$ while keeping the field amplitude $\varepsilon_\alpha = 2\epsilon_{\vb{k}}\alpha$ constant, which is appropriate in the context of strong-field physics where driving fields contain an extremely large number of photons.~The results are displayed in Fig.~\ref{Fig:Liss:and:Ellip}~(d), where the mean ellipticity is shown with the solid curve and its fluctuations with the shaded region, both plotted against the amount of squeezing, represented in terms of $I_{\text{squ}}$, i.e., its contribution to the total field intensity; for DSV states, the intensity is given by $I = \epsilon_{\vb{k}}^2 \abs{\alpha}^2 + \epsilon_{\vb{k}}^2 \sinh[2](r) \equiv I_{\text{coh}} + I_{\text{squ}}$.~The mean ellipticity decreases to $\mathsf{E} < 1$ as $I_{\text{squ}}$ increases, while fluctuations grow, both of these quantities showing an emergence of a non-classical induced ellipticity.~Interestingly, the rate of fluctuation growth differs between phase- and amplitude-squeezed states, with more pronounced effects observed for amplitude squeezing.

Despite inducing an effective ellipticity in the field, these results show that, on average, the modifications do not lead to vanishing field amplitudes for both polarization components simultaneously.~This suggests that, under the considered field configurations, observation of harmonic radiation despite circularly polarized conditions cannot simply be attributed to the mean field passively guiding electrons back to the parent ion, as occurs with bicircular fields~\cite{fleischer_spin_2014,pisanty_spin_2014}.~Instead, it suggests an active modification of electron trajectories driven by field fluctuations, which can be interpreted as generating an effective force that steers electrons towards recombination~\cite{even_tzur_photon-statistics_2023}. %In the following, we explore how and under what conditions these induced field fluctuations become sufficient to trigger HHG events. 

\emph{HHG driven by non-classical structured light.---}To characterize how non-classically induced field fluctuations influences HHG, we extend the theoretical approach from Refs.~\cite{gorlach_high-harmonic_2023,even_tzur_photon-statistics_2023} to arbitrarily polarized initial fields. This approach, suitable for driving fields with arbitrary photon statistics, represents the driver's initial state using the generalized $P$-representation $
	\hat{\rho}_{\vb{k}_0}(t_0)
		= \bigotimes_{\vb{\mu}}
				\int \dd^2 \alpha_{\mu} \int \dd^2 \beta_{\mu}
					P(\alpha_{\mu},\beta_{\mu}^*)
					\dyad{\alpha_{\mu}}{\beta_{\mu}}/\braket{\beta_{\mu}}{\alpha_{\mu}}
$~\cite{drummond_generalised_1980},~where the initial state of the total field reads $\hat{\rho}_{\text{field}}(t_0) = \hat{\rho}_{\vb{k}_0}(t_0) \bigotimes_{\vb{k}\neq\vb{k}_0,\mu} \lvert 0_{\vb{k},\mu}\rangle\!\langle0_{\vb{k},\mu}\rvert$, with the atom initially in its ground state $\ket{\text{g}}$.

Assuming that the applied field does not substantially deplete the atom~\cite{lewenstein_generation_2021,rivera-dean_strong_2022,stammer_quantum_2023,stammer_squeezing_2023}, the final state of the light-matter system, after tracing out the driving field modes, can be expressed as~\cite{Supplementary}
\begin{equation}\label{Eq:Final:state}
	\begin{aligned}
	\hat{\rho}(t)
		= 
                \prod_{\mu}
			\int \dd^2 \alpha_\mu
				\int \dd^2 \beta_\mu
					&P(\alpha_\mu,\beta_\mu^*)
						\dyad{\phi_{\boldsymbol{\alpha}}(t)}{\phi_{\boldsymbol{\beta}}(t)}
					\\&
					\bigotimes_{\vb{k}\neq\vb{k}_0,\mu}
						\lvert\chi_{\vb{k},\mu}^{(\boldsymbol{\alpha})}(t)\rangle\!\langle\chi_{\vb{k},\mu}^{(\boldsymbol{\beta})}(t)\rvert,
	\end{aligned}
\end{equation}
where $\boldsymbol{\alpha} = (\alpha_{\parallel},\alpha_{\perp})$.~After interaction, the optical modes that were initially in a vacuum state become populated by coherent states with amplitudes
$
	\chi^{(\boldsymbol{\alpha})}_{\vb{k},\mu}(t)
		= \epsilon_{\vb{k}}
			\int^t_{t_0} \dd \tau
				\mel{\text{g}}{\hat{d}^{(\boldsymbol{\alpha})}_{\mu}(\tau)}{\text{g}}e^{i\omega_{\vb{k}}\tau}
$~\cite{lewenstein_generation_2021,rivera-dean_strong_2022,stammer_quantum_2023}, where $\hat{d}_{\mu}^{(\boldsymbol{\alpha})}(t)$ is the time-dependent dipole moment operator along the $\mu$-polarization component, in the interaction picture with respect to $\hat{H}^{(\boldsymbol{\alpha})}_{\text{sc}}(t) = \hat{H}_{\text{at}} + \hat{d}_{\parallel}E_{\alpha_{\parallel}}(t) + \hat{d}_{\perp}E_{\alpha_\perp}(t)$.~Here, $\hat{H}_{\text{at}}$ denotes the atomic Hamiltonian, $E_{\alpha_\mu}(t) = \langle \alpha_\mu\vert \hat{E}_{\vb{k}_0,\mu}(t) \vert \alpha_\mu \rangle$ the classical field amplitude, and $\ket{\phi_{\boldsymbol{\alpha}}(t)}$ the solution to the semiclassical Schrödinger equation
$
	i\hbar \pdv*{\ket{\phi_{\boldsymbol{\alpha}}(t)}}{t}
		= \hat{H}^{(\boldsymbol{\alpha})}_{\text{sc}}(t)
			\ket{\phi_{\boldsymbol{\alpha}}(t)}.
$

\begin{figure}
	\centering
	\includegraphics[width=1\columnwidth]{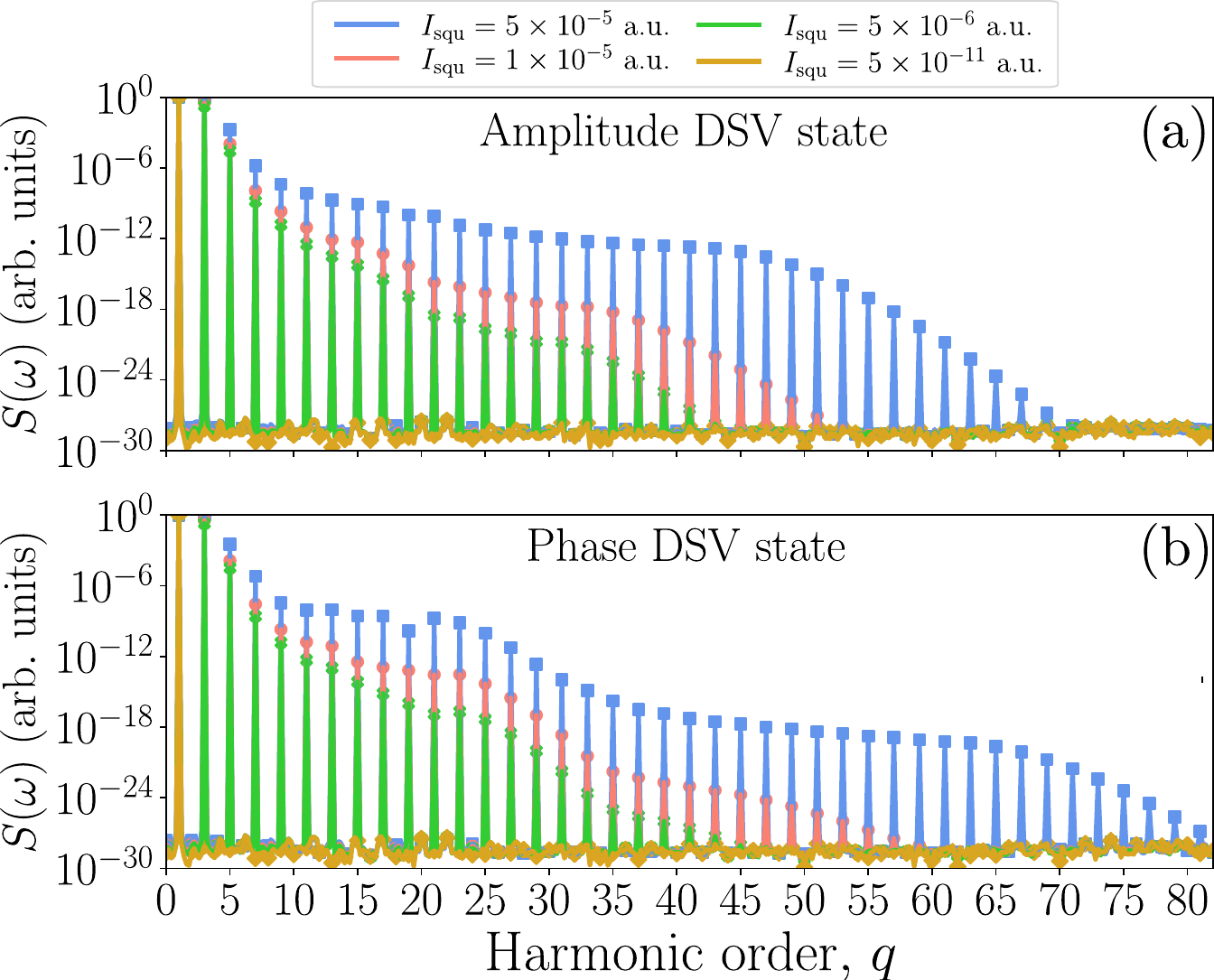}
	\caption{HHG spectra computed for (a) amplitude and (b) phase squeezing, along the $\perp$-polarization direction, respectively. Here, we set $E_{\parallel}(t) = \lvert\bar{\varepsilon}_{\parallel}^{(x)}\rvert \cos(\omega t)$ and $E_{\perp}(t) = \lvert\bar{\varepsilon}_{\perp}^{(y)}\rvert\sin(\omega t)$, with $\lvert\bar{\varepsilon}_{\parallel}^{(x)}\rvert=\lvert\bar{\varepsilon}_{\perp}^{(y)}\rvert = 0.053$ a.u. ($I = 10^{14}$ W/cm$^2$) and $\omega = 0.057$ a.u. ($\lambda = 800$ nm). The spectra were evaluated over 5 optical cycles ($\Delta t = 13$ fs) with hydrogen ($I_p = 0.5$ a.u.)~as the atomic system. All spectra are normalized to their maximum value.} 
	\label{Fig:Spec:and:Ellip}
\end{figure}

Using Eq.~\eqref{Eq:Final:state}, we compute the HHG spectra as $
S(\omega)
= \lim_{\epsilon_{\vb{k}}\to0}
\pdv*{\mathcal{E}(t)}{\omega}
$~\cite{gorlach_quantum-optical_2020,rivera-dean_strong_2022,gorlach_high-harmonic_2023}, with $\mathcal{E}(t) = \sum_{\vb{k},\mu}\text{tr}[\hat{a}^\dagger_{\vb{k},\mu}\hat{a}_{\vb{k},\mu}\hat{\rho}(t)]$.~For the fields considered in this work, we find~\cite{Supplementary}
\begin{equation}\label{Eq:HHG:Spec}
	\begin{aligned}
	S(\omega)
		\propto \dfrac{\omega^4}{\sqrt{8\pi I_{\text{squ}}}}
			&\int \dd \varepsilon^{(i)}_{\alpha,\perp}
				\Big\{\!
					e^{-\frac{[\varepsilon^{(i)}_{\alpha,\perp} -\bar{\varepsilon}_{\perp}^{(i)}]^2}{8 I_{\text{squ}}}}
					\\&\times
					\big[
						\lvert d_{\boldsymbol{\varepsilon_{\alpha}},\perp}(\omega) \rvert^2
						+	\lvert d_{\boldsymbol{\varepsilon_{\alpha}},\parallel}(\omega) \rvert^2
					\big]
				\Big\},
	\end{aligned}
\end{equation}
where $i \in \{x,y\}$ indicates the optical quadrature along which squeezing takes place, and $\bar{\varepsilon}_{\perp} = \bar{\varepsilon}^{(x)}_{\perp} + i\bar{\varepsilon}^{(y)}_{\perp}$ the mean amplitude and phase of the electric field along the $\perp$-polarization. Here, $\bar{\varepsilon}_{\perp} = i\bar{\varepsilon}_{\parallel}$. When $I_{\text{squ}}\to 0$, corresponding to an initial coherent state in both polarizations, the Gaussian prefactor in Eq.~\eqref{Eq:HHG:Spec} tends to a Dirac delta, recovering the semiclassical expression for the HHG spectra~\cite{lewenstein_theory_1994,lewenstein_generation_2021,rivera-dean_strong_2022,stammer_squeezing_2023,Supplementary}.

Figure~\ref{Fig:Spec:and:Ellip} shows $S(\omega)$ for amplitude-squeezed (panel (a)) and phase-squeezed states (panel (c)) for different $I_{\text{squ}}$.~Contrarily to what is found with coherent states, the introduction of high-enough field fluctuations through squeezing results in pronounced spectral features despite the circularly polarized arrangement, with characteristics depending on the specific type of squeezing.~These features diminish as $I_{\text{squ}}$ decreases, with the spectrum vanishing entirely when $I_{\text{squ}}$ becomes small enough ($I_{\text{squ}} = 5 \times 10^{-11}$ a.u.), in agreement with semiclassical theories~\cite{antoine_theory_1996}.

For amplitude squeezing, the spectrum exhibits a single plateau structure for odd harmonic orders, with a well-defined cutoff frequency that varies with $I_{\text{squ}}$.~Meanwhile, phase squeezing results in a double plateau structure: the first cutoff appears independent of the squeezing strength, whereas the second plateau extends further as $I_{\text{squ}}$ increases, generally exceeding the cutoff observed for amplitude squeezing.~In both cases the cutoff extends beyond what is observed for linearly polarized drivers of comparable intensities ($I = 10^{14}$ W/cm$^2$ leading to $q_c \approx 21$ for hydrogen).~While such an extension is expected~\cite{gorlach_high-harmonic_2023}, it can be attributed to how field fluctuations induced by squeezing modify the effective field amplitude experienced by the electron.~For amplitude squeezing, the electric field vector seen by the electron takes the form $\boldsymbol{\varepsilon}_0 = (\varepsilon_{\parallel},i \varepsilon_{\perp} + \Delta \varepsilon)$, leading to a total intensity $I_{\text{tot}}\propto \lvert\boldsymbol{\varepsilon}\rvert^2 = \lvert\varepsilon_{\parallel}\rvert^2 + \lvert\varepsilon_{\perp}\rvert^2 + \Delta \varepsilon^2 $.~For phase squeezing, we have $\boldsymbol{\varepsilon}_0 = (\varepsilon_{\parallel},i (\varepsilon_{\perp} + \Delta \varepsilon))$ resulting in $I_{\text{tot}}\propto \lvert\boldsymbol{\varepsilon}\rvert^2 = \lvert\varepsilon_{\parallel}\rvert^2 + \lvert\varepsilon_{\perp}\rvert^2 + \Delta \varepsilon^2 + 2 \varepsilon_{\perp}\Delta \varepsilon$, exceeding the amplitude-squeezed case by $2 \varepsilon_{\perp}\Delta \varepsilon$. In both cases, the field intensity explicitly depends on the fluctuations introduced by squeezing, explaining the cutoff extension as a function of $I_{\text{squ}}$.

\emph{Analysis of the electron dynamics.---}While qualitative features in the harmonic spectra---such as cutoff enhancement---can be understood through how field fluctuations perturb the average driver's intensity, specific characteristics like multiple plateaus or varying harmonic yields require a deeper analysis of how these fluctuations modify the electron dynamics.~In strong-field physics, such dynamics can be accessed through the Fourier transform of the induced electronic dipole~\cite{lewenstein_theory_1994,even_tzur_photon-statistics_2023}, given by $\langle \hat{\vb{d}}(\omega)\rangle = \int \dd t_2 e^{i\omega t_2} \text{tr}[\hat{\vb{d}}\hat{\rho}(t_2)]$, which encapsulates the three-step HHG mechanism. In the classical limit, $\langle \hat{\vb{d}}(\omega)\rangle$ reduces to~\cite{Supplementary}
\begin{equation}\label{Eq:Mel:Dipole}
	\begin{aligned}
	\langle \hat{\vb{d}}(\omega)\rangle
		&= \dfrac{1}{\sqrt{8\pi I_{\text{squ}}}}\!
			\int\! \dd t_2\!
				 \int \!\dd \varepsilon_{\alpha,\perp}^{(i)}
						\!e^{-\frac{(\varepsilon^{(i)}_{\alpha,\perp+i\omega t_2} -\bar{\varepsilon}_{\perp}^{(i)})^2}{8 I_{\text{squ}}}}
						\vb{d}_{\boldsymbol{\varepsilon}_{\boldsymbol{\alpha}}}(t_2)
	\end{aligned}
\end{equation}
where $\vb{d}_{\boldsymbol{\varepsilon_{\alpha}}}(t) = \langle \phi_{\boldsymbol{\alpha}}(t)\vert \hat{\vb{d}}\vert \phi_{\boldsymbol{\alpha}}(t)\rangle$ represents the dipole moment expectation value.~Due to the highly oscillatory nature of $\vb{d}_{\boldsymbol{\varepsilon_{\alpha}}}(t)$, we can evaluate Eq.~\eqref{Eq:Mel:Dipole} using the saddle-point approximation, with the saddle-points encoding key electron dynamic features such as ionization and recombination times~\cite{lewenstein_theory_1994,amini_symphony_2019,even_tzur_photon-statistics_2023}.

\begin{figure}
	\centering
	\includegraphics[width=1\columnwidth]{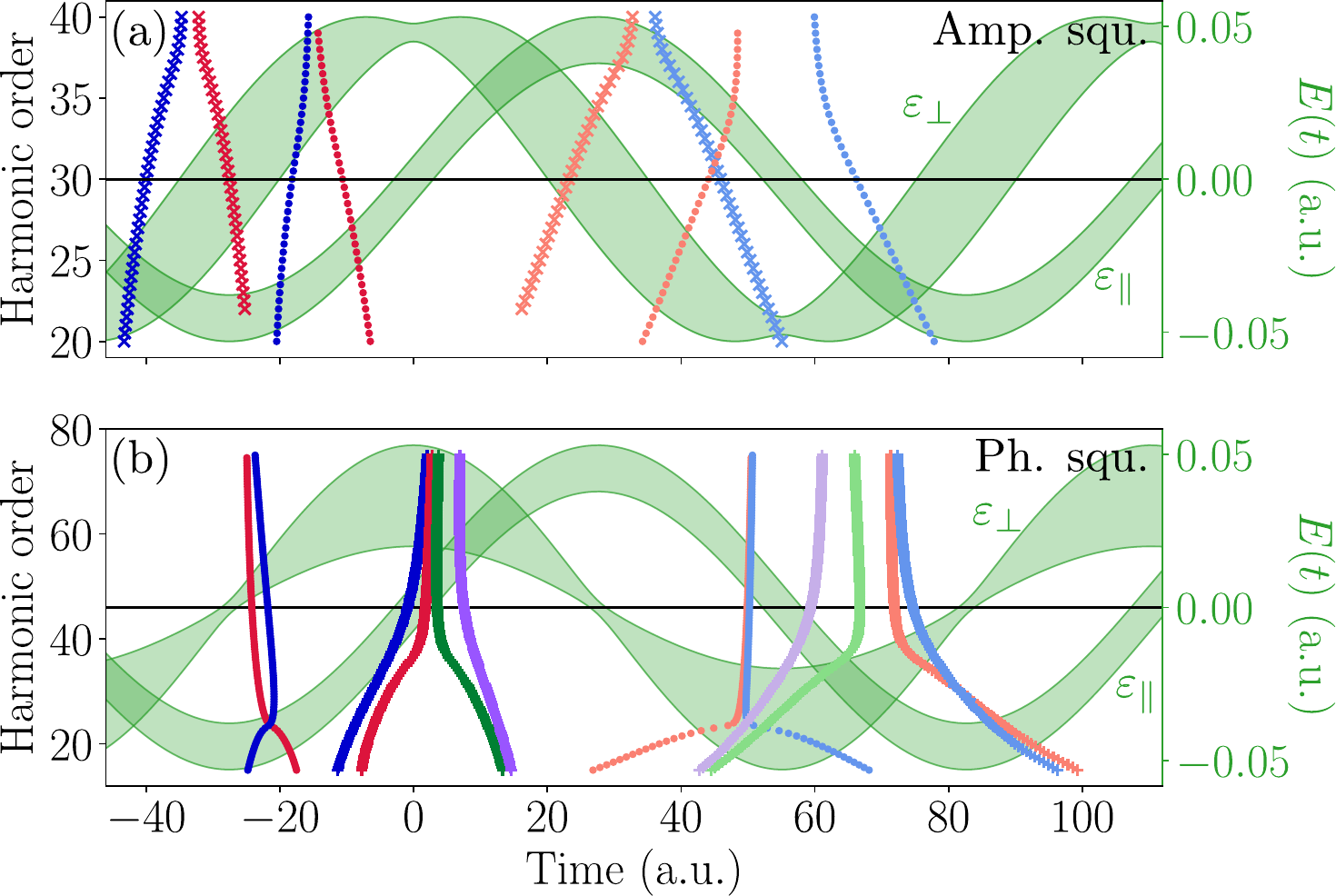}
	\caption{Real part of the ionization (bright colors) and recombination (soft colors) times, with the electric field represented for both polarizations in green.~Markers of the same type are used to associate ionization and recombination events that belong to the same electron trajectory.~The $\perp$-field component is (a) amplitude squeezed and (b) phase-squeezed with $I_{\text{squ}} = 5 \times 10^{-5} \ \text{a.u.}$ for both cases.~The mean field is circularly polarized with strength $\lvert\bar{\varepsilon}_{\mu}\rvert = 0.053$ a.u. ($I = 10^{14}$ W/cm$^2$) for each polarization component.}
	\label{Fig:Traj:Field}
\end{figure}

Figure \ref{Fig:Traj:Field} shows the real parts of ionization (bright colors) and recombination times (soft colors), with markers linking corresponding ionization-recombination events. Panel (a) uses an amplitude-squeezed driver along the $\perp$-direction, while panel (b) a phase-squeezed driver, both adding an squeezed-intensity contribution $I_{\text{squ}} = 5 \times 10^{-5}$ a.u.~to the coherent part.~When squeezing features along the $\perp$-polarization component are absent or sufficiently weak, no saddle-point solutions exist---unambiguously tying the observed dynamics to the additional field fluctuations induced by squeezed drivers.~These squeezing-induced field fluctuations act as a competing force~\cite{even_tzur_photon-statistics_2023,Supplementary} against the coherent $\perp$-component that typically suppresses electron-ion recombination~\cite{corkum_plasma_1993,antoine_theory_1996}.

Comparing the trajectories in Fig.~\ref{Fig:Traj:Field} with the HHG spectra in Fig.~\ref{Fig:Spec:and:Ellip}, we relate the spectral features to the electron dynamics induced by the different types of squeezing.~For amplitude squeezing [Fig.~\ref{Fig:Traj:Field}~(a)], we observe two sets of pairs of trajectories within a single half-cycle of the field.~Each set consists of two-well defined trajectories---long (blue) and short (red)---which become increasingly similar at higher harmonic orders.~Both sets share a common cutoff frequency around $q_c \approx 40$, in qualitative agreement with that observed in Fig.~\ref{Fig:Spec:and:Ellip}~(a) (blue curve).~Notably, harmonics in this region originate when ionized electrons interact with both field polarization components simultaneously.~This dual interaction occurs for all trajectories in this regime, meaning electrons immediately experience both field components after ionization.~Consequently, the squeezed component counteracts the $\perp$-coherent contribution and modifies the electron's energy at ionization, leading to a $I_{\text{squ}}$-dependence of $q_c$. 

For phase squeezing, the trajectories exhibit notable differences compared to the amplitude-squeezed ones [Fig.~\ref{Fig:Traj:Field}~(b)]. While two trajectory sets still emerge within a single half-cycle, their timing differs significantly---they occur either at maxima (minima) of the $\parallel$-coherent ($\perp$-squeezed) field component or vice versa.~This distinction critically impacts electron dynamics and dictates the cutoff frequencies.~The first trajectory set generates the lowest cutoff ($q_c \approx 20$ in qualitative agreement with Fig.~\ref{Fig:Spec:and:Ellip}~(b)), matching the first plateau in the spectrum. These trajectories primarily depend on the coherent state component of the field, making their cutoff independent of $I_{\text{squ}}$, and therefore alike to that found with coherent, linearly polarized fields.~Here, squeezing-induced field fluctuations have the unique role of counteracting the $\perp$-coherent component's drift effect, but do not alter the electron's energy at recombination for trajectories leading to the cutoff frequencies, as this event occurs near maxima of the $\perp$-field (minima of the vector potential).~Conversely, the second trajectory set leads to a higher cutoff ($q_c \approx 60$, matching qualitatively Fig.~\ref{Fig:Traj:Field}~(b)). For these trajectories, squeezing fluctuations modify the electron paths, enabling recombination and actively contributing to the recombination energy, as cutoff harmonics arise near the minima of the $\perp$-field (maxima of the vector potential).~Increasing $I_{\text{squ}}$ enhances the kinetic energy converted to harmonic radiation, explaining the second cutoff's $I_{\text{squ}}$-dependence.

Notably, the second set contains additional trajectory types beyond the standard short/long paths (red/green curves), presented with the purple and blue curves. This feature emerges only for high-enough $I_{\text{squ}}$~\cite{Supplementary}, as squeezing-enhanced field fluctuations create new trajectory possibilities that lead to electron recombination with the parent ion~\cite{Supplementary}.

\emph{Role of field fluctuations.---}While our results illustrate how non-classical light challenges established paradigms in strong-field physics~\cite{stammer_limitations_2024}, the question of whether non-classicality is strictly necessary for HHG under circular polarization remains open.~To address this, we replace squeezed states in our scheme with displaced thermal states, a classical state where thermal noise increases fluctuations homogeneously in both optical quadratures.~This allows understanding whether field fluctuations alone, irrespective of their classical or non-classical origin, are sufficient to enable HHG with circular polarized drivers, or if the asymmetric fluctuation distribution from squeezing turns essential.

\begin{figure}
	\centering
	\includegraphics[width=1\columnwidth]{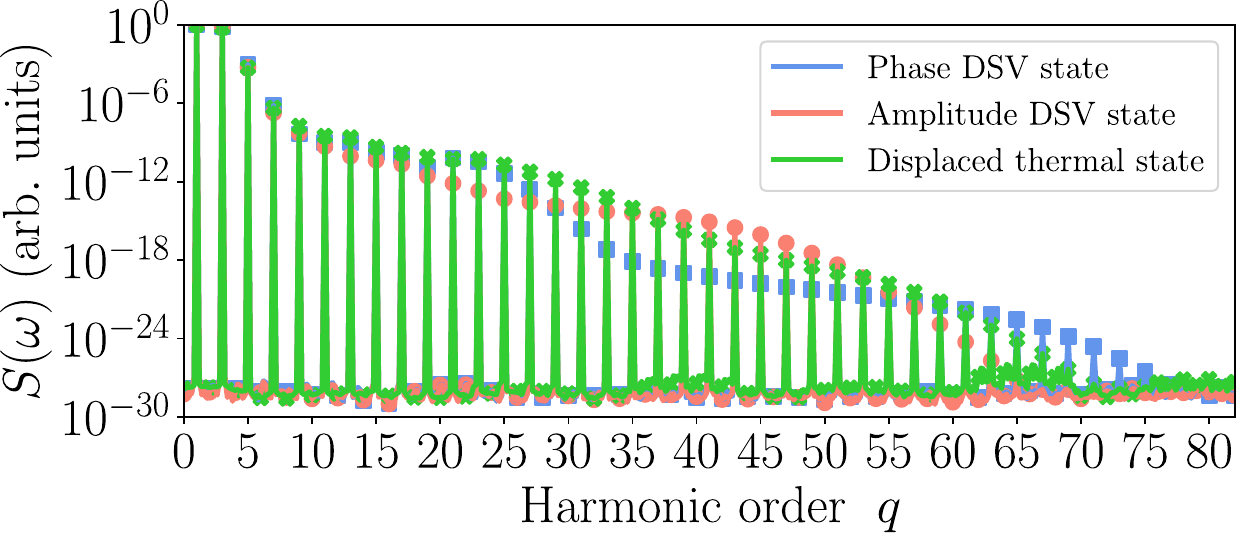}
	\caption{HHG spectra computed when adding squeezed (blue and red curves) and thermal (green curve) light to the $\perp$-polarization direction. For the squeezed cases, we set $I_{\text{squ}} = 2.5 \times 10^{-5}$ a.u., while for the thermal $I_{\text{th}} = 5 \times 10^{-5}$ a.u., as provides a comparable amount of field fluctuations.~All spectra are normalized to their maximum value.} 
	\label{Fig:Spec:Thermal}
\end{figure}

Following an analysis analogous to Eq.~\eqref{Eq:HHG:Spec}~\cite{Supplementary}, Fig.~\ref{Fig:Spec:Thermal} compares the HHG spectra of squeezed states (blue and red curves) with those from displaced thermal states (green curve).~To ensure a fair comparison, the total field fluctuations are matched by adjusting the average intensities of the squeezed and thermal components to $I_{\text{squ}} = 2.5 \times 10^{-5}$ a.u.~and $I_{\text{th}} = 5 \times 10^{-5}$ a.u., respectively~\cite{Supplementary}.~As observed, displaced thermal states produce non-negligible harmonic radiation, demonstrating that non-classicality is sufficient but not necessary for HHG to occur under circular polarized drivers; large enough strong-field fluctuations are the critical factor.~Specifically, we observe that the overall harmonic yield matches that of amplitude squeezing across the whole spectral range, but with larger cutoff frequencies---nearly comparable to those of the second plateau obtained with phase-squeezed drivers.~This hybrid behavior arises because thermal states display homogeneously distributed noise across both quadratures; consequently, their HHG spectral features combine characteristics of both amplitude and phase squeezing.

\emph{Conclusions.---}Our results demonstrate that injecting sufficient field fluctuations---such as through non-classical light---fundamentally challenges long-standing notions in strong-field physics~\cite{stammer_limitations_2024}, particularly the belief that circularly polarized light cannot generate harmonic radiation.~Yet, non-classical light is not strictly required to enable HHG; instead, strong field fluctuations, regardless of their origin, impose the sufficient conditions for harmonic generation.~However, a key distinction of using non-classical structured drivers lies in the emergence of non-trivial HHG spectral features [Fig.~\ref{Fig:Spec:and:Ellip}] compared to fully classical counterparts [Fig.~\ref{Fig:Spec:Thermal}], as well as distinct spatial characteristics in the driving field [Fig.~\ref{Fig:Liss:and:Ellip}]. 

This latter aspect presents exciting opportunities for further exploration, particularly in solid-state~\cite{ghimire_observation_2011,you_anisotropic_2017,you_probing_2018} and molecular systems~\cite{smirnova_high_2009,ayuso_synthetic_2019,mayer_chiral_2024}, where HHG is highly sensitive to the relative orientation between the material and the driver’s polarization.~This sensitivity suggests that the interplay between non-classical structured light and strong-field physics could serve as a powerful tool for controlling and probing ultrafast electron dynamics in complex systems~\cite{bhattacharya_stronglaserfield_2023,cruz-rodriguez_quantum_2024}.

%TC:ignore
\emph{Acknowledgements.---}ICFO-QOT group acknowledges support from:
European Research Council AdG NOQIA; MCIN/AEI (PGC2018-0910.13039/501100011033,  CEX2019-000910-S/10.13039/501100011033, Plan National FIDEUA PID2019-106901GB-I00, Plan National STAMEENA PID2022-139099NB, I00, project funded by MCIN/AEI/10.13039/501100011033 and by the “European Union NextGenerationEU/PRTR" (PRTR-C17.I1), FPI); QUANTERA DYNAMITE PCI2022-132919, QuantERA II Programme co-funded by European Union’s Horizon 2020 program under Grant Agreement No 101017733; Ministry for Digital Transformation and of Civil Service of the Spanish Government through the QUANTUM ENIA project call - Quantum Spain project, and by the European Union through the Recovery, Transformation and Resilience Plan - NextGenerationEU within the framework of the Digital Spain 2026 Agenda; Fundació Cellex; Fundació Mir-Puig; Generalitat de Catalunya (European Social Fund FEDER and CERCA program; Barcelona Supercomputing Center MareNostrum (FI-2023-3-0024); Funded by the European Union. Views and opinions expressed are however those of the author(s) only and do not necessarily reflect those of the European Union, European Commission, European Climate, Infrastructure and Environment Executive Agency (CINEA), or any other granting authority.  Neither the European Union nor any granting authority can be held responsible for them (HORIZON-CL4-2022-QUANTUM-02-SGA  PASQuanS2.1, 101113690, EU Horizon 2020 FET-OPEN OPTOlogic, Grant No 899794, QU-ATTO, 101168628),  EU Horizon Europe Program (This project has received funding from the European Union’s Horizon Europe research and innovation program under grant agreement No 101080086 NeQSTGrant Agreement 101080086 — NeQST); ICFO Internal ``QuantumGaudi'' project;

P. Stammer acknowledges support from: The European Union’s Horizon 2020 research and innovation programme under the Marie Skłodowska-Curie grant agreement No 847517.

M.~F.~C.~acknowledges support from: the National Key Research and Development Program of China (Grant No. 2023YFA1407100), the Guangdong Province
Science and Technology Major Project (Future functional materials under extreme conditions - 2021B0301030005) and the Guangdong Natural Science Foundation (General Program project No. 2023A1515010871).
%TC:endignore
%\bibliographystyle{apsrev4-2}
\bibliography{References.bib}{}

\include{Supplementary_for_arXiv}
%TC:endignore
\end{document}

%% file: Supplementary_for_arXiv.tex
% Some commands
\newcommand{\alphapar}{\alpha_{\parallel}}
\newcommand{\betapar}{\beta_{\parallel}}
\newcommand{\alphaperp}{\alpha_{\perp}}
\newcommand{\betaperp}{\beta_{\perp}}
\newcommand{\boldpar}{\boldsymbol{\parallel}}
\newcommand{\boldperp}{\boldsymbol{\perp}}

% THE TEXT
\onecolumngrid
\begin{center}
    {\large \textbf{\textsc{Supplementary Material}}}
\end{center}

\section{Analytical description of the light-matter interaction}\label{App:LM:int}
Our analytical derivation closely follows the approach outlined in Ref.~\cite{gorlach_high-harmonic_2023}, which describes the interaction of high-intensity linearly polarized light with atomic systems, considering arbitrary photon statistics in the driving field. In this work, we consider the case of arbitrary driving field polarization and explicitly distinguish between the contributions from the horizontal ($\parallel$) and vertical ($\perp$) polarization components of the driving field. Atomic units are used throughout ($\hbar = m_{\mathsf{e}} = \abs{\mathsf{e}} = 1$ with $m_\mathsf{e}$ and $\mathsf{e}$ the electronic mass and charge, respectively).

In general, the dynamics resulting from the interaction of light with an atomic system is given by the Liouville-von Neumann equation (L-vNE)
\begin{equation}
	i \pdv{\hat{\rho}(t)}{t}
	= \comm{\hat{H}}{\hat{\rho}(t)},
\end{equation}
where $\hat{\rho}(t)$ denotes the density matrix of the joint light-matter system. Under the dipole approximation and single-active electron approximation, while working in the length gauge, the Hamiltonian characterizing the interaction can be expressed as~\cite{stammer_quantum_2023}
\begin{equation}
	\hat{H}
	= \hat{H}_{\text{at}} + \hat{\vb{d}}\cdot \hat{\vb{E}} + \hat{H}_{\text{field}},
\end{equation}
where $\hat{H}_{\text{at}}$ denotes the atomic Hamiltonian and $\hat{\vb{d}}$ is the electric dipole operator. In this expression, the electric field operator is given by $\hat{\vb{E}} = -i\sum_{\vb{k},\mu} \vb{u}_{\mu} \epsilon_{\vb{k}}(\hat{a}_{\vb{k},\mu}^\dagger - \hat{a}_{\vb{k},\mu})$, where $\mu \in \{\perp, \parallel\}$ represents the polarization direction, $\vb{u}_\mu$ is a unit vector pointing along the polarization direction $\mu$, and $\epsilon_{\vb{k}} = \sqrt{\omega_{\vb{k}}/(2 V \epsilon_0)}$ is a factor arising from the expansion of the electric field operator in terms of its normal modes, with $V$ denoting the quantization volume. The operator $\hat{a}_{\vb{k},\mu}$ ($\hat{a}^\dagger_{\vb{k},\mu}$) is the annihilation (creation) operator acting on the optical mode $(\vb{k},\mu)$. Finally, the Hamiltonian for the electromagnetic field is given by $\hat{H}_{\text{field}} = \sum_{\vb{k},\mu} \omega_{\vb{k}} \hat{a}^\dagger_{\vb{k},\mu} \hat{a}_{\vb{k},\mu}$.

In the interaction picture with respect to $\hat{H}_{\text{field}}$, the L-vNE transforms as
\begin{equation}\label{Eq:TDL:transf}
	i\pdv{\hat{\bar{\rho}}(t)}{t}
	= \comm{\hat{H}_{\text{at}} + \hat{\vb{d}}\cdot \hat{\vb{E}}(t)}{\hat{\bar{\rho}}(t)},
\end{equation}
where $\hat{\bar{\rho}}(t) \equiv e^{-i \hat{H}_{\text{field}}t}\hat{\rho}(t)e^{i \hat{H}_{\text{field}}t}$. Due to this transformation, the electric field operator acquires a time-dependence, leading to the substitution $\hat{a}_{\vb{k},\mu} \to \hat{a}_{\vb{k},\mu}e^{-i\omega_{\vb{k}} t}$. For simplicity, we denote the Hamiltonian in this picture as $\hat{H}(t) = \hat{H}_{\text{at}} + \hat{\vb{d}}\cdot \hat{\vb{E}}(t)$.

As initial conditions, we consider the electron initially occupies the ground state of the atom ($\ket{\text{g}}$), and the field is in some arbitrary state described by the density matrix $\hat{\bar{\rho}}_f(t_0)$, where in our case both polarization directions are taken into account. To describe the generic initial state of the field, we use the positive generalized $P$-representation~\cite{drummond_generalised_1980}
\begin{equation}
	\hat{\bar{\rho}}_f(t_0)
	= \int \dd^2 \alpha_\parallel
	\int \dd^2 \beta_\parallel
	P(\alpha_\parallel,\beta_\parallel^*)
	\dfrac{\dyad{\alpha_\parallel}{\beta_\parallel}}{\braket{\alpha_\parallel}{\beta_\parallel}}
	\otimes 
	\int \dd^2 \alpha_\perp
	\int \dd^2 \beta_\perp
	P(\alpha_\perp,\beta_\perp^*)
	\dfrac{\dyad{\alpha_\perp}{\beta_\perp}}{\braket{\alpha_\perp}{\beta_\perp}}
	\bigotimes_{\vb{k}\neq \vb{k}_0,\mu} \dyad{0},
\end{equation}
where we consider that the driving field mode $\vb{k}_0$ is in an arbitrary non-vacuum state, while all other modes $\vb{k}\neq \vb{k}_0$ are in the vacuum state. Additionally, we assume that the two polarization modes of $\vb{k}_0$ are in distinct states, meaning $P(\alpha_\parallel,\beta_\parallel^*) \neq P(\alpha_\perp,\beta_\perp^*)$.

To solve the differential equation above, we consider the following ansatz
\begin{equation}
	\hat{\bar{\rho}}(t)
	= \int \dd^2 \alphapar\int\dd^2 \betapar
	\int \dd^2\alphaperp\int\dd^2\betaperp
	P_{\parallel}(\alphapar,\betapar^*)
	P_{\perp}(\alphaperp,\betaperp^*)
	\hat{\rho}_{\parallel,\perp}(t),
\end{equation}
where $\hat{\bar{\rho}}_{\parallel,\perp}(t)$ is a function of the integration variables, though we omit them here for simplicity. Substituting this ansatz into Eq.~\eqref{Eq:TDL:transf}, we obtain
\begin{equation}
	\int \dd^2 \alphapar\int\dd^2 \betapar
	\int \dd^2\alphaperp\int\dd^2\betaperp
	\bigg\{
	i\hbar \pdv{\hat{\bar{\rho}}_{\parallel,\perp}(t)}{t}
	- \comm{\hat{H}(t)}{\hat{\bar{\rho}}_{\parallel,\perp}(t)}
	\bigg\}
	= 0,
\end{equation}
allowing us to solve each term in the sum independently. Consequently, we can write
\begin{equation}\label{Eq:TDLvN:par:perp}
	i \pdv{\hat{\bar{\rho}}_{\parallel,\perp}}{t}
	= \comm{\hat{H}(t)}{\hat{\bar{\rho}}_{\parallel,\perp}(t)}.
\end{equation}

At this point, we express the density matrix as
\begin{equation}
	\hat{\bar{\rho}}_{\parallel,\perp}(t)
	= \big[
	\hat{D}_{\parallel}(\alphapar)
	\otimes \hat{D}_{\perp}(\alphaperp)
	\big]
	\hat{\tilde{\rho}}_{\parallel,\perp}(t)
	\big[
	\hat{D}^\dagger_{\parallel}(\betapar)
	\otimes \hat{D}^\dagger_{\perp}(\betaperp)
	\big],
\end{equation}
where $\hat{D}_{\mu}(\alpha) = \text{exp}[\alpha\hat{a}_{\vb{k},\mu}^\dagger - \alpha^* \hat{a}_{\vb{k},\mu}]$ is the displacement operator acting on the optical mode $(\vb{k}_0,\mu)$. Substituting this expression in Eq.~\eqref{Eq:TDLvN:par:perp}, we can rewrite the latter as
\begin{equation}\label{Eq:TDLvN:par:perp:transf}
	i \pdv{\hat{\tilde{\rho}}_{\parallel,\perp}(t)}{t}
	= \hat{H}_{\boldsymbol{\alpha}}(t)\hat{\bar{\rho}}_{\parallel,\perp}(t)
	- \hat{\bar{\rho}}_{\parallel,\perp}(t) \hat{H}_{\boldsymbol{\beta}}(t),
\end{equation}
where we have defined
\begin{equation}
	\hat{H}_{\boldsymbol{\alpha}}
	= \hat{H}_{\text{at}}
	+ \hat{d}_{\parallel} E_{\alpha_{\parallel}}(t)
	+ \hat{d}_{\perp} E_{\alpha_{\perp}}(t)
	+ \hat{\vb{d}}\cdot \hat{\vb{E}}(t),
	\quad\quad
	\hat{H}_{\boldsymbol{\beta}}
	= \hat{H}_{\text{at}}
	+ \hat{d}_{\parallel} E_{\beta_{\parallel}}(t)
	+ \hat{d}_{\perp} E_{\beta_{\perp}}(t)
	+ \hat{\vb{d}}\cdot \hat{\vb{E}}(t),
\end{equation}
with $E_{\alpha_{\mu}}(t) = \text{tr}[\hat{E}_{\mu}(t) \dyad{\alpha_{\mu}}\otimes \dyad{\bar{0}}]$ (same for $\beta_\mu$), which represents the expected value of the electric field operator (along a specific polarization direction) with respect to a coherent state of amplitude $\alpha_{\mu}$ ($\beta_\mu$) for the driving field mode $(\vb{k}_0,\mu)$. In this context, a solution to Eq.~\eqref{Eq:TDLvN:par:perp:transf} can always be written as
\begin{equation}\label{Eq:sol:trans}
	\hat{\tilde{\rho}}_{\parallel,\perp}(t)
	= \hat{U}_{\boldsymbol{\alpha}}(t,t_0)
	\hat{\tilde{\rho}}_{\parallel,\perp}(t_0)
	\hat{U}^\dagger_{\boldsymbol{\beta}}(t,t_0),
\end{equation}
with $\hat{U}_{\boldsymbol{\alpha}}(t)$ satisfying the following equation
\begin{equation}
	i \pdv{\hat{U}_{\boldsymbol{\alpha}}(t)}{t}
	= \hat{H}_{\boldsymbol{\alpha}}(t) \hat{U}_{\boldsymbol{\alpha}}(t).
\end{equation}

After substituting Eq.~\eqref{Eq:sol:trans} into our ansatz, and taking into account the intermediate transformation we have performed, together with the initial condition, we obtain
\begin{equation}
	\begin{aligned}
		\hat{\rho}(t)
		&= \int \dd^2\alpha_{\parallel}\int\dd^2\beta_{\parallel}
		\int \dd^2\alpha_{\perp}\int\dd^2\beta_{\perp}
		\dfrac{P_{\parallel}(\alphapar,\betapar^*)}{\braket{\betapar}{\alphapar}}
		\dfrac{P_{\perp}(\alphaperp,\betaperp^*)}{\braket{\betaperp}{\alphaperp}}
		\\&\hspace{3cm}\times
		\big[
		\hat{D}_{\parallel}(\alphapar)
		\otimes \hat{D}_{\perp}(\alphaperp)
		\big]
		\dyad{\psi_{\boldsymbol{\alpha}}(t)}{\psi_{\boldsymbol{\beta}}(t)}
		\big[
		\hat{D}^\dagger_{\parallel}(\betapar)
		\otimes \hat{D}^\dagger_{\perp}(\betaperp)
		\big],
	\end{aligned}
\end{equation}
where $\ket{\psi_{\boldsymbol{\alpha}}(t)}$ satisfies the time-dependent Schrödinger equation
\begin{equation}
	i \pdv{\ket{\psi_{\boldsymbol{\alpha}}(t)}}{t}
	= \hat{H}_{\boldsymbol{\alpha}}(t)
	\ket{\psi_{\boldsymbol{\alpha}}(t)},
\end{equation}
which in the interaction picture with respect to the semiclassical Hamiltonian $\hat{H}^{(\boldsymbol{\alpha})}_{\text{sc}}(t) \equiv \hat{H}_{\text{at}}+ \hat{d}_{\parallel}E_{\alpha_{\parallel}}(t)+ \hat{d}_{\perp} E_{\alpha_{\perp}}(t)$, can be written as
\begin{equation}
	i\pdv{\ket{\bar{\psi}_{\boldsymbol{\alpha}}(t)}}{t}
	= \hat{\boldsymbol{d}}_{\boldsymbol{\alpha}}(t)
	\cdot \hat{\boldsymbol{E}}(t)
	\ket{\bar{\psi}_{\boldsymbol{\alpha}}(t)}.
\end{equation}

Here, $\hat{\boldsymbol{d}}_{\boldsymbol{\alpha}}(t)$ is the time-dependent dipole operator in the  interaction picture with respect to $\hat{H}^{(\boldsymbol{\alpha})}_{\text{sc}}(t)$. In the regime of extremely weak depletion~\cite{stammer_quantum_2023,stammer_squeezing_2023}, the final state can be well approximated (in the laboratory frame) as
\begin{equation}\label{Eq:App:HHG:state}
	\ket{\bar{\psi}_{\boldsymbol{\alpha}}(t)}
	\approx \ket{\phi_{\boldsymbol{\alpha}}(t)}\bigotimes_{\vb{k}}
	\Big[
	\lvert \chi^{(\boldsymbol{\alpha})}_{\vb{k},\parallel}(t)\rangle
	\otimes
	\lvert\chi^{(\boldsymbol{\alpha})}_{\vb{k},\perp}(t)\rangle
	\Big],
\end{equation}
where $\ket{\phi_{\boldsymbol{\alpha}}(t)}$ represents the solution to
\begin{equation}
	i\pdv{\ket{\phi_{\boldsymbol{\alpha}}(t)}}{t}
	= \hat{H}^{(\boldsymbol{\alpha})}_{\text{sc}}(t) \ket{\phi_{\boldsymbol{\alpha}}(t)}.
\end{equation}

Equation~\eqref{Eq:App:HHG:state} represents a product state between the electronic and quantum optical degrees of freedom, with the latter being excited to a set of coherent states~\cite{lewenstein_generation_2021,rivera-dean_strong_2022,stammer_quantum_2023}. Specifically, this yields a product state structure involving various coherent states: the coherent state in mode $\vb{k}_0$ represents a depletion of the driving field mode, while those in modes $\vb{k}\neq \vb{k}_0$ signify excitations of the different harmonic modes. These coherent state amplitudes are given by
\begin{align}\label{Eq:Coh:State:Amplit}
	&\chi^{(\boldsymbol{\alpha})}_{\vb{k},\mu}(t)
	= \epsilon_{\vb{k}}
	\int^{t}_{t_0} \dd \tau 
	\bra{\text{g}}{\hat{d}_{\boldsymbol{\alpha},\mu}(\tau)}\ket{\text{g}}e^{-i\omega_{\vb{k}} \tau},
\end{align}
while for the electronic state we have
\begin{equation}
	\ket{\phi_{\boldsymbol{\alpha}}(t)} = \hat{U}_{\text{sc},\boldsymbol{\alpha}}(t)\ket{\text{g}}.    
\end{equation}

Therefore, the final state of the joint electron-light system is given by
\begin{equation}
	\begin{aligned}
		\hat{\bar{\rho}}(t)
		&= \int \dd^2\alpha_{\parallel}\int \dd^2\beta_{\parallel}
		\int \dd^2\alpha_{\perp}\int\dd^2\beta_{\perp}
		\dfrac{P_{\parallel}(\alphapar,\betapar^*)}{\braket{\betapar}{\alphapar}}
		\dfrac{P_{\perp}(\alphaperp,\betaperp^*)}{\braket{\betaperp}{\alphaperp}}
		\\&\hspace{3cm}\times
		\dyad{\phi_{\boldsymbol{\alpha}}(t)}{\phi_{\boldsymbol{\beta}}(t)}
		\big[
		\hat{D}_{\parallel}(\alphapar)
		\otimes \hat{D}_{\perp}(\alphaperp)
		\big]
		\bigotimes_{\vb{k}}
		\dyad{\chi_{\boldsymbol{\alpha},\vb{k}}(t)}{\chi_{\boldsymbol{\beta},\vb{k}}(t)}
		\big[
		\hat{D}^\dagger_{\parallel}(\betapar)
		\otimes \hat{D}^\dagger_{\perp}(\betaperp)
		\big].
	\end{aligned}
\end{equation}
where we define $\ket{\chi_{\boldsymbol{\alpha},\vb{k}}(t)} \equiv 	\lvert\chi^{(\boldsymbol{\alpha})}_{\vb{k},\parallel}(t)\rangle \otimes\lvert\chi^{(\boldsymbol{\alpha})}_{\vb{k},\perp}(t)\rangle$ as a shorthand notation.~Finally, and before proceeding further, we trace out the $\vb{k}_0$ degrees of freedom of the driving field. Assuming that the depletion of the fundamental mode is weak enough to approximate $\bra{\chi_{\boldsymbol{\beta},\vb{k}_0}(t)}[\hat{D}_\parallel^\dagger(\beta_{\parallel})\otimes\hat{D}_\perp^\dagger(\beta_{\perp})][\hat{D}_\parallel(\alpha_{\parallel})\otimes\hat{D}_\perp(\alpha_{\perp})] \ket{\chi_{\boldsymbol{\alpha},\vb{k}_0}(t)} \approx  \braket{\beta_\parallel}{\alpha_\parallel}\braket{\beta_\perp}{\alpha_\perp}$, we can write the resulting state
\begin{equation}
	\begin{aligned}
		\hat{\bar{\rho}}_{\vb{k}\neq \vb{k}_0}(t)
		&= \int \dd^2\alpha_{\parallel}\int \dd^2\beta_{\parallel}
		\int \dd^2\alpha_{\perp}\int\dd^2\beta_{\perp}
		P_{\parallel}(\alphapar,\betapar^*)
		P_{\perp}(\alphaperp,\betaperp^*)
		\dyad{\phi_{\boldsymbol{\alpha}}(t)}{\phi_{\boldsymbol{\beta}}(t)}
		\bigotimes_{\vb{k}\neq \vb{k}_0}
		\dyad{\chi_{\boldsymbol{\alpha},\vb{k}}(t)}{\chi_{\boldsymbol{\beta},\vb{k}}(t)},
	\end{aligned}
\end{equation}
an expression that will become useful for the analysis performed in the next section.

\section{The high-harmonic generation spectrum and the classical limit}\label{Sec:HHG:Spec}
To compute the spectrum of the emitted light, we begin by evaluating its expected energy, given by the Hamiltonian $\hat{H}_{\text{em}} = \sum_{\vb{k}\neq\vb{k}_0,\mu} \omega_{\vb{k}} \hat{a}^\dagger_{\vb{k},\mu} \hat{a}_{\vb{k},\mu}$ (in the following we omit the ``$\neq\vb{k}_0$'' for clarity). This results in
\begin{equation}
	\begin{aligned}
		\mathcal{E} = \tr(\hat{\bar{\rho}}_{\vb{k}\neq \vb{k}_0}(t) \hat{H}_{\text{em}})
		&= \sum_{\vb{k},\mu}
		\omega_{\vb{k}}
		\tr(\hat{\bar{\rho}}_{\vb{k}\neq \vb{k}_0}(t)
		\hat{a}_{\vb{k},\mu}^\dagger \hat{a}_{\vb{k},\mu})
		\\&= \sum_{\vb{k},\mu}
		\omega_{\vb{k}}
		\int \dd^2\alpha_{\parallel}\int \dd^2\beta_{\parallel}
		\int \dd^2\alpha_{\perp}\int\dd^2\beta_{\perp}
		P_{\parallel}(\alphapar,\betapar^*)
		P_{\perp}(\alphaperp,\betaperp^*)
		\\&\hspace{6.2cm}\times
		\braket{\phi_{\boldsymbol{\beta}}(t)}{\phi_{\boldsymbol{\alpha}}(t)}
		\mel{\chi_{\boldsymbol{\beta},\vb{k}}(t)}{\hat{a}_{\vb{k},\mu}^\dagger \hat{a}_{\vb{k},\mu}}{\chi_{\boldsymbol{\alpha},\vb{k}}(t)},		
	\end{aligned}											
\end{equation}
and by defining $\ket{\chi_{\boldsymbol{\alpha},\vb{k}'\neq \vb{k}}(t)} \equiv \bigotimes_{\vb{k}'\neq\vb{k}} \ket{\chi_{\boldsymbol{\alpha},\vb{k}'}(t)}$, the expression above can be rewritten as 
\begin{equation}
	\begin{aligned}
		\mathcal{E} 
		&= \sum_{\vb{k}}
		\omega_{\vb{k}}
		\int \dd^2\alpha_{\parallel}\int \dd^2\beta_{\parallel}
		\int \dd^2\alpha_{\perp}\int\dd^2\beta_{\perp}
		P_{\parallel}(\alphapar,\betapar^*)
		P_{\perp}(\alphaperp,\betaperp^*)
		\Big\{\big[\chi^{(\boldsymbol{\beta})}_{\vb{k},\parallel}(t)]^*
		\chi^{(\boldsymbol{\alpha})}_{\vb{k},\parallel}(t)
		+\big[\chi^{(\boldsymbol{\beta})}_{\vb{k},\perp}(t)]^*
		\chi^{(\boldsymbol{\alpha})}_{\vb{k},\perp}(t)
		\Big\}
		\\&\hspace{6.3cm}\times
		\braket{\phi_{\boldsymbol{\beta}}(t)}{\phi_{\boldsymbol{\alpha}}(t)}
		\braket{\chi_{\boldsymbol{\beta},\vb{k'}\neq \vb{k}}(t)}{\chi_{\boldsymbol{\alpha},\vb{k'}\neq \vb{k}}(t)}.		
	\end{aligned}											
\end{equation}

To proceed further, we now consider the continuous-frequency limit by replacing the summation with respect to $\vb{k}$ by
\begin{equation}
	\sum_{\vb{k}} \to
	\dfrac{V}{(2\pi)^3}\int \dd \omega\ \omega^2\int \dd \Omega,
\end{equation}
where $\dd\Omega$ represents the infinitesimal solid angle element. After doing so, and expanding the coherent state amplitudes using the definition in Eq.~\eqref{Eq:Coh:State:Amplit}
\begin{equation}
	\begin{aligned}
		\mathcal{E} 
		&\propto  \int \dd \omega\  \omega^4
		\int \dd^2\alpha_{\parallel}\int \dd^2\beta_{\parallel}
		\int \dd^2\alpha_{\perp}\int\dd^2\beta_{\perp}
		P_{\parallel}(\alphapar,\betapar^*)
		P_{\perp}(\alphaperp,\betaperp^*)
		\big\{
		[d_{\boldsymbol{\beta},\parallel}(\omega)]^*
		d_{\boldsymbol{\alpha},\parallel}(\omega)
		+ 	[d_{\boldsymbol{\beta},\perp}(\omega)]^*
		d_{\boldsymbol{\alpha},\perp}(\omega)
		\big\}
		\\&\hspace{6.7cm}\times
		\braket{\phi_{\boldsymbol{\beta}}(t)}{\phi_{\boldsymbol{\alpha}}(t)}
		\braket{\chi_{\boldsymbol{\beta},\vb{k'}\neq \vb{k}}(t)}{\chi_{\boldsymbol{\alpha},\vb{k'}\neq \vb{k}}(t)},		
	\end{aligned}
\end{equation}
where $d_{\boldsymbol{\alpha},\mu}(\omega)$ denotes the Fourier transform of $ \langle \text{g}\vert \hat{d}_{\boldsymbol{\alpha},\mu}(t)\vert \text{g} \rangle$.~Therefore, the HHG spectrum $S(\omega) = \dv*{\mathcal{E}}{\omega}$ is given by
\begin{equation}\label{Eq:HHG:Spec}
	\begin{aligned}
		S(\omega)
		&\propto   \omega^4
		\int \dd^2\alpha_{\parallel}\int \dd^2\beta_{\parallel}
		\int \dd^2\alpha_{\perp}\int\dd^2\beta_{\perp}
		P_{\parallel}(\alphapar,\betapar^*)
		P_{\perp}(\alphaperp,\betaperp^*)
		\big\{
		[d_{\boldsymbol{\beta},\parallel}(\omega)]^*
		d_{\boldsymbol{\alpha},\parallel}(\omega)
		+ 	[d_{\boldsymbol{\beta},\perp}(\omega)]^*
		d_{\boldsymbol{\alpha},\perp}(\omega)
		\big\}
		\\&\hspace{5.7cm}\times
		\braket{\phi_{\boldsymbol{\beta}}(t)}{\phi_{\boldsymbol{\alpha}}(t)}
		\braket{\chi_{\boldsymbol{\beta},\vb{k'}\neq \vb{k}}(t)}{\chi_{\boldsymbol{\alpha},\vb{k'}\neq \vb{k}}(t)}.		
	\end{aligned}
\end{equation}

While this expression is quite cumbersome, its evaluation can be significantly simplified when working in the classical limit, also referred to as \emph{small single photon field amplitude} limit in Ref.~\cite{gorlach_high-harmonic_2023}. In this limit, we express the coherent state amplitude $\alpha$ for mode $\vb{k}_0$ in terms of the field amplitude $\varepsilon_{\alpha} = 2 \epsilon_{\vb{k}_0} \alpha \equiv 2 \epsilon \alpha $, while taking the limit of $\epsilon \to 0$, corresponding to $V\to \infty$---given that we consider electromagnetic fields propagating in free space~\cite{tannoudji_classical_1997}---and $\alpha \to \infty$---since we are dealing with fields containing and extremely large number of photons---in such a way that $\varepsilon_\alpha$ remains constant. It is important to note that this limit only affects the $P_{\parallel}(\alpha_{\parallel},\beta_{\parallel}^*)$ and $P_{\perp}(\alpha_{\perp},\beta_{\perp}^*)$, as both the quantum state of the electron $\ket{\phi_{\boldsymbol{\alpha}}(t)}$ and the Fourier transform of the time-dependent dipole moment depend on the field strength $\varepsilon_{\alpha}$.

With this in mind, and noting that for an arbitrary quantum state of light the $P_{\mu}(\alpha_\mu,\beta_\mu^*)$ functions can be expressed as
\begin{equation}
	P_{\mu}(\alpha_\mu,\beta_\mu^*)
	= \dfrac{1}{4\pi}
	\exp[-\dfrac{\abs{\alpha - \beta^*}^2}{4}]
	Q\bigg(\dfrac{\alpha+\beta^*}{2}\bigg),
\end{equation}
where $Q(\cdot)$ is the Husimi function of our quantum optical state~\cite{husimi_formal_1940,ScullyBookCh3}, Eq.~\eqref{Eq:HHG:Spec} can be rewritten as
\begin{equation}\label{Eq:HHG:Spec:Fields}
	\begin{aligned}
		S(\omega)
		&\propto  
		\dfrac{\omega^4}{(16\epsilon^4)^2}
		\int \dd^2\varepsilon_{\alpha,\parallel}\int \dd^2\varepsilon_{\beta,\parallel}
		\int \dd^2\varepsilon_{\alpha,\perp}\int \dd^2\varepsilon_{\beta,\perp}
		\bigg[	
		P_{\parallel}(\varepsilon_{\alpha,\parallel},\varepsilon_{\beta,\parallel}^*)
		P_{\perp}(\varepsilon_{\alpha,\perp},\varepsilon_{\beta,\perp}^*)
		\\&\hspace{2cm}\times
		\big\{
		[d_{\boldsymbol{\beta},\parallel}(\omega)]^*
		d_{\boldsymbol{\alpha},\parallel}(\omega)
		+ 	[d_{\boldsymbol{\beta},\perp}(\omega)]^*
		d_{\boldsymbol{\alpha},\perp}(\omega)
		\big\}
		\braket{\phi_{\boldsymbol{\beta}}(t)}{\phi_{\boldsymbol{\alpha}}(t)}
		\braket{\chi_{\boldsymbol{\beta},\vb{k'}\neq \vb{k}}(t)}{\chi_{\boldsymbol{\alpha},\vb{k'}\neq \vb{k}}(t)}
		\bigg].
	\end{aligned}
\end{equation}

In this case we have
\begin{equation}
	P_{\mu}(\varepsilon_{\alpha,\mu},\varepsilon_{\beta,\mu}^*)
	= \dfrac{1}{4\pi}
	\exp[-\dfrac{\big\lvert\varepsilon_{\alpha,\mu} -\varepsilon_{\beta,\mu}^*\big\rvert^2}{16\epsilon^2}
	]
	Q\bigg(\dfrac{\varepsilon_{\alpha,\mu} +\varepsilon_{\beta,\mu}^*}{4\epsilon}\bigg),
\end{equation}
where, without loss of generality, we assume that $\epsilon \in \mathbbm{R}$, as any phase associated with this quantity can always be absorbed into $\alpha$.

Next, we evaluate how the classical limit affects the functions $	P_{\parallel}(\varepsilon_{\alpha,\parallel},\varepsilon_{\beta,\parallel}^*)$ and $P_{\perp}(\varepsilon_{\alpha,\perp},\varepsilon_{\beta,\perp}^*)$. In our case, the quantum optical states we are dealing with have Husimi function with a Gaussian form---specifically coherent, squeezed and thermal states---which are well-behaved functions.~Therefore, the limit of interest can be evaluated separately for the vertical and horizontal polarization components.~In the following subsections, we evaluate how this limit applies to each case, and conclude by showing how the limit impacts the HHG spectrum depending on the initial state considered. More specifically, we evaluate 
\begin{align}
	S(\omega) &\propto
	\lim_{\epsilon \to 0}
	\dfrac{\omega^4}{(16\epsilon^4)^2}
	\int \dd^2\varepsilon_{\alpha,\parallel}\int \dd^2\varepsilon_{\beta,\parallel}
	\int \dd^2\varepsilon_{\alpha,\perp}\int \dd^2\varepsilon_{\beta,\perp}
	\bigg[	
	P_{\parallel}(\varepsilon_{\alpha,\parallel},\varepsilon_{\beta,\parallel}^*)
	P_{\perp}(\varepsilon_{\alpha,\perp},\varepsilon_{\beta,\perp}^*)\nonumber
	\\&\hspace{2cm}\times
	\big\{
	[d_{\boldsymbol{\beta},\parallel}(\omega)]^*
	d_{\boldsymbol{\alpha},\parallel}(\omega)
	+ 	[d_{\boldsymbol{\beta},\perp}(\omega)]^*
	d_{\boldsymbol{\alpha},\perp}(\omega)
	\big\}
	\braket{\phi_{\boldsymbol{\beta}}(t)}{\phi_{\boldsymbol{\alpha}}(t)}
	\braket{\chi_{\boldsymbol{\beta},\vb{k'}\neq \vb{k}}(t)}{\chi_{\boldsymbol{\alpha},\vb{k'}\neq \vb{k}}(t)}
	\bigg]. 
	\\&=
	\dfrac{\omega^4}{(16\epsilon^4)^2}
	\int \dd^2\varepsilon_{\alpha,\parallel}\int \dd^2\varepsilon_{\beta,\parallel}
	\int \dd^2\varepsilon_{\alpha,\perp}\int \dd^2\varepsilon_{\beta,\perp}
	\lim_{\epsilon\to 0}
	\bigg[	
	P_{\parallel}(\varepsilon_{\alpha,\parallel},\varepsilon_{\beta,\parallel}^*)
	P_{\perp}(\varepsilon_{\alpha,\perp},\varepsilon_{\beta,\perp}^*)\nonumber
	\\&\hspace{2cm}\times
	\big\{
	[d_{\boldsymbol{\beta},\parallel}(\omega)]^*
	d_{\boldsymbol{\alpha},\parallel}(\omega)
	+ 	[d_{\boldsymbol{\beta},\perp}(\omega)]^*
	d_{\boldsymbol{\alpha},\perp}(\omega)
	\big\}
	\braket{\phi_{\boldsymbol{\beta}}(t)}{\phi_{\boldsymbol{\alpha}}(t)}
	\braket{\chi_{\boldsymbol{\beta},\vb{k'}\neq \vb{k}}(t)}{\chi_{\boldsymbol{\alpha},\vb{k'}\neq \vb{k}}(t)}
	\bigg].
\end{align}

As mentioned earlier, both the dipoles and the electronic wavefunction properties depend on the field strength, and therefore will not be affected by the classical limit. For this reason, in the following we evaluate how this limit affects the behavior of the $P$-functions. Thus, in the next sections we evaluate
\begin{equation}
	L = \lim_{\epsilon\to 0}
	\Big\{
	P_{\parallel}(\varepsilon_{\alpha,\parallel},\varepsilon_{\beta,\parallel}^*)
	P_{\perp}     (\varepsilon_{\alpha,\perp},\varepsilon_{\beta,\perp}^*)
	\Big\}.
\end{equation}

\subsection{Classical limit applied to coherent states}
In general, the Husimi function of a given quantum optical state $\hat{\rho}$ is given by $Q(\alpha) = (1/\pi) \mel{\alpha}{\hat{\rho}}{\alpha}$. For the case of a coherent state of light with amplitude $\bar{\alpha}$, this results in
\begin{equation}\label{Eq:Husimi:func:coh}
	Q(\alpha) = \dfrac{1}{\pi} \exp[-\abs{\alpha - \bar{\alpha}}^2],
\end{equation}
such that the limit we are interested in becomes
\begin{equation}
	\begin{aligned}
		L^{(\text{coh})}
		&= \lim_{\epsilon \to 0}
		\Bigg\{
		\dfrac{1}{16 \epsilon^4}
		\dfrac{1}{4\pi^2}
		\exp[-\dfrac{\lvert\varepsilon_{\alpha,\mu} -\varepsilon_{\beta,\mu}^*\rvert^2}{16\epsilon^2}]
		\exp[
		-\dfrac{\lvert\varepsilon_{\alpha,\mu} +\varepsilon_{\beta,\mu}^* - 2\bar{\varepsilon}_{\mu}\rvert^2}{16\epsilon^2}
		]
		\Bigg\}
		\\&=  \lim_{\epsilon \to 0}
		\Bigg\{
		\dfrac{1}{16 \epsilon^4}
		\dfrac{1}{4\pi^2}
		\exp[-\dfrac{1}{8\epsilon^2}
		\Big(
		\lvert\varepsilon_{\alpha,\mu} - \bar{\varepsilon}_{\mu}\rvert^2
		+ \lvert\varepsilon_{\beta,\mu} - \bar{\varepsilon}_{\mu}\rvert^2
		\Big)]
		\Bigg\},
	\end{aligned}
\end{equation}
where, in going from the first to the second equality, we have algebraically expanded and combined the two exponents. In this expression, we have defined $\bar{\varepsilon}_\mu = 2\epsilon\bar{\alpha}_{\mu}$ as the mean electric field amplitude of the driving field. In the following, we denote $\bar{\varepsilon}_{\mu} = \bar{\varepsilon}^{(x)}_{\mu} + i\bar{\varepsilon}^{(y)}_{\mu}$.

At this point, it is important to note that $P_{\mu}(\varepsilon_{\alpha,\mu}, \varepsilon_{\beta,\mu}^*)$ is a four-dimensional function, as both $\varepsilon_{\alpha,\mu}$ and $\varepsilon_{\beta,\mu}$ are complex-valued quantities. This motivates us to define $\sigma = 4 \epsilon^2$, allowing us to express the limit above as
\begin{equation}
	L^{(\text{coh})}
	= \lim_{\epsilon \to 0}
	\Bigg\{
	\dfrac{1}{4\pi^2\sigma^2}
	\exp[-\dfrac{1}{2\sigma}
	\Big(
	\lvert\varepsilon_{\alpha,\mu} - \bar{\varepsilon}_{\mu}\rvert^2
	+ \lvert\varepsilon_{\beta,\mu} - \bar{\varepsilon}_{\mu}\rvert^2
	\Big)]
	\Bigg\}
	= \delta(\varepsilon_{\alpha,\mu} - \bar{\varepsilon}_{\mu})
	\delta(\varepsilon_{\beta,\mu} - \bar{\varepsilon}_{\mu}),
\end{equation}
where $\delta(\cdot)$ denotes the Dirac delta function. By introducing this result into Eq.~\eqref{Eq:HHG:Spec:Fields}, and considering the coherent$_\parallel$ +  coherent$_\perp$ driving field configuration, we obtain
\begin{equation}
	S(\omega)
	\propto \omega^4
	\big[
	\lvert d_{\bar{\boldsymbol{\varepsilon}},\parallel}(\omega)\rvert^2
	+ \lvert d_{\bar{\boldsymbol{\varepsilon}},\perp}(\omega)\rvert^2
	\big].
\end{equation}

\begin{figure}
	\centering
	\includegraphics[width=0.9\textwidth]{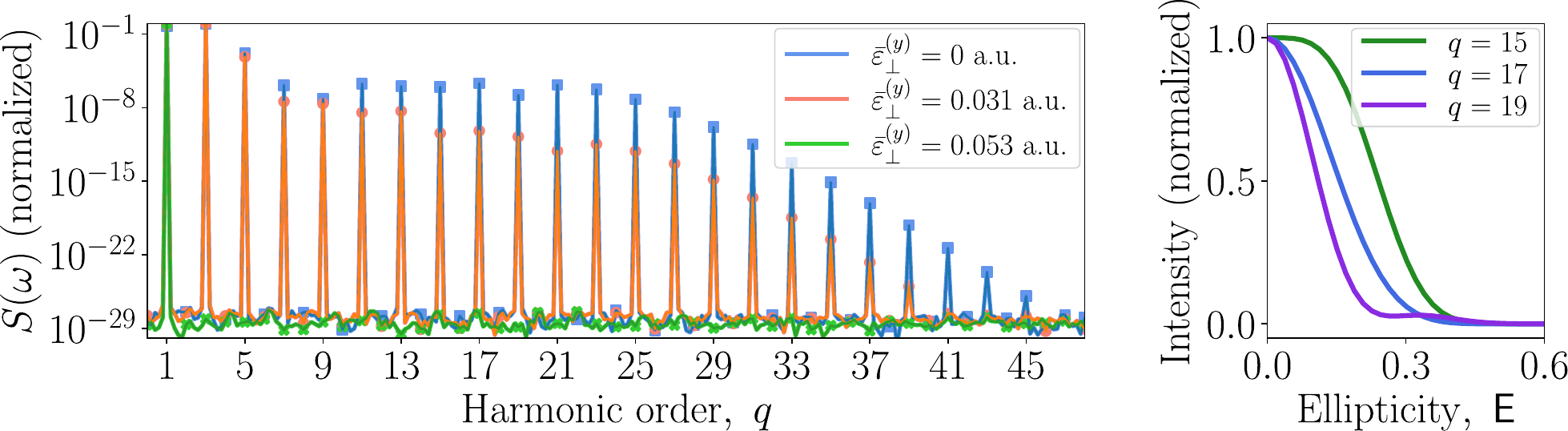}
	\caption{HHG spectrum for a coherent$_{\parallel}$ + coherent$_{\perp}$ driving field, where the coherent state amplitude of the $\perp$-polarization component is modified, ranging from linear polarization ($\bar{\varepsilon}^{(y)}_{\perp} = 0$ a.u., $\mathsf{E} = 0$) to circular polarization ($\bar{\varepsilon}^{(y)}_{\perp} = 0.053$ a.u., $\mathsf{E} = 1$). The other field components are set to $\bar{\varepsilon}^{(x)}_{\parallel} = 0.053$ a.u., $\bar{\varepsilon}^{(y)}_{\parallel} = 0$ a.u. and $\bar{\varepsilon}^{(x)}_{\perp} = 0$ a.u., consequently. }
	\label{Fig:App:HHG:Spec:Coh}
\end{figure}

This corresponds to the semiclassical expression for the HHG spectrum, neglecting contributions from dipole moment correlations~\cite{sundaram_high-order_1990}. For a circularly polarized laser field, both vertical and horizontal components vanish for all harmonic modes, resulting in the disappearance of the HHG spectrum. This is shown in Fig.~\ref{Fig:App:HHG:Spec:Coh}, where the left subplot displays HHG spectra for various values of $\bar{\varepsilon}^{(y)}_{\perp}$ while setting $\bar{\varepsilon}^{(x)}_{\parallel} = 0.053$ a.u., $\bar{\varepsilon}^{(y)}_{\parallel} = 0$ a.u. and $\bar{\varepsilon}^{(x)}_{\perp} = 0$ a.u. for the other field components. In the right subplot, we examine the dependence of high-harmonic orders on the ellipticity $\mathsf{E}$, with $\mathsf{E} = 0$ representing linear polarization, and $\mathsf{E} \in (0,1]$ denoting elliptical polarization.

\subsection{Classical limit applied to displaced squeezed vacuum states}\label{Sec:DSV:th:lim}
Here, we consider the case of displaced squeezed vacuum (DSV) states, i.e., $\ket{r,\bar{\alpha}} =  \hat{D}(\bar{\alpha}) \hat{S}(r) \ket{0}$, where $r$ denotes the squeezing parameter and $\hat{S}(r) = \exp[r^*\hat{a}^2 - r\hat{a}^{\dagger 2}]$ is the squeezing operator. In general, $r \in \mathbbm{C}$ and depending on its specific value, we would obtain squeezing along different possible directions in phase space. In our analysis, we restrict to $r\in \mathbbm{R}$, with the specific value of $r$---relative to the coherent state phase---determining the different kind of squeezing, that is, whether in phase or amplitude. For this kind of states, the Husimi function reads
\begin{equation}
	Q(\alpha)
	= \dfrac{1}{\pi \cosh(r)}
	\exp[- \dfrac{2(\alpha_x - \bar{\alpha}_x)^2}{1+e^{2r}} - \dfrac{2(\alpha_y - \bar{\alpha}_y)^2}{1+e^{-2r}}],
\end{equation}
where we denote $\alpha = \alpha_x + i \alpha_y$.

When dealing with coherent states, the classical limit basically resulted in that the variance along both optical quadratures tend to zero. For the case of DSV states, we instead have that the variance along both optical quadratures behave differently depending on the value of $r$. This difference must be reflected in the analytical expressions through a dependence of the squeezing parameter on $\epsilon$. Otherwise, if $r$ is treated as a constant, then in the classical limit coherent states and DSV states would be essentially the same, which is not the case. Thus, to relate the squeezing parameter with the quantization volume, we compute the intensity of a DSV, with the intensity operator given by $\hat{I} = \epsilon^2 \hat{a}^\dagger \hat{a}$. Note that the introduction of $\epsilon$ here is needed to provide the intensity its proper units. This results in 
\begin{equation}
	\expval{I} = \epsilon^2 \abs{\bar{\alpha}}^2 + \epsilon^2 \sinh[2](r) \equiv I_{\text{coh}} + I_{\text{squ}},
\end{equation}
where $I_{\text{squ}}$ represents the intensity contribution to the total intensity due to the squeezing. From this, we can express the squeezing parameter as
\begin{equation}
	r = \sinh[-1](\pm \dfrac{\sqrt{I_{\text{squ}}}}{\epsilon}),
\end{equation} 
with the $\pm$ sign denoting the optical quadrature along which squeezing takes place. For the sake of clarity, in the following we consider $r>0$ and just keep the positive sign, though we generalize the results later to $r<0$ as well.

Before proceeding further, let us first rewrite the Husimi function as
\begin{align}
	Q\bigg(
	\dfrac{\varepsilon_{\alpha,\mu} + \varepsilon^*_{\beta,\mu}}{4 \epsilon}
	\bigg)
	&= \dfrac{1}{\pi\cosh r}
	\exp[-\dfrac{1}{2(1+e^{2r})\epsilon^2}
	\bigg(
	\dfrac{\varepsilon^{(x)}_{\alpha,\mu} + [\varepsilon^{(x)}_{\beta,\mu}]^* }{2}
	- \bar{\varepsilon}_{\mu}^{(x)}
	\bigg)^2]\nonumber
	\\&\quad\times
	\exp[-\dfrac{1}{2(1+e^{-2r})\epsilon^2}
	\bigg(
	\dfrac{\varepsilon^{(y)}_{\alpha,\mu} + [\varepsilon^{(y)}_{\beta,\mu}]^* }{2}
	- \bar{\varepsilon}_{\mu}^{(y)}
	\bigg)^2]
	\\&= \dfrac{1}{\pi\cosh r}
	\exp[-\dfrac{1}{4e^{r}\cosh(r)\epsilon^2}
	\bigg(
	\dfrac{\varepsilon^{(x)}_{\alpha,\mu} + [\varepsilon^{(x)}_{\beta,\mu}]^* }{2}
	- \bar{\varepsilon}_{\mu}^{(x)}
	\bigg)^2
	]\nonumber
	\\&\quad\times
	\exp[-\dfrac{1}{4e^{-r}\cosh(r)\epsilon^2}
	\bigg(
	\dfrac{\varepsilon^{(y)}_{\alpha,\mu} + [\varepsilon^{(y)}_{\beta,\mu}]^* }{2}
	- \bar{\varepsilon}_{\mu}^{(y)}
	\bigg)^2]
	\\&= 
	\dfrac{2\epsilon^2}{\pi \sqrt{\sigma_x\sigma_y}}
	\exp[-\dfrac{1}{2\sigma_x}
	\bigg(
	\dfrac{\varepsilon^{(x)}_{\alpha,\mu} + [\varepsilon^{(x)}_{\beta,\mu}]^* }{2}
	- \bar{\varepsilon}_{\mu}^{(x)}
	\bigg)^2
	]
	\exp[-\dfrac{1}{2\sigma_y}
	\bigg(
	\dfrac{\varepsilon^{(y)}_{\alpha,\mu} + [\varepsilon^{(y)}_{\beta,\mu}]^* }{2}
	- \bar{\varepsilon}_{\mu}^{(y)}
	\bigg)^2],
\end{align}
where in the last equality we defined $\sigma_x = 2 e^r \cosh(r) \epsilon^2$ and $\sigma_y = 2e^{-r}\cosh(r)\epsilon^2$. Now, taking into account the following mathematical relations
\begin{equation}
	e^{\pm \sinh^{-1}(x)} = \sqrt{x^2+1} \pm x,
	\quad
	\cosh(\sinh[-1](x))	= \sqrt{x^2+1},
\end{equation}
we can express the $\sigma_x$ and $\sigma_y$ as
\begin{align}
	&\sigma_x 
	= 2 \Big[
	\sqrt{I_{\text{squ}}+\epsilon^2} + \sqrt{I_{\text{squ}}}
	\Big]
	\sqrt{I_{\text{squ}}+\epsilon^2},
	\\&\sigma_y 
	= 2 \Big[
	\sqrt{I_{\text{squ}}+\epsilon^2} - \sqrt{I_{\text{squ}}}
	\Big]
	\sqrt{I_{\text{squ}}+\epsilon^2}.
\end{align}

Thus, in the classical limit, we observe that $\sigma_x \to 4I_{\text{squ}}$ while $\sigma_y \to 0$. Consequently, we formally arrive to
\begin{align}
	L^{(\text{squ})}_{r>0}
	&=  \lim_{\epsilon \to 0}
	\Bigg\{
	\dfrac{1}{2\pi\sigma}
	\exp[-\dfrac{\lvert\varepsilon_{\alpha,\mu} -\varepsilon_{\beta,\mu}^*\rvert^2}{2\sigma}]
	\dfrac{1}{\sqrt{2\pi\sigma_x}}
	\exp[-\dfrac{1}{2\sigma_x}
	\bigg(
	\dfrac{\varepsilon^{(x)}_{\alpha,\mu} + [\varepsilon^{(x)}_{\beta,\mu}]^* }{2}
	- \bar{\varepsilon}_{\mu}^{(x)}
	\bigg)^2\nonumber
	]
	\\&\hspace{1.5cm}\times
	\dfrac{1}{\sqrt{2\pi\sigma_y}}
	\exp[-\dfrac{1}{2\sigma_y}
	\bigg(
	\dfrac{\varepsilon^{(y)}_{\alpha,\mu} + [\varepsilon^{(y)}_{\beta,\mu}]^* }{2}
	- \bar{\varepsilon}_{\mu}^{(y)}
	\bigg)^2]
	\Bigg\},
	\\&= \delta(\varepsilon_{\alpha,\mu}-\varepsilon^*_{\beta,\mu})
	\dfrac{1}{\sqrt{2\pi\sigma_x}}
	\exp[-\dfrac{1}{2\sigma_x}
	\bigg(
	\dfrac{\varepsilon^{(x)}_{\alpha,\mu} + [\varepsilon^{(x)}_{\beta,\mu}]^* }{2}
	- \bar{\varepsilon}_{\mu}^{(x)}
	\bigg)^2
	]
	\delta
	\bigg(
	\dfrac{\varepsilon^{(y)}_{\alpha,\mu} + [\varepsilon^{(y)}_{\beta,\mu}]^* }{2}
	- \bar{\varepsilon}_{\mu}^{(y)}
	\bigg),
\end{align} 
where we defined $\sigma = 8 \epsilon^2$. Similarly, for the $r<0$ case we observe 
\begin{equation}
	L^{(\text{squ})}_{r<0}
	=  \delta(\varepsilon_{\alpha,\mu}-\varepsilon^*_{\beta,\mu})
	\delta
	\bigg(
	\dfrac{\varepsilon^{(x)}_{\alpha,\mu} + [\varepsilon^{(x)}_{\beta,\mu}]^* }{2}
	- \bar{\varepsilon}_{\mu}^{(x)}
	\bigg)^2
	\dfrac{1}{\sqrt{2\pi\sigma_y}}
	\exp[-\dfrac{1}{2\sigma_y}
	\bigg(
	\dfrac{\varepsilon^{(y)}_{\alpha,\mu} + [\varepsilon^{(y)}_{\beta,\mu}]^* }{2}
	- \bar{\varepsilon}_{\mu}^{(y)}
	\bigg)^2
	].
\end{equation}

If we assume that along the $\parallel$-polarization component we have a coherent state, while along the $\perp$-polarization we have a squeezed state, we arrive to the following expressions for the HHG spectrum
\begin{align}
	&S_{r>0}(\omega)
	\propto 
	\dfrac{\omega^4}{\sqrt{2\pi\sigma_x}}
	\int \dd \varepsilon_{\alpha,\perp}^{(x)}
	\Bigg\{
	\exp[-\dfrac{\big(\varepsilon^{(x)}_{\alpha,\perp} - \bar{\varepsilon}_{\perp}^{(x)}\big)^2}{2\sigma_x}]
	\big[
	\lvert d_{\bar{\boldsymbol{\varepsilon}},\parallel}(\omega)\rvert^2
	+ \lvert d_{\bar{\boldsymbol{\varepsilon}},\perp}(\omega)\rvert^2
	\big]
	\Bigg\},\label{Eq:App:Spectrum:x:DSV}
	\\&
	S_{r<0}(\omega)
	\propto 
	\dfrac{\omega^4}{\sqrt{2\pi\sigma_y}}
	\int \dd \varepsilon_{\alpha,\perp}^{(y)}
	\Bigg\{
	\exp[-\dfrac{\big(\varepsilon^{(y)}_{\alpha,\perp} - \bar{\varepsilon}_{\perp}^{(y)}\big)^2}{2\sigma_y}]
	\big[
	\lvert d_{\bar{\boldsymbol{\varepsilon}},\parallel}(\omega)\rvert^2
	+ \lvert d_{\bar{\boldsymbol{\varepsilon}},\perp}(\omega)\rvert^2
	\big]
	\Bigg\},\label{Eq:App:Spectrum:y:DSV}
\end{align}
which can be understood as an average of the HHG spectrum over a marginal of the Husimi function for squeezed states. It is important to note that, for Eq.~\eqref{Eq:App:Spectrum:x:DSV}, $\bar{\boldsymbol{\varepsilon}} = (\bar{\varepsilon}^{(x)}_\parallel + i \bar{\varepsilon}^{(y)}_\parallel,\varepsilon^{(x)}_{\alpha,\perp} + i \bar{\varepsilon}^{(y)}_\perp)$, which depends on the integration variable and therefore makes the HHG spectrum change along the whole range of values considered in the integral. Similarly, for Eq.~\eqref{Eq:App:Spectrum:y:DSV} we find $\bar{\boldsymbol{\varepsilon}} = (\bar{\varepsilon}^{(x)}_\parallel + i \bar{\varepsilon}^{(y)}_\parallel,\bar{\varepsilon}^{(x)}_{\perp} + i \varepsilon^{(y)}_{\alpha,\perp})$. In the context of our work, Eq.~\eqref{Eq:App:Spectrum:x:DSV} represents the spectrum for an amplitude-squeezed state, and Eq.~\eqref{Eq:App:Spectrum:y:DSV} for a phase-squeezed state.

\subsection{Classical limit applied to displaced thermal states}
Displaced thermal states are defined through the following statistical mixture
\begin{equation}
	\hat{\rho}_{\text{th}}
	= \sum_{n=0}^{\infty}
	\dfrac{n_{\text{th}}^n}{(n_{\text{th}}+1)^{n+1}}
	\hat{D}(\bar{\alpha})
	\dyad{n}
	\hat{D}^\dagger(\bar{\alpha}),
\end{equation}
where $n_{\text{th}}$ denotes the thermal fluctuations. As $n_{\text{th}}\to 0$, this expression converges to that of a coherent state of light. Similar to the DSV states, but unlike coherent states, thermal states with $n_{\text{th}}>0$ do not saturate the uncertainty product between optical quadratures. Consequently, when evaluating the classical limit, it is necessary to first establish a relation between $n_{\text{th}}$ and $\epsilon$. To do so, we proceed as in Sec.~\ref{Sec:DSV:th:lim} and calculate the average intensity of displaced thermal states
\begin{equation}
	\langle \hat{I} \rangle
	= \epsilon^2\abs{\bar{\alpha}}^2
	+ \epsilon^2 n_{\text{th}}
	\equiv I_{\text{coh}} + I_{\text{th}},
\end{equation}
from which we find the relation $I_{\text{th}} = \epsilon^2 n_{\text{th}}$. In the classical limit, we aim to keep the intensity contribution of the thermal part constant.

For displaced thermal states, the Husimi function is given by
\begin{equation}
	Q(\alpha)
	= \dfrac{1}{\pi(n_{\text{th}}+1)}
	\exp[-\dfrac{\abs{\alpha - \bar{\alpha}}^2}{n_{\text{th}}+1}],
\end{equation} 
which, expressed in terms of the thermal intensity $I_{\text{th}}$ and field amplitudes $\varepsilon_{\alpha}$, specifically using the kernel that appears in the limit we want to evaluate, we have
\begin{equation}
	\begin{aligned}
		Q\bigg(
		\dfrac{\varepsilon_{\alpha,\mu} + \varepsilon^*_{\beta,\mu}}{4 \epsilon}
		\bigg)
		&= \dfrac{\epsilon^2}{\pi(I_{\text{th}}+\epsilon^2)}
		\exp[-\dfrac{1}{4(I_{\text{th}} + \epsilon^2)}
		\abs{\dfrac{\varepsilon_{\alpha,\mu} - \varepsilon_{\beta,\mu}^*}{2} - \bar{\varepsilon}_\mu}^2]
		\\&
		= \dfrac{2\epsilon^2}{\pi\sigma_{\text{th}}}
		\exp[-\dfrac{1}{2\sigma_{\text{th}}}
		\abs{\dfrac{\varepsilon_{\alpha,\mu} - \varepsilon_{\beta,\mu}^*}{2} - \bar{\varepsilon}_\mu}^2]
	\end{aligned}	
\end{equation}
where in moving from the first to the second equality, we defined $\sigma_{\text{th}} = 2 (I_{\text{th}} + \epsilon^2)$.

In the thermodynamics limit, $\sigma_{\text{th}} \to 2 I_{\text{th}}$. Consequently, we formally obtain
\begin{align}
	L^{(\text{th})}
	&=  \lim_{\epsilon \to 0}
	\Bigg\{
	\dfrac{1}{2\pi\sigma}
	\exp[-\dfrac{\lvert\varepsilon_{\alpha,\mu} -\varepsilon_{\beta,\mu}^*\rvert^2}{2\sigma}]
	\dfrac{1}{2\pi\sigma_{\text{th}}}
	\exp[-\dfrac{1}{2\sigma_{\text{th}}}
	\abs{
		\dfrac{\varepsilon_{\alpha,\mu} - \varepsilon_{\beta,\mu}^*}{2}
		- \bar{\varepsilon}_{\mu}
	}^2\nonumber
	]
	\\&= 
	\delta(\varepsilon_{\alpha,\mu}-\varepsilon^*_{\beta,\mu})
	\dfrac{1}{4\pi I_{\text{th}}}
	\exp[-\dfrac{1}{4I_{\text{th}}}
	\abs{
		\dfrac{\varepsilon_{\alpha,\mu} - \varepsilon_{\beta,\mu}^*}{2}
		- \bar{\varepsilon}_{\mu}			
	}^2]
	\Bigg\},
\end{align}
such that in this case the HHG spectrum is given by
\begin{equation}\label{Eq:App:Spectrum:thermal}
	S(\omega)
	\propto	
	\dfrac{1}{4\pi I_{\text{th}}}
	\int \dd^2 \varepsilon_{\alpha,\perp}
	\Bigg\{
	\exp[-\dfrac{1}{4I_{\text{th}}}
	\abs{
		\varepsilon_{\alpha,\perp}
		- \bar{\varepsilon}_{\perp}			
	}^2]
	\big[
	\lvert d_{\boldsymbol{\alpha},\parallel}(\omega)\rvert^2
	+ \lvert d_{\boldsymbol{\alpha},\perp}(\omega)\rvert^2
	\big]
	\Bigg\},
\end{equation}
where it is worth noting that in the limit $I_{\text{th}}\to0$, we recover the spectrum derived earlier for a coherent state of light, as expected.

\subsection{Numerical analysis}
To evaluate the HHG spectra presented in the main text, as well as those shown in Sec.~\ref{Sec:Thermal:states:Spec}, we discretized the integral(s) with respect to $\varepsilon_{\alpha,\perp}^{(i)}$. For each field amplitude included in the sum, the HHG spectrum was computed using the \texttt{RB-SFA} \texttt{Mathematica} package~\cite{RBSFA}. For squeezed states of light, we used a 1D grid of 241 field values, though we observed that 122 values were sufficient to achieve convergence. In the case of thermal states, we computed a total of 5576 harmonic spectra to build a $76 \times 76$ grid. The obtained spectra were saved and later processed in \texttt{Python} to compute the HHG spectra for the considered states of light.

\section{Ellipticity analysis}\label{App:Ellip:Analysis}
In this section, we provide a detailed description of the ellipticity analysis for the initial driver used in the main text, consisting of a coherent state $\ket{0,\alpha_1}_{\parallel}$ along the $\parallel$-polarization and a DSV $\ket{r,\alpha_2}_{\perp}$ along the $\perp$-polarization.  To achieve this, we introduce the quantum optical analogue of the Stoke parameters
\begin{equation}
	\hat{S}_0 
	=\epsilon^2
	\big(
	\hat{a}^\dagger_{\perp}\hat{a}_{\perp}
	+ \hat{a}^\dagger_{\parallel} \hat{a}_{\parallel}
	\big),
	\quad
	\hat{S}_3 
	= i \epsilon^2
	\big(
	\hat{a}^\dagger_{\parallel} \hat{a}_{\perp}
	- \hat{a}^\dagger_{\perp} \hat{a}_{\parallel}
	\big),
\end{equation}
where the presence of the $\epsilon$ parameter provides the proper units to the intensity. This is motivated by the requirement that, for a fully coherent-state driver, $\hat{S}_0$ should represent the total field intensity, which---based on the definition of the electric field amplitude $\varepsilon_{\alpha} = 2\epsilon \alpha$---is proportional to $\epsilon^2 \abs{\alpha}^2$. 

Using these operators, we can define the ellipticity $\mathsf{E}$ of the driving field as
\begin{equation}
	\textsf{E}
	= \dfrac{\langle \hat{S}_3\rangle}{\langle \hat{S}_0\rangle}
	\equiv \dfrac{S_3}{S_0}.
\end{equation}

The ellipticity for the initial configuration introduced at the beginning of this section gives us
\begin{equation}
	\textsf{E}
	= i\dfrac{\alpha_{\parallel}^*\alpha_{\perp} - \alpha_{\parallel}\alpha_{\perp}^*}{\abs{\alpha_{\perp}}^2 + \abs{\alpha_{\parallel}}^2 + \sinh^2(r)}
	=  i\dfrac{\bar{\varepsilon}_{\parallel}^*\bar{\varepsilon}_{\perp} - \bar{\varepsilon}_{\parallel}\bar{\varepsilon}_{\perp}^*}{\abs{\bar{\varepsilon}_{\perp}}^2 + \abs{\bar{\varepsilon}_{\parallel}}^2 + 4I_\text{squ}},
\end{equation}
where it can be explicitly observed that squeezing indeed impacts the mean ellipticity. In the case where $r=0$, we recover the classical-optics expression of the field ellipticity. Furthermore, we observe that achieving circularly polarized light requires both components to have a phase difference of $\pi/2$. However, the fluctuations in ellipticity are just as important as its mean value. Given that our initial state is a product state, with expected values that can be treated as independent random variables, we can compute the variance of \textsf{E} using the variance formula. Thus, we obtain
\begin{equation}
	(\Delta \textsf{E})^2
	= \bigg(\pdv{\textsf{E}}{S_3}\bigg)^2
	(\Delta S_3)^2
	+ \bigg(\pdv{\textsf{E}}{S_0}\bigg)^2
	(\Delta S_0)^2
	= \dfrac{(\Delta S_3)^2}{S_0^2}
	+ \dfrac{S_3^2}{S_0^4}
	(\Delta S_0)^2.
\end{equation}

In the following, we evaluate the expressions of $(\Delta S_3)^2$ and $(\Delta S_0)^2$ individually. These expressions are used to assess the ellipticity and its fluctuations in Fig.~3 of the main text.

\subsubsection{Computing the variance of $\hat{S}_0$}
To start, we express $\hat{S}_0^2$ in normal order
\begin{align}
	\hat{S}_0^2
	&= \epsilon^4
	\Big[
	\big(\hat{a}^\dagger_{\perp}\hat{a}_{\perp}\big)^2
	+ 2 \big(\hat{a}^\dagger_{\perp} \hat{a}_{\perp}\big)
	\big(\hat{a}^\dagger_{\parallel} \hat{a}_{\parallel}\big)
	+ \big(\hat{a}^\dagger_{\parallel}\hat{a}_{\parallel}\big)^2
	\Big]
	\nonumber
	\\&
	=\epsilon^4
	\Big[
	\hat{a}^{\dagger 2}_{\perp} \hat{a}^2_{\perp}
	+\hat{a}^{\dagger 2}_{\parallel} \hat{a}^2_{\parallel}
	+ 2 \big(\hat{a}^\dagger_{\perp} \hat{a}_{\perp}\big)
	\big(\hat{a}^\dagger_{\parallel} \hat{a}_{\parallel}\big)
	+ \hat{a}^\dagger_{\perp}\hat{a}_{\perp}
	+ \hat{a}^\dagger_{\parallel} \hat{a}_{\parallel}
	\Big],
\end{align}
and evaluate its mean value. Let us first examine how the first term in this expression appears for a DSV state, i.e., $\hat{D}(\alpha)\hat{S}(r)\ket{0}$. To proceed, we evaluate how this term is modified under the action of the displacement and squeezing operators
\begin{align}
	\hat{S}^\dagger(r)\hat{D}^\dagger(\alpha)
	\hat{a}^{\dagger 2}_{\perp} \hat{a}^2_{\perp}
	\hat{D}(\alpha)\hat{S}(r)
	&= \hat{S}^\dagger(r)
	\big( \hat{a}^\dagger_{\perp} + \alpha^*\big)^2
	\big(\hat{a}_{\perp} + \alpha \big)^2
	\hat{S}(r)
	\\&
	= \big(
	\hat{a}^\dagger_{\perp} \cosh(r)
	- \hat{a}_{\perp}\sinh(r)
	+ \alpha_{\perp}^*
	\big)^2
	\big(
	\hat{a}_{\perp}\cosh(r)
	- \hat{a}_{\perp}^\dagger \sinh(r)
	+\alpha_{\perp}
	\big)^2,
\end{align}
and, to continue, we evaluate the action of the leftmost bracket on a vacuum state
\begin{align}
	\big(
	\hat{a}_{\perp}\cosh(r)
	- \hat{a}_{\perp}^\dagger \sinh(r)
	+\alpha_{\perp}
	\big)^2 \ket{0}
	&= \big(
	\hat{a}_{\perp}\cosh(r)
	- \hat{a}_{\perp}^\dagger \sinh(r)
	+\alpha_{\perp}
	\big)
	\big[
	- \sinh(r) \ket{1} + \alpha_{\perp} \ket{0}
	\big]
	\nonumber
	\\&
	= \big[\!
	-\cosh(r)\sinh(r)
	+  \alpha_{\perp}^2
	\big]\ket{0}
	- 2\alpha_{\perp} \sinh(r)\ket{1}
	+ \sqrt{2}\sinh^2(r) \ket{2},
\end{align}
from which we obtain
\begin{align}
	\bra{0}
	\hat{S}^\dagger(r)\hat{D}^\dagger(\alpha)
	\hat{a}^{\dagger 2}_{\perp} \hat{a}^2_{\perp}
	\hat{D}(\alpha)\hat{S}(r)
	\ket{0}
	&=
	\abs{
		-\cosh(r)\sinh(r)
		+  \alpha_{\perp}^2
	}^2
	+ 4 \abs{\alpha_{\perp}}^2\sinh^2(r)
	+ 2 \sinh^4(r)
	\nonumber
	\\&=
	\abs{\alpha_{\perp}}^4
	+ \cosh^2(r)\sinh^2(r)
	- \big(\alpha_{\perp}^2+\alpha_{\perp}^{*2}\big) \cosh(r)\sinh(r)\nonumber
	\\&\quad + 4 \abs{\alpha_{\perp}}^2\sinh^2(r)
	+ 2 \sinh^4(r).
\end{align}

Consequently, the expected value of $\hat{S}_0^2$ reads
\begin{align}
	\langle \hat{S}_0^2\rangle
	&= \epsilon^4
	\Big[
	\abs{\alpha_{\parallel}}^4
	+\abs{\alpha_{\perp}}^4
	+ \cosh^2(r)\sinh^2(r)
	- \big(\alpha_{\perp}^2+\alpha_{\perp}^{*2}\big) \cosh(r)\sinh(r)
	+ 4 \abs{\alpha_{\perp}}^2\sinh^2(r)
	+ 2 \sinh^4(r)\nonumber
	\\&\quad+ 2 \abs{\alpha_{\parallel}}^2
	\big[
	\abs{\alpha_{\perp}}^2
	+ \sinh^2(r)
	\big]
	+ \abs{\alpha_{\perp}}^2 + \sinh^2(r)
	+ \abs{\alpha_{\parallel}}^2
	\Big],
\end{align}
while, on the other hand, we have
\begin{align}
	\langle \hat{S}_0\rangle^2
	&= \epsilon^4\big[
	\abs{\alpha_\perp}^2
	+ \abs{\alpha_\parallel}^2
	+ \sinh^2(r)
	\big]^2\nonumber
	\\&=
	\epsilon^4
	\Big[
	\abs{\alpha_{\parallel}}^4 
	+2 \abs{\alpha_{\parallel}}^2
	\big[
	\abs{\alpha_{\perp}}^2
	+ \sinh^2(r)
	\big]
	+ \abs{\alpha_{\perp}}^4
	+ \sinh^{4}(r)
	+ 2\abs{\alpha_{\perp}}^2\sinh^2(r)
	\Big],
\end{align}
such that we find for the variance
\begin{equation}
	\begin{aligned}
		\big(\Delta S_0\big)^2
		&= \epsilon^4
		\Big[
		\cosh^2(r)\sinh^2(r)
		- \big(\alpha_{\perp}^2+\alpha_{\perp}^{*2}\big) \cosh(r)\sinh(r)
		\\&\hspace{1cm}
		+ 2 \abs{\alpha_{\perp}}^2\sinh^2(r)
		+ \sinh^4(r)
		+ \abs{\alpha_{\perp}}^2 + \sinh^2(r)
		+ \abs{\alpha_{\parallel}}^2
		\Big].
	\end{aligned}
\end{equation}

Taking into account the mathematical relation $\cosh^2(r) = 1 + \sinh^2(r)$, we can rewrite the previous expression as
\begin{align}
	\big(\Delta S_0\big)^2
	&= \epsilon^4
	\bigg[
	\abs{\alpha_{\perp}}^2
	+ \abs{\alpha_{\parallel}}^2
	+ 2 \sinh^2(r)
	+ 2 \sinh^4(r)
	+ 2 \abs{\alpha_{\perp}}^2\sinh^2(r)
	- \big(
	\alpha_{\perp}^2
	+\alpha_{\perp}^{*2}
	\big)
	\sinh(r)
	\sqrt{1+\sinh^2(r)}
	\bigg],
\end{align}
which in terms of $\varepsilon_{\perp},\varepsilon_{\parallel}$ and $I_{\text{squ}}$ reads
\begin{align}
	\big(\Delta S_0\big)^2
	&= \epsilon^4 
	\bigg[
	\dfrac{\abs{\varepsilon_{\perp}}^2}{4\epsilon^2}
	+ \dfrac{\abs{\varepsilon_{\parallel}}^2}{4\epsilon^2}
	+ 2 \dfrac{I_{\text{squ}}}{\epsilon^2}
	+ 2 \dfrac{I_{\text{squ}}^2}{\epsilon^4}
	+ 2 \dfrac{\abs{\varepsilon_{\perp}}^2I_{\text{squ}}}{4\epsilon^4}
	\mp \dfrac{1}{4\epsilon^4}
	\big(
	\varepsilon_{\perp}^2
	+\varepsilon_{\perp}^{*2}
	\big)
	\sqrt{I_{\text{squ}}\epsilon^2+I^2_{\text{squ}}}
	\bigg],
\end{align}
and in the limit $\epsilon\to 0$ we arrive to
\begin{equation}
	\big(\Delta S_0\big)^2
	= 2I_{\text{squ}}^2	
	+ \dfrac12\abs{\varepsilon_{\perp}}^2 I_{\text{squ}}
	\mp \dfrac14
	\big(
	\varepsilon_{\perp}^2
	+\varepsilon_{\perp}^{*2}
	\big)
	I_{\text{squ}},
\end{equation}
with the $\mp$ depending on whether $r$ is either positive or negative, corresponding in our scenario to amplitude and phase squeezing, respectively.

\subsubsection{Computing the variance of $\hat{S}_3$}
In this case, we can write for $\hat{S}_3^2$
\begin{align}
	\hat{S}^2_3
	&= - 
	\epsilon^4
	\big(
	\hat{a}^\dagger_{\parallel} \hat{a}_{\perp}
	- \hat{a}^\dagger_{\perp} \hat{a}_{\parallel}
	\big)^2
	= \epsilon^4
	\Big[
	-\hat{a}_\parallel^{\dagger 2} \hat{a}^2_\perp
	- \hat{a}_\perp^{\dagger 2} \hat{a}_\parallel^2
	+ \big(\hat{a}^\dagger_{\parallel}\hat{a}_\parallel\big)
	\big(\hat{a}_\perp\hat{a}^\dagger_\perp\big)
	+ \big(\hat{a}^\dagger_\perp\hat{a}_\perp\big)
	\big(\hat{a}_\parallel\hat{a}^\dagger_\parallel\big)
	\Big]
	\nonumber
	\\&= 
	\epsilon^4
	\Big[
	-\hat{a}_\parallel^{\dagger 2} \hat{a}^2_\perp
	- \hat{a}_\perp^{\dagger 2} \hat{a}_\parallel^2
	+ \big(\hat{a}^\dagger_{\parallel}\hat{a}_\parallel\big)
	+ \big(\hat{a}^\dagger_\perp\hat{a}_\perp\big)
	+ 2 \big(\hat{a}^\dagger_{\parallel}\hat{a}_\parallel\big)
	\big(\hat{a}^\dagger_\perp\hat{a}_\perp\big)
	\Big],
\end{align}
an operator whose expected value reads
\begin{equation}
	\begin{aligned}
		\langle\hat{S}^2_3\rangle
		&= \epsilon^4
		\Big[
		\abs{\alpha_{\parallel}}^2
		+ \abs{\alpha_{\perp}}^2
		+ \sinh^2(r)
		+ 2 \abs{\alpha_{\parallel}}^2
		\big[
		\abs{\alpha_\perp}^2
		+ \sinh^2(r)
		\big]
		\\&
		\hspace{1cm}- \alpha_{\parallel}^{*2}
		\big[
		\alpha_{\perp}^2
		-\cosh(r)\sinh(r)
		\big]
		- \alpha_{\parallel}^{2}
		\big[
		\alpha_{\perp}^{*2}
		-\cosh(r)\sinh(r)
		\big]
		\Big],
	\end{aligned}
\end{equation}
while for the squared of the mean value we find
\begin{equation}
	\langle \hat{S}_3\rangle^2
	= \epsilon^4
	\Big[
	-\alpha_\parallel^{*2}\alpha_\perp^2
	- \alpha_\parallel^{2}\alpha_\perp^{*2}
	+ 2 \abs{\alpha_\parallel}^2\abs{\alpha_\perp}^2
	\Big],
\end{equation}
such that the variance reads
\begin{equation}
	\big(\Delta S_3\big)^2
	= \epsilon^4
	\Big[
	\abs{\alpha_{\parallel}}^2
	+ \abs{\alpha_{\perp}}^2
	+ \sinh^2(r)
	+ 2 \abs{\alpha_{\parallel}}^2\sinh^2(r)
	+ \big(
	\alpha_{\parallel}^{*2}
	+ \alpha_{\parallel}^{2}
	\big)
	\cosh(r)\sinh(r)
	\Big].
\end{equation}

Finally, expressing this quantity in terms of field amplitudes and intensities, it reads
\begin{equation}
	\big(\Delta S_3\big)^2
	=\epsilon^4
	\bigg[
	\dfrac{\abs{\varepsilon_{\parallel}}^2}{4\epsilon^2}
	+ \dfrac{\abs{\varepsilon_{\perp}}^2}{4\epsilon^2}
	+ \dfrac{I_{\text{squ}}}{\epsilon^2}
	+ 2 \dfrac{\abs{\varepsilon_{\parallel}}^2 I_{\text{squ}}}{4\epsilon^4}
	\pm \dfrac{1}{4\epsilon^4}\big(
	\varepsilon_{\parallel}^{*2}
	+ \varepsilon_{\parallel}^{2}
	\big)
	\sqrt{I_{\text{squ}}\epsilon^2+I^2_{\text{squ}}}
	\bigg],
\end{equation}
and in the limit $\epsilon\to0$ we arrive to
\begin{equation}
	\big(\Delta S_3\big)^2
	=
	\dfrac12\abs{\varepsilon_{\parallel}}^2 I_{\text{squ}}
	\pm \dfrac{1}{4}\big(
	\varepsilon_{\parallel}^{*2}
	+ \varepsilon_{\parallel}^{2}
	\big)
	I_{\text{squ}}.
\end{equation}

\begin{figure}
	\centering
	\includegraphics[width=1\textwidth]{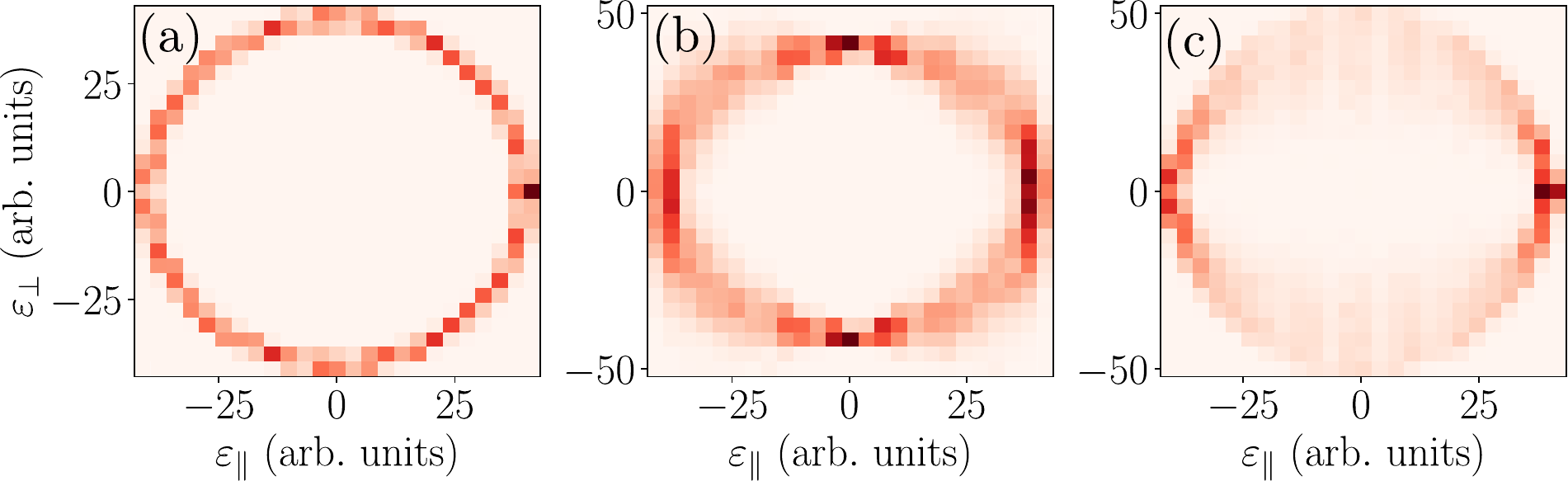}
	\caption{2D histogram of the Lissajous figures presented in Fig.~1 of the main text: (a) coherent state, (b) amplitude-squeezed state and (c) phase-squeezed states.}
	\label{Fig:2d:histogram}
\end{figure}

Figure~\ref{Fig:2d:histogram} presents a 2D histogram analogue of Fig.~2~(a–c) from the main text.~The histogram clearly shows that regions with weaker field fluctuations correspond to higher point density, whereas regions with weaker fluctuations appear more dense. For coherent states, the density remains constant for all combinations of $(\varepsilon_{\parallel},\varepsilon_{\perp})$.

\section{Modified electronic quantum orbits}\label{App:Mod:Qu:Or}
We observe that the use of photon statistics differing from those of a coherent state of light leads to HHG spectra that, in certain cases as discussed in the main text, result in non-vanishing HHG radiation even when employing circularly polarized fields. From a quantum-orbit-based perspective, this suggests that non-Poissonian photon number statistics induce recombination events, even when considering circularly polarized light drivers. In this section, following a similar approach to that in Ref.~\cite{even_tzur_photon-statistics_2023} and incorporating the expressions derived in Section~\ref{Sec:HHG:Spec}, we explicitly compute the equations that describe the electronic quantum orbits as influenced by the arbitrary photon-statistics of the driver.

\subsection{Quantum orbits when using displaced squeezed states along one of the polarization axes}
In Section~\ref{Sec:HHG:Spec}, we computed the HHG spectrum in the classical limit.~However, the approach we followed is not only applicable to the evaluation of this observable but can also be used to obtain the reduced density matrix of the electronic state. Specifically, by tracing out the quantum optical degrees of freedom and considering the classical limit, we find the reduced density matrix of the electronic state to be
\begin{equation}
	\hat{\rho}_{\text{elec}}(t)
	= \dfrac{1}{\sqrt{2\pi \sigma_i}}
	\int \dd \varepsilon^{(i)}_{\alpha,\perp}  
	\Bigg\{
	\exp[-\dfrac{\big(\varepsilon^{(i)}_{\alpha,\perp} - \bar{\varepsilon}_{\perp}^{(i)}\big)^2}{2\sigma_i}]
	\dyad{\phi_{\boldsymbol{\varepsilon_\alpha}}(t)}
	\Bigg\},
\end{equation}
where $i \in \{x,y\}$. To derive this expression, we focused on the case described in the main text, where the $\parallel$-mode is excited by a coherent state, while the $\perp$-mode by a phase- or amplitude-squeezed state. 

The concept of quantum orbits can be rigorously defined by applying the saddle-point approximation to the Fourier transform of the time-dependent dipole moment's mean value. In our case, this quantity can be expressed as
\begin{align}
	\langle \hat{\vb{d}}(\Omega)\rangle
	&= \dfrac{1}{\sqrt{2\pi \sigma_i}}
	\int \dd t_2\
	e^{i \Omega t_2}\int \dd \varepsilon^{(i)}_{\alpha,\perp}  
	\Bigg\{
	\exp[-\dfrac{\big(\varepsilon^{(i)}_{\alpha,\perp} - \bar{\varepsilon}_{\perp}^{(i)}\big)^2}{2\sigma_i}]
	\mel{\phi_{\boldsymbol{\varepsilon_\alpha}}(t)}{\vb{\hat{d}}}{\phi_{\boldsymbol{\varepsilon_\alpha}}(t)}
	\Bigg\}
	\\&\approx
	\dfrac{1}{\sqrt{2\pi \sigma_i}}
	\int \dd t_2
	\int \dd \varepsilon^{(i)}_{\alpha,\perp}  
	\Bigg\{
	\exp[-\dfrac{\big(\varepsilon^{(i)}_{\alpha,\perp} - \bar{\varepsilon}_{\perp}^{(i)}\big)^2}{2\sigma_i}]\nonumber
	\\&\hspace{2.5cm}\times
	\Bigg[
	i\int \dd \vb{p}\int^{t_2}_{t_0} \dd t_1\
	\boldsymbol{d}^*\big(\vb{p} + \vb{A}_{\boldsymbol{\varepsilon_\alpha}}(t_2)\big)
	e^{-iS_{\boldsymbol{\varepsilon_\alpha}}(\vb{p},t_2,t_1)+i\Omega t_2}
	\vb{E}_{\boldsymbol{\varepsilon_\alpha}}(t_1)
	\cdot 
	\boldsymbol{d}\big(\vb{p} + \vb{A}_{\boldsymbol{\varepsilon_\alpha}}(t_1)\big)
	+ \text{c.c.}
	\Bigg]
	\Bigg\},\label{Eq:App:QSFA}
\end{align}
where, in going from the first to the second equality, we applied the semiclassical Strong-Field Approximation~\cite{lewenstein_theory_1994,amini_symphony_2019}. Here, $\vb{A}_{\boldsymbol{\varepsilon}_\alpha}(t)$ represents the vector potential, related to the electric field operator by $\vb{E}_{\boldsymbol{\varepsilon_\alpha}}(t) = -\pdv*{\vb{A}_{\boldsymbol{\varepsilon_\alpha}}(t)}{t}$. Additionally, $\boldsymbol{d}(\vb{p}) = \mel{\vb{p}}{\hat{\vb{d}}}{\text{g}}$ denotes the transition matrix element of the dipole moment operator between the electronic ground state $\ket{\text{g}}$ and a continuum state $\ket{\vb{p}}$. Finally, $S_{\boldsymbol{\varepsilon_\alpha}}(\vb{p},t_2,t_1)$ is the semiclassical action, representing the phase accumulated by the electron between times $t_1$ and $t_2$ in the absence of interaction with the atomic core, and is given by
\begin{equation}
	S_{\boldsymbol{\varepsilon_\alpha}}(\vb{p},t_2,t_1)
	= \dfrac12 \int^{t_2}_{t_1} \dd \tau 
	\big[
	\vb{p} + \vb{A}_{\boldsymbol{\varepsilon_\alpha}}(\tau)
	\big]^2
	+ I_p (t_2 - t_1).
\end{equation}

The saddle-point method is a mathematical approach used to simplify the evaluation of integrals involving rapidly oscillating functions by approximating them as a sum over selected points, called saddle-points~(see Appendix A of Ref.~\cite{olga_simpleman} and Ref.~\cite{nayak_saddle_2019} for an introduction). This method is particularly useful for Eq.~\eqref{Eq:App:QSFA} due to the presence of high-harmonic modes in the semiclassical action. In this case, the saddle-points correspond to those of $S_{\boldsymbol{\varepsilon_\alpha}}(\vb{p},t_2,t_1)$. Beyond being an effective mathematical tool for evaluating the HHG spectrum, the saddle-point method provides valuable analytical insights into the electron dynamics, especially given that Eq.~\eqref{Eq:App:QSFA} can be viewed as an interference of different electron trajectories. Within this interpretation, saddle-points represent the electron trajectories that contribute most significantly to the HHG dynamics.

To explore how photon statistics differing from those of coherent states affect the quantum orbits, we apply the saddle-point approximation to the integral over $\varepsilon_{\alpha,\perp}^{(i)}$ by rewriting Eq.~\eqref{Eq:App:QSFA} as
\begin{align}
	\langle \hat{\vb{d}}(\Omega)\rangle
	= \dfrac{i}{\sqrt{2\pi \sigma_i}}
	\int \dd t_2
	\int \dd \varepsilon^{(i)}_{\alpha,\perp}  
	\int \dd \vb{p}
	\int_{t_0}^{t_2} \dd t_1
	\Big[
	\boldsymbol{d}^*\big(\vb{p} + \vb{A}_{\boldsymbol{\varepsilon_\alpha}}(t_2)\big)
	e^{-iS^{(\text{QO})}_{\boldsymbol{\varepsilon_\alpha}}(\vb{p},t_2,t_1)}
	\vb{E}_{\boldsymbol{\varepsilon_\alpha}}(t_1)
	\cdot 
	\boldsymbol{d}\big(\vb{p} + \vb{A}_{\boldsymbol{\varepsilon_\alpha}}(t_1)\big)
	+ \text{c.c.}
	\Big],
\end{align}
where we define
\begin{equation}\label{Eq:App:QSFA:Action}
	S^{(\text{QO})}_{\boldsymbol{\varepsilon_\alpha}}(\vb{p},t_2,t_1)
	= \dfrac12 \int^{t_2}_{t_1} \dd \tau 
	\big[
	\vb{p} + \vb{A}_{\boldsymbol{\varepsilon_\alpha}}(\tau)
	\big]^2
	+ I_p (t_2 - t_1)
	- \Omega t_2
	- \dfrac{i}{2\sigma_i}
	\Big(
	\varepsilon^{(i)}_{\alpha,\perp} - \bar{\varepsilon}_{\perp}^{(i)}
	\Big)^2,
\end{equation}
and observe that the photon statistics effectively modify the saddle-point locations, differing from those obtained in the semiclassical regime. In the following, we separately analyze the impact of amplitude squeezing (obtained by setting $i=x$ using the field parameters considered throughout this work) and phase squeezing (obtained by setting $i=y$), as they influence the electronic quantum orbits differently. Hereupon, we express the vector potential as
\begin{equation}
	\vb{A}_{\boldsymbol{\varepsilon_\alpha}}(t)
	= A_{\perp}(t) \vb{u}_{\perp} 
	+ A_{\parallel}(t)\vb{u}_{\parallel}, 
	\quad \text{with} \
	A_{\mu}(t)
	= \dfrac{\varepsilon^{(x)}_{\mu}}{\omega}
	\cos(\omega t)
	+ \dfrac{\varepsilon^{(y)}_{\mu}}{\omega}
	\sin(\omega t),
\end{equation}
where here we represent with $\omega$ the driving field frequency.

\subsection{Modified quantum orbits when considering amplitude squeezed states}
\begin{figure}[h!]
	\centering
	\includegraphics[width=0.7\textwidth]{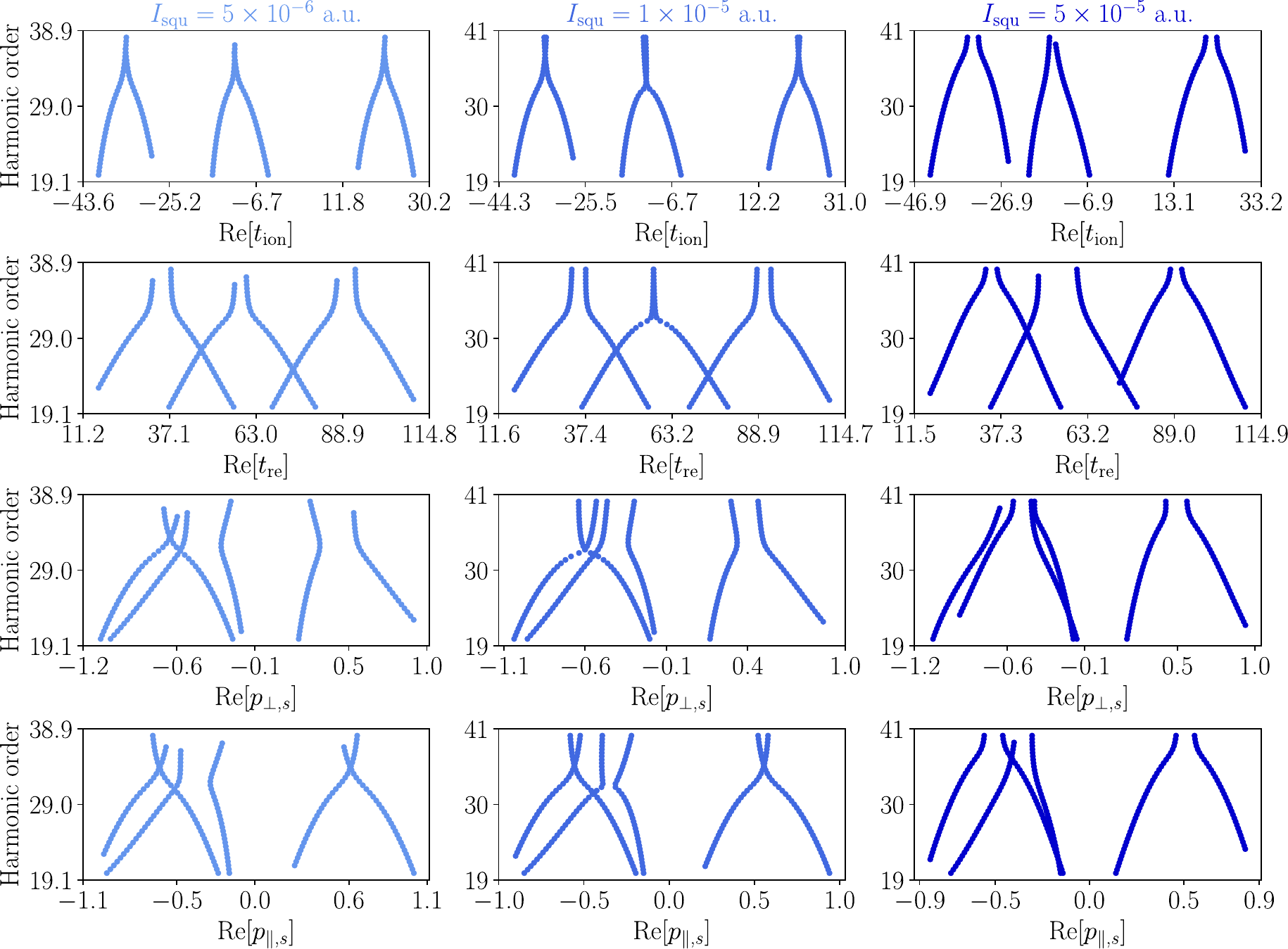}
	\caption{Real part of the saddle-point solutions when considering coherent$_{\parallel}$ + amplitude-squeezed$_{\perp}$ drivers. The field frequency has been set to $\omega = 0.057$ a.u., while the amplitudes to $\bar{\varepsilon}^{(x)}_{\parallel} = 0.053$ a.u., $\bar{\varepsilon}^{(y)}_{\parallel} = 0$ a.u., $\bar{\varepsilon}^{(x)}_{\perp} = 0$ a.u., $\bar{\varepsilon}^{(y)}_{\perp} = 0.053$ a.u., and the squeezing intensities to $I_{\text{squ}} = 5\times 10^{-6}$ a.u., $I_{\text{squ}} = 1\times 10^{-5}$ a.u., and $I_{\text{squ}} = 5\times 10^{-5}$ a.u., from left to right, respectively.}
	\label{Fig:App:saddle:points:x:BSV:Re}
\end{figure}

\begin{figure}[h!]
	\centering
	\includegraphics[width=0.7\textwidth]{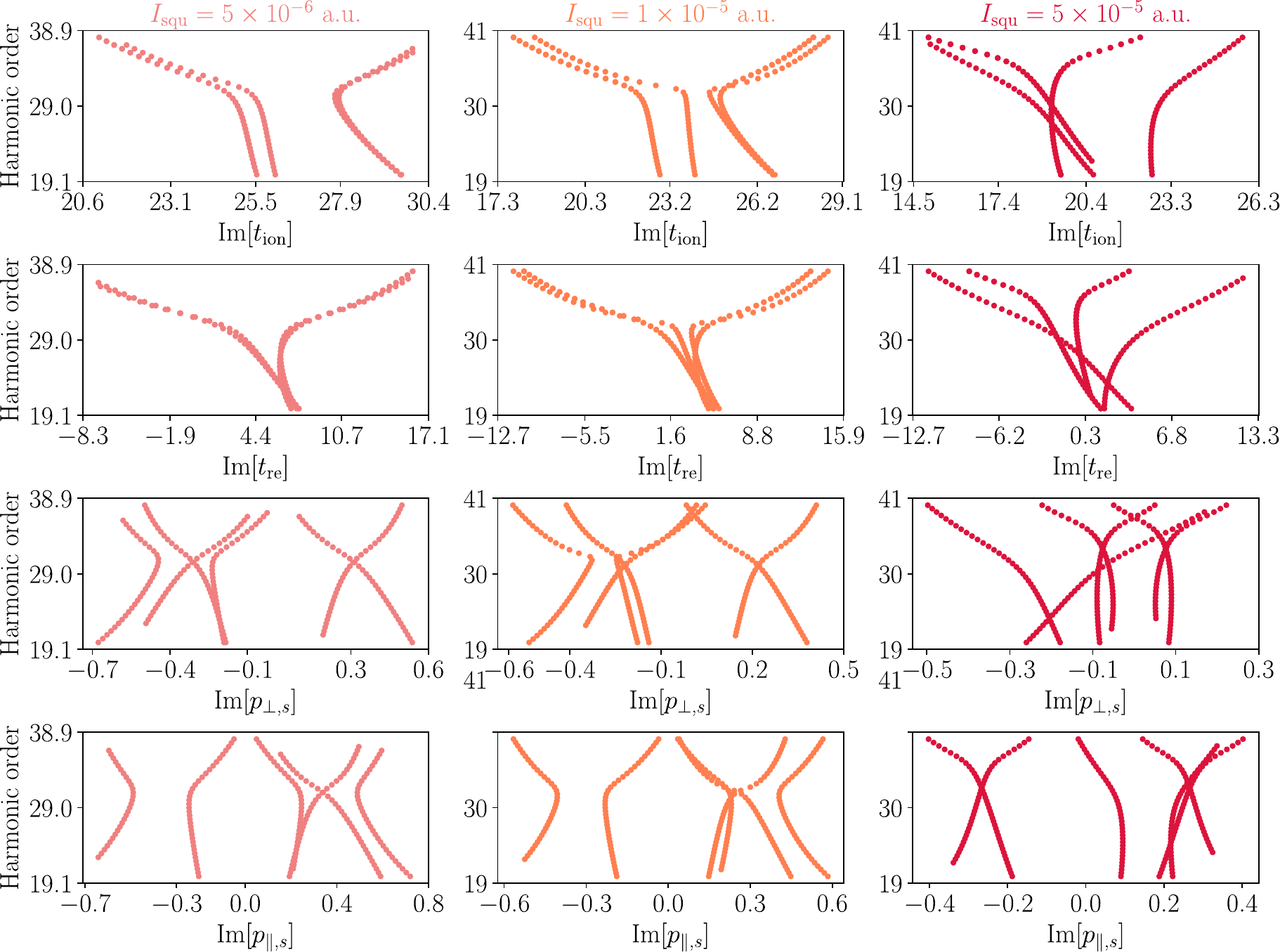}
	\caption{Imaginary part of the saddle-point solutions when considering coherent$_{\parallel}$ + amplitude-squeezed$_{\perp}$ drivers. The field frequency has been set to $\omega = 0.057$ a.u., while the amplitudes to $\bar{\varepsilon}^{(x)}_{\parallel} = 0.053$ a.u., $\bar{\varepsilon}^{(y)}_{\parallel} = 0$ a.u., $\bar{\varepsilon}^{(x)}_{\perp} = 0$ a.u., $\bar{\varepsilon}^{(y)}_{\perp} = 0.053$ a.u., and the squeezing intensities to $I_{\text{squ}} = 5\times 10^{-6}$ a.u., $I_{\text{squ}} = 1\times 10^{-5}$ a.u., and $I_{\text{squ}} = 5\times 10^{-5}$ a.u., from left to right, respectively.}
	\label{Fig:App:saddle:points:x:BSV:Im}
\end{figure}

In this subsection, we set $i=x$ in Eq.~\eqref{Eq:App:QSFA:Action} and compute the saddle-points of the action with respect to the five variables under consideration, i.e., $\boldsymbol{\theta} = (t_1,t_2,p_{\perp},p_{\parallel},\varepsilon^{(x)}_{\perp})$. Note that in this setup, $\varepsilon^{(y)}_{\perp} = \bar{\varepsilon}^{(y)}_{\perp}$. By setting the partial derivatives of Eq.~\eqref{Eq:App:QSFA:Action} with respect to each of these variables to zero , we obtain the following set of saddle-point equations
\begin{align}
	&\pdv{S_{\boldsymbol{\varepsilon_\alpha}}^{(\text{QO})}(\boldsymbol{\theta})}{t_1}\bigg\rvert_{\boldsymbol{\theta} = \boldsymbol{\theta_s}} = 0 
	\Rightarrow
	\dfrac{[p_{\perp,s} + A_{\perp}(t_{\text{ion}})]^2}{2}
	+ 	\dfrac{[p_{\parallel,s} + A_{\parallel}(t_{\text{ion}})]^2}{2}
	+ I_p = 0,
	\\&
	\pdv{S_{\boldsymbol{\varepsilon_\alpha}}^{(\text{QO})}(\boldsymbol{\theta})}{p_{\perp}}\bigg\rvert_{\boldsymbol{\theta} = \boldsymbol{\theta_s}} = 0 
	\Rightarrow
	\int^{t_{\text{re}}}_{t_{\text{i}}} \dd \tau \big[ p_{\perp,s} + A_\perp(\tau)\big] = 0,
	\label{Eq:SP:xBSV:pperp}
	\\&
	\pdv{S_{\boldsymbol{\varepsilon_\alpha}}^{(\text{QO})}(\boldsymbol{\theta})}{p_{\parallel}}\bigg\rvert_{\boldsymbol{\theta} = \boldsymbol{\theta_s}} = 0 
	\Rightarrow
	\int^{t_{\text{re}}}_{t_{\text{ion}}} \dd \tau \big[ p_{\parallel,s} + A_\parallel(\tau)\big] = 0,
	\\&
	\pdv{S_{\boldsymbol{\varepsilon_\alpha}}^{(\text{QO})}(\boldsymbol{\theta})}{t_2}\bigg\rvert_{\boldsymbol{\theta} = \boldsymbol{\theta_s}} = 0 
	\Rightarrow
	\dfrac{[p_{\perp,s} + A_{\perp}(t_{\text{re}})]^2}{2}
	+ 	\dfrac{[p_{\parallel,s} + A_{\parallel}(t_{\text{re}})]^2}{2}
	+ I_p = \Omega,	
	\\&
	\pdv{S_{\boldsymbol{\varepsilon_\alpha}}^{(\text{QO})}(\boldsymbol{\theta})}{\varepsilon_{\perp}^{(x)}}\bigg\rvert_{\boldsymbol{\theta} = \boldsymbol{\theta_s}} = 0 
	\Rightarrow
	\dfrac{1}{\omega}
	\int^{t_{\text{re}}}_{t_{\text{ion}}} \dd \tau
	\Big[
	p_{\perp,s} 
	+ \dfrac{\varepsilon^{(x)}_{\perp,s}}{\omega} \cos(\omega \tau)
	+ \dfrac{\bar{\varepsilon}^{(y)}_{\perp}}{\omega} \sin(\omega \tau)
	\Big] \cos(\omega \tau)
	- \dfrac{i}{\sigma_x} \Big[\varepsilon_{\perp,s}^{(x)}- \bar{\varepsilon}_{\perp}^{(x)}\Big] = 0\label{Eq:SP:xBSV:field},
\end{align}
where we refer to the solutions of these equations, i.e., the saddle-points, as $\boldsymbol{\theta_s} = (t_{\text{ion}},t_{\text{re}},p_{\perp,s},p_{\parallel,s},\varepsilon_{\perp,s}^{(x)})$. These are represented in Figs.~\ref{Fig:App:saddle:points:x:BSV:Re} and \ref{Fig:App:saddle:points:x:BSV:Im}, showing the real and imaginary parts, respectively, within the time-span $t \in [-\pi/\omega,2\pi/\omega]$ and for the squeezing parameters discussed in the main text. Notably, within this range and with coherent state drivers for both polarization directions, corresponding to circular polarization, no trajectories are found. 

The presence of non-vanishing electron trajectories can be attributed to an effective force induced by the non-Poissonian photon statistics of the driver, which distorts electron trajectories during propagation in the continuum. This concept was initially introduced in Ref.~\cite{even_tzur_photon-statistics_2023} for linearly polarized drivers with non-classical photon statistics. Here, we derive the expression for this force in the case of a coherent state in the $\parallel$-polarization combined with an amplitude squeezed state in the $\perp$-polarization.~To accomplish this, we first seek a solution to Eq.~\eqref{Eq:SP:xBSV:field}, which, after straightforward algebraic manipulations, can be reformulated as
\begin{equation}
	\begin{aligned}
		\varepsilon^{(x)}_{\perp,s}
		&= \dfrac{\bar{\varepsilon}^{(x)}_{\perp} - \frac{i\sigma_x}{\omega}\int^{t_{\text{re}}}_{t_{\text{ion}}}\dd \tau\bigg[p_{\perp,s} + \frac{\bar{\varepsilon}^{(y)}_{\perp}}{\omega}\sin(\omega \tau)\bigg]\cos(\omega \tau)}{1 + \frac{i\sigma_x}{\omega^2}\int^{t_{\text{re}}}_{t_{\text{ion}}} \dd \tau \cos[2](\omega\tau)}
		\\&\approx
		\bar{\varepsilon}^{(x)}_{\perp}
		\bigg[
		1-\dfrac{i\sigma_x}{\omega^2}
		\int^{t_{\text{re}}}_{t_{\text{ion}}}
		\dd \tau \cos[2](\omega \tau)
		\bigg]
		- i \dfrac{\sigma_x}{\omega}
		\int^{t_{\text{re}}}_{t_{\text{ion}}}
		\dd \tau
		\bigg[
		p_{\perp,s}
		+ \dfrac{\varepsilon^{(y)}_{\perp}}{\omega}\sin(\omega \tau)
		\bigg]\cos(\omega \tau),
	\end{aligned},
\end{equation}
where in transitioning from the first to the second line, we have applied a Taylor expansion to the denominator provided that $\sigma_x/\omega^2 \ll 1$ regime, valid here since the squeezing intensity is at most two orders of magnitude smaller than $\omega$. Noting further than in our setup, the amplitude squeezed states have associated $\bar{\varepsilon}_{\perp,x} = 0$, this expression simplifies to
\begin{equation}
	\varepsilon_{\perp,x}
	\approx 
	- i \dfrac{\sigma_x}{\omega}
	\int^{t_{\text{re}}}_{t_{\text{ion}}}
	\dd \tau
	\bigg[
	p_{\perp,s}
	+ \dfrac{\varepsilon^{(y)}_{\perp}}{\omega}\sin(\omega \tau)
	\bigg]\cos(\omega \tau).
\end{equation}

\begin{figure}
	\centering
	\includegraphics[width = 0.7\textwidth]{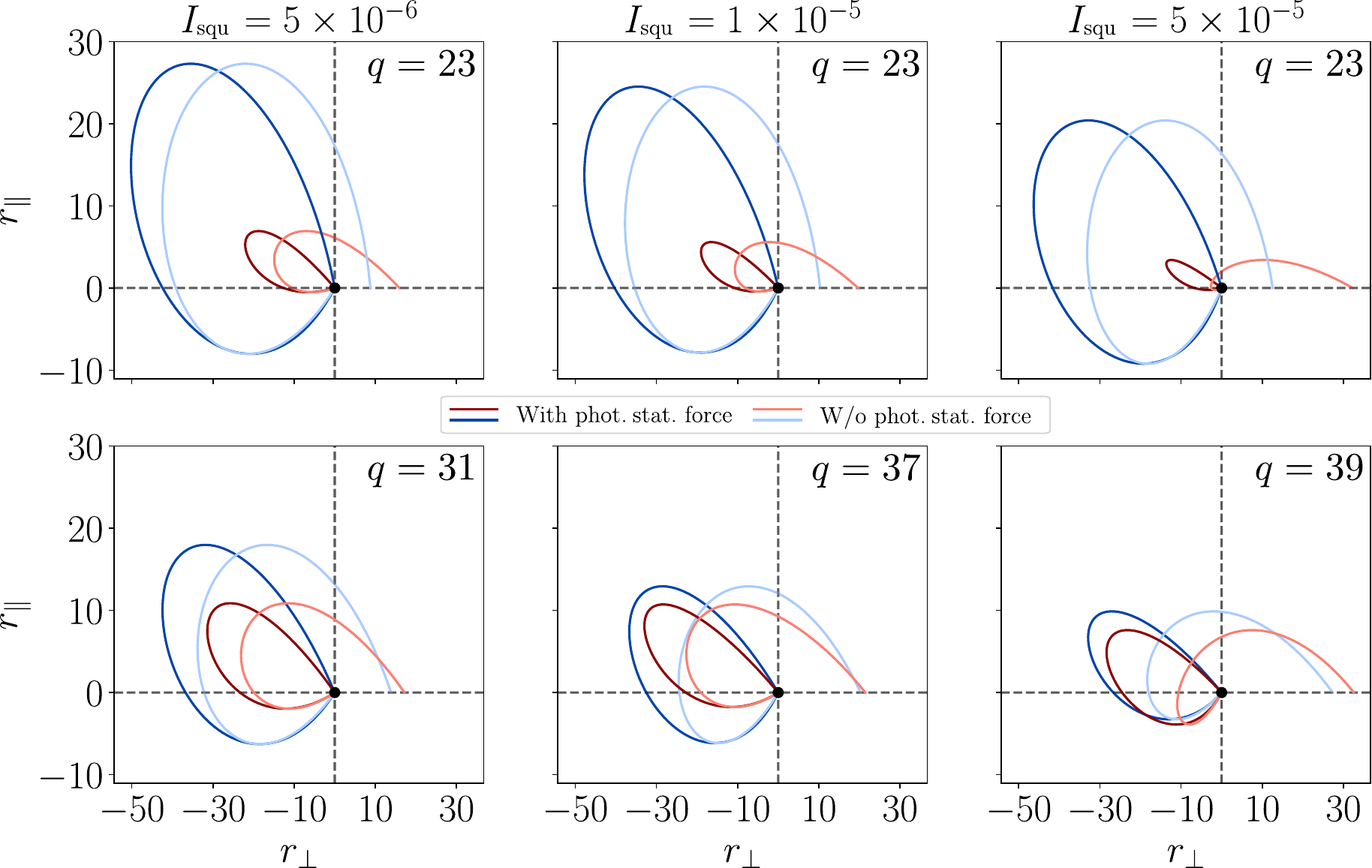}
	\caption{Representation of the electronic trajectories in real space for various harmonic modes and different squeezing intensities. The field frequency has been set to $\omega = 0.057$ a.u., while the amplitudes to $\bar{\varepsilon}^{(x)}_{\parallel} = 0.053$ a.u., $\bar{\varepsilon}^{(y)}_{\parallel} = 0$ a.u., $\bar{\varepsilon}^{(x)}_{\perp} = 0$ a.u., $\bar{\varepsilon}^{(y)}_{\perp} = 0.053$ a.u., and the squeezing intensities to $I_{\text{squ}} = 5\times 10^{-6}$ a.u., $I_{\text{squ}} = 1\times 10^{-5}$ a.u., and $I_{\text{squ}} = 5\times 10^{-5}$ a.u., from left to right, respectively.}
	\label{Fig:x:DSV:trajectories}
\end{figure}

Next, we substitute this solution into Eq.~\eqref{Eq:SP:xBSV:pperp}, which describes the electron's trajectory along the perpendicular direction. This yields
\begin{equation}\label{Eq:SP:xBSV:pperp:v2}
	\int^{t_{\text{re}}}_{t_{\text{ion}}}
	\dd \tau_1
	\Bigg\{
	p_{\perp,s}
	+ \dfrac{\varepsilon_{\perp}^{(y)}}{\omega}\sin(\omega \tau)
	- i \dfrac{\sigma_x}{\omega^2}\cos(\omega \tau)
	\int^{\tau_1}_{t_{\text{ion}}}
	\dd \tau_2
	\bigg[
	p_{\perp,s}
	+ \dfrac{\varepsilon_{\perp}^{(y)}}{\omega}\sin(\omega \tau_2)
	\bigg]\cos(\omega \tau_2)
	\Bigg\} = 0,
\end{equation}
where we observe the influence of the electric field of amplitude $\bar{\varepsilon}^{(y)}$ along with an additional term. This extra term, known as the \emph{photon-statistics force}~\cite{even_tzur_photon-statistics_2023}, is what effectively bends the electron's trajectory, enabling it to recombine with the parent ion. Figure~\ref{Fig:x:DSV:trajectories} illustrates these trajectories for $t_{\text{ion}}< -25$ a.u., considering various squeezing intensities and harmonic modes. From this figure, we observe that in the absence of this additional force---when setting $\varepsilon^{(x)}_{\perp,s} = 0$---the electron cannot return to the origin or recombine with the parent ion. Consequently, this added force effectively bends the electron trajectories, facilitating recombination events.

\subsection{Modified quantum orbits when considering phase squeezed states}
\begin{figure}
	\centering
	\includegraphics[width=0.7\textwidth]{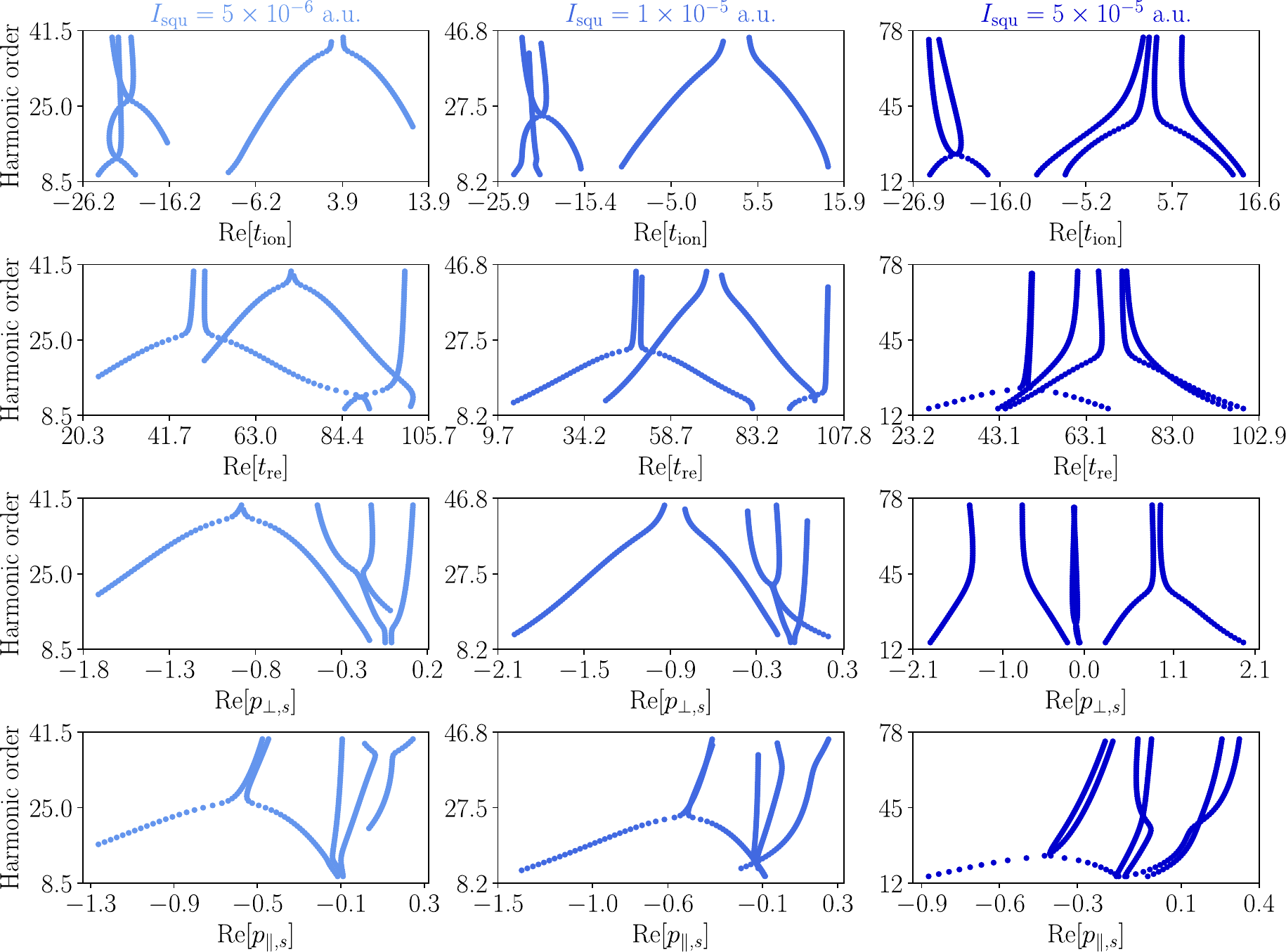}
	\caption{Real part of the saddle-point solutions when considering coherent$_{\parallel}$ + phase-squeezed$_{\perp}$ drivers. The field frequency has been set to $\omega = 0.057$ a.u., while the amplitudes to $\bar{\varepsilon}^{(x)}_{\parallel} = 0.053$ a.u., $\bar{\varepsilon}^{(y)}_{\parallel} = 0$ a.u., $\bar{\varepsilon}^{(x)}_{\perp} = 0$ a.u., $\bar{\varepsilon}^{(y)}_{\perp} = 0.053$ a.u., and the squeezing intensities to $I_{\text{squ}} = 5\times 10^{-6}$ a.u., $I_{\text{squ}} = 1\times 10^{-5}$ a.u., and $I_{\text{squ}} = 5\times 10^{-5}$ a.u., from left to right, respectively.}
	\label{Fig:App:saddle:points:y:BSV:Re}
\end{figure}

\begin{figure}
	\centering
	\includegraphics[width=0.7\textwidth]{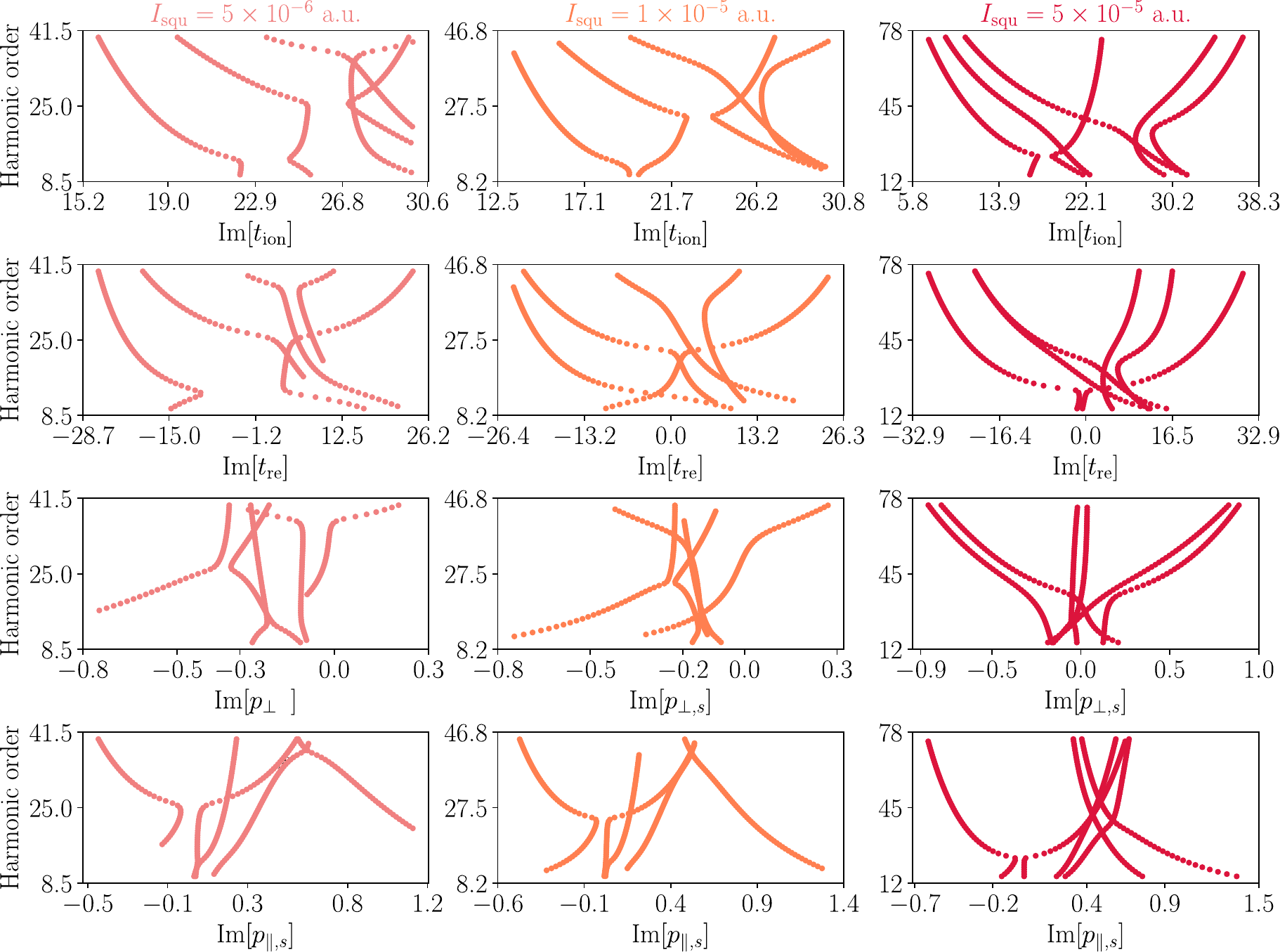}
	\caption{Imaginary part of the saddle-point solutions when considering coherent$_{\parallel}$ + phase-squeezed$_{\perp}$ drivers. The field frequency has been set to $\omega = 0.057$ a.u., while the amplitudes to $\bar{\varepsilon}^{(x)}_{\parallel} = 0.053$ a.u., $\bar{\varepsilon}^{(y)}_{\parallel} = 0$ a.u., $\bar{\varepsilon}^{(x)}_{\perp} = 0$ a.u., $\bar{\varepsilon}^{(y)}_{\perp} = 0.053$ a.u., and the squeezing intensities to $I_{\text{squ}} = 5\times 10^{-6}$ a.u., $I_{\text{squ}} = 1\times 10^{-5}$ a.u., and $I_{\text{squ}} = 5\times 10^{-5}$ a.u., from left to right, respectively.}
	\label{Fig:App:saddle:points:y:BSV:Im}
\end{figure}

In this subsection, we set $i=y$ in Eq.~\eqref{Eq:App:QSFA:Action} and apply the same approach as in the previous subsection. This results in the same saddle-point equations, but now with the variables $\boldsymbol{\theta} = (t_1,t_2,p_{\perp},p_{\parallel},\varepsilon^{(y)}_{\perp})$ and with $\varepsilon_{\perp}^{(x)} = \bar{\varepsilon}_{\perp}^{(x)}$. Consequently Eq.~\eqref{Eq:SP:xBSV:field} is substituted by
\begin{equation}
	\pdv{S_{\boldsymbol{\varepsilon_\alpha}}^{(\text{QO})}(\boldsymbol{\theta})}{\varepsilon_{\perp}^{(y)}}\bigg\rvert_{\boldsymbol{\theta} = \boldsymbol{\theta_s}} = 0 
	\Rightarrow
	\dfrac{1}{\omega}
	\int^{t_{\text{re}}}_{t_{\text{ion}}} \dd \tau
	\Big[
	p_{\perp,s} 
	+ \dfrac{\varepsilon^{(x)}_{\perp,s}}{\omega} \cos(\omega \tau)
	+ \dfrac{\bar{\varepsilon}^{(y)}_{\perp}}{\omega} \sin(\omega \tau)
	\Big] \sin(\omega \tau)
	- \dfrac{i}{\sigma_x} \Big[\varepsilon_{\perp,s}^{(y)}- \bar{\varepsilon}_{\perp}^{(y)}\Big] = 0\label{Eq:SP:yBSV:field},
\end{equation}
where the saddle-points are $\boldsymbol{\theta_s} = (t_{\text{ion}},t_{\text{re}},p_{\perp,s},p_{\parallel,s},\varepsilon_{\perp,s}^{(y)})$. Solutions to these equations are shown in Figs.~\ref{Fig:App:saddle:points:y:BSV:Re} and \ref{Fig:App:saddle:points:y:BSV:Im}, representing the real and imaginary components, respectively, over the time range $t\in [-\pi/\omega,2\pi/\omega]$ with the squeezing parameters discussed in the main text.

Notably, the trajectories obtained here significantly differ from those derived for amplitude squeezed states. Specifically, we observe ionization events occurring between $t_{\text{ion}}\in[-6.2,13.9]$ a.u.---a period in which no ionization events are seen for amplitude squeezed drivers. This already suggests that the dynamics influencing recombination events are intrinsically dependent on the type of squeezing applied. Furthermore, examining $\text{Re}[t_{\text{re}}]$ as a function of the harmonic order reveals two distinct cutoffs that vary with the squeezing intensity, and that originate from seemingly different trajectories.

Following a similar approach as in the amplitude squeezed case, we now compute the expression for the photon-statistics force for phase squeezed drivers. Starting from Eq.~\eqref{Eq:SP:yBSV:field} and applying the same steps, we obtain
\begin{equation}
	\varepsilon^{(y)}_{\perp}
	\approx
	\bar{\varepsilon}^{(y)}_{\perp}
	\bigg[ 
	1 
	- \dfrac{i\sigma_y}{\omega^2}
	\int^{t_{\text{re}}}_{t_{\text{ion}}} \dd \tau \sin[2](\omega \tau)
	\bigg]
	- \dfrac{i\sigma_y}{\omega}
	\int^{t_{\text{re}}}_{t_{\text{ion}}}
	\dd \tau
	p_{\perp,s}\sin(\omega \tau),
\end{equation}
where we have set $\varepsilon^{(x)}_{\perp} = 0$. By introducing this result into the saddle-point equation describing the electronic trajectories along the perpendicular ($\perp$) direction, we derive
\begin{equation}
	\begin{aligned}
		\int \dd \tau_2
		\bigg[
		p_\perp + \dfrac{\bar{\varepsilon}_{\perp}^{(y)}}{\omega}\sin(\omega\tau_2)
		- \dfrac{i\bar{\varepsilon}^{(y)}_{\perp}\sigma_y}{\omega^3}
		\sin(\omega \tau_2)
		\int^{\tau_2}_{t_{\text{ion}}}
		\dd \tau_1
		\sin[2](\omega\tau_1)
		- \dfrac{ip_{\perp,s}\sigma_y}{\omega^2}
		\sin(\omega \tau_2)
		\int^{\tau_2}_{t_{\text{ion}}}
		\dd \tau_2 \sin(\omega \tau_2)
		\bigg]=0.
	\end{aligned}
\end{equation}

\begin{figure}[h!]
	\centering
	\includegraphics[width = 0.7\textwidth]{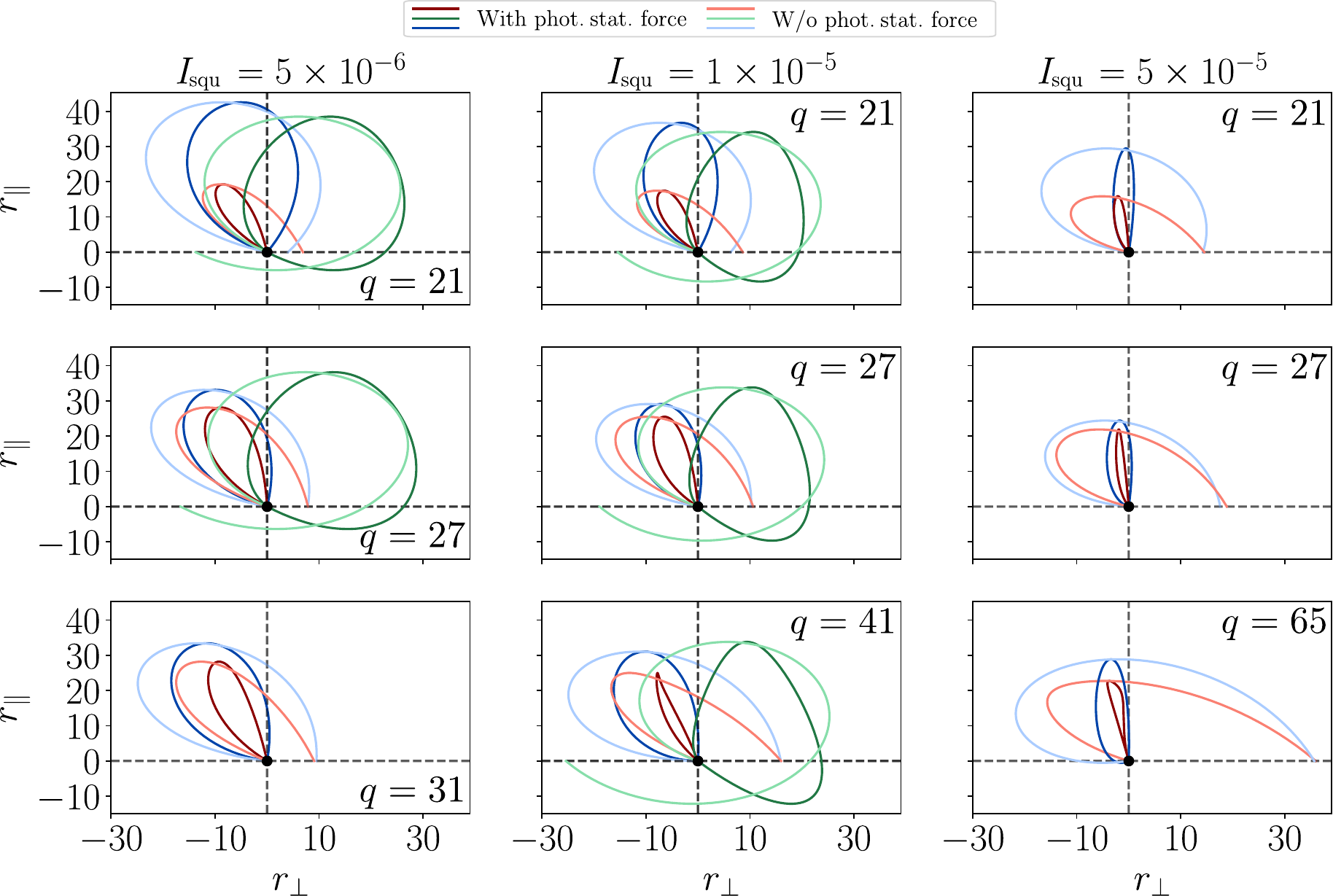}
	\caption{Representation of the electronic trajectories in real space for various harmonic modes and different squeezing intensities. Here, the results for $t_{\text{ion}}< -13$ a.u. are shown. The field frequency has been set to $\omega = 0.057$ a.u., while the amplitudes to $\bar{\varepsilon}^{(x)}_{\parallel} = 0.053$ a.u., $\bar{\varepsilon}^{(y)}_{\parallel} = 0$ a.u., $\bar{\varepsilon}^{(x)}_{\perp} = 0$ a.u., $\bar{\varepsilon}^{(y)}_{\perp} = 0.053$ a.u., and the squeezing intensities to $I_{\text{squ}} = 5\times 10^{-6}$ a.u., $I_{\text{squ}} = 1\times 10^{-5}$ a.u., and $I_{\text{squ}} = 5\times 10^{-5}$ a.u., from left to right, respectively.}
	\label{Fig:y:DSV:trajectories:tm13}
\end{figure}

\begin{figure}[h!]
	\centering
	\includegraphics[width = 0.8\textwidth]{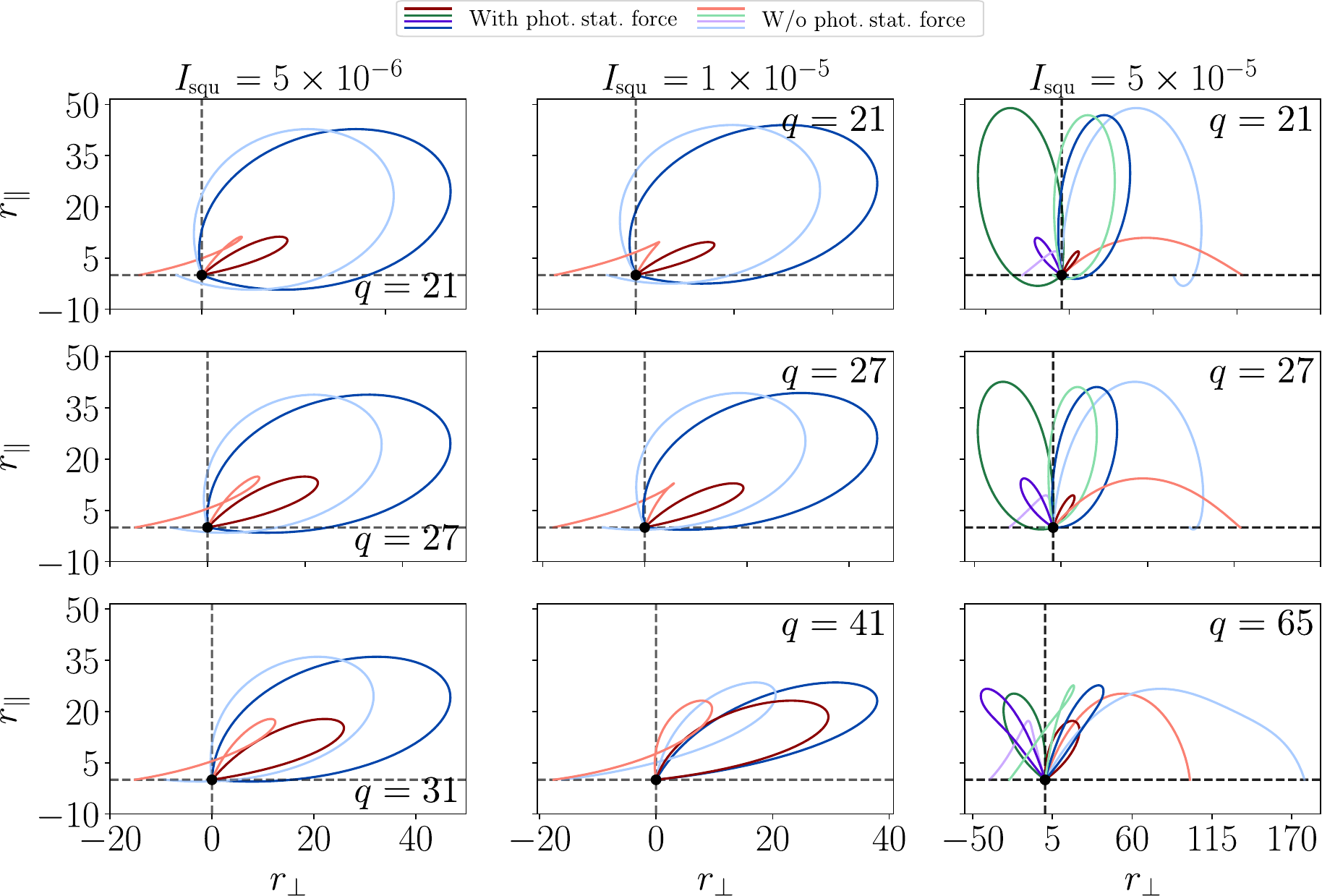}
	\caption{Representation of the electronic trajectories in real space for various harmonic modes and different squeezing intensities. Here, the results for $t_{\text{ion}}> -13$ a.u. are shown. The field frequency has been set to $\omega = 0.057$ a.u., while the amplitudes to $\bar{\varepsilon}^{(x)}_{\parallel} = 0.053$ a.u., $\bar{\varepsilon}^{(y)}_{\parallel} = 0$ a.u., $\bar{\varepsilon}^{(x)}_{\perp} = 0$ a.u., $\bar{\varepsilon}^{(y)}_{\perp} = 0.053$ a.u., and the squeezing intensities to $I_{\text{squ}} = 5\times 10^{-6}$ a.u., $I_{\text{squ}} = 1\times 10^{-5}$ a.u., and $I_{\text{squ}} = 5\times 10^{-5}$ a.u., from left to right, respectively.}
	\label{Fig:y:DSV:trajectories:tgtr13}
\end{figure}

Unlike the case of amplitude squeezed states, where only a single term added to the $(\bar{\varepsilon}^{(y)}_{\perp}/\omega)\sin(\omega \tau_2)$ component, we observe here the influence from two distinct forces, $F_1$ and $F_2$, specifically
\begin{align}
	&F_1 
	= \dfrac{i\bar{\varepsilon}^{(y)}_{\perp}\sigma_y}{\omega^3}
	\sin(\omega \tau_2)
	\int^{\tau_2}_{t_{\text{ion}}}\dd \tau_1
	\sin[2](\omega\tau_1)\label{Eq:App:F1}
	\\&F_2
	= \dfrac{ip_{\perp,s}\sigma_y}{\omega^2}
	\sin(\omega \tau_2)
	\int^{\tau_2}_{t_{\text{ion}}}\dd \tau_2
	\sin(\omega \tau_2)\label{Eq:App:F2},
\end{align}
which collectively determine how the electron trajectories are modified. We have shown the joint impact of these forces as a function of harmonic order in the main text. Here, to supplement those findings, we present the electron trajectories in real space. Unlike in the previous subsection, we include results for both $t_{\text{ion}} < -13$ a.u. (Fig.~\ref{Fig:y:DSV:trajectories:tm13}) and $t > -13$ a.u. (Fig.~\ref{Fig:y:DSV:trajectories:tgtr13}). This selection is motivated by the results shown in Fig.~\ref{Fig:App:saddle:points:x:BSV:Re}, which suggest that different trajectories may lead to seemingly distinct plateaus.

Similar to the amplitude squeezed case, Figs.~\ref{Fig:y:DSV:trajectories:tm13} and \ref{Fig:y:DSV:trajectories:tgtr13} show that these forces tend to bend the electronic trajectories, guiding them back towards the atomic core. However, with multiple forces at play, a wider variety of recombination effects emerge---some of which involve more than two trajectories. This enables a potential distinction between contributions from spatially ``ultra-long'' trajectories, such as the green curves in Fig.~\ref{Fig:y:DSV:trajectories:tm13}), and spatially ``ultra-short'' trajectories, like the red curves in Fig.~\ref{Fig:y:DSV:trajectories:tgtr13} when $I_{\text{squ}} = 5 \times 10^{-5}$ a.u., for instance.

\subsection{Expanded description of Figure 4 in the main text}
\begin{figure}[h!]
	\centering
	\includegraphics[width=0.7\columnwidth]{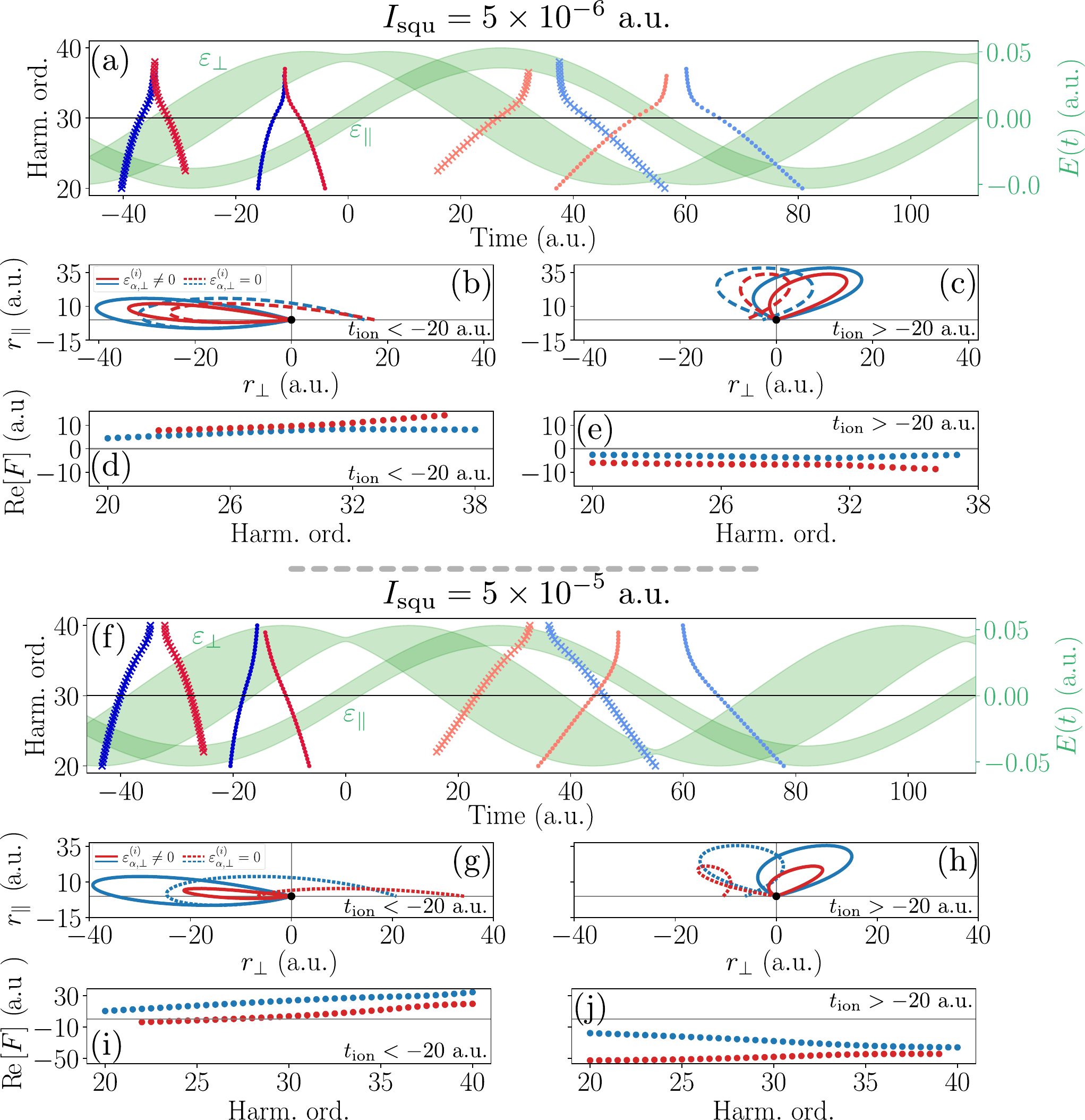}
	\caption{(a), (f) Real part of the ionization (bright colors) and recombination (soft colors) times, with the electric field represented for both polarizations in green; (b), (c), (g), (h) Electron trajectories in real space for the 33rd harmonic order in the absence (dashed curves) and presence (solid curves) of field fluctuations. The colors match with the trajectories in (a) and (f); (d), (e), (i), (j) Real part of the photon-statistical force. The mean field is circularly polarized with strength $\lvert\bar{\varepsilon}_{\mu}\rvert = 0.053$ a.u. ($I = 10^{14}$ W/cm$^2$).}
	\label{Fig:Traj:Field:xDSV}
\end{figure}

Altogether, Fig.~\ref{Fig:Traj:Field:xDSV}~(a) and (f) illustrate the emergence of ionization (bright colors) and recombination (soft colors) times when applying amplitude squeezing along the $\perp$-polarization direction, for $I_{\text{squ}} = 5 \times 10^{-6}$ a.u.~and $I_{\text{squ}} = 5 \times 10^{-5}$ a.u., respectively. In both cases, ionization events occur on either side of the $\lvert\varepsilon_{\parallel}\rvert$ maxima, coinciding with the time span where fluctuations in $\varepsilon_{\perp}$ are the greatest.~These ionization events give rise to two well-defined trajectories, long (in blue) and short (in red), which become increasingly similar as the harmonic order increases.~The cutoff harmonics $q_c$ remains the same regardless of whether ionization occurs on the left or right of the $\lvert\varepsilon_{\parallel}\rvert$ maxima, and increases with higher values of $I_{\text{squ}}$. Specifically, $q_{c} \approx 33$ for panel~(a) while $q_c \approx 40$ for panel~(f).

\begin{figure}[h!]
	\centering
	\includegraphics[width=0.7\columnwidth]{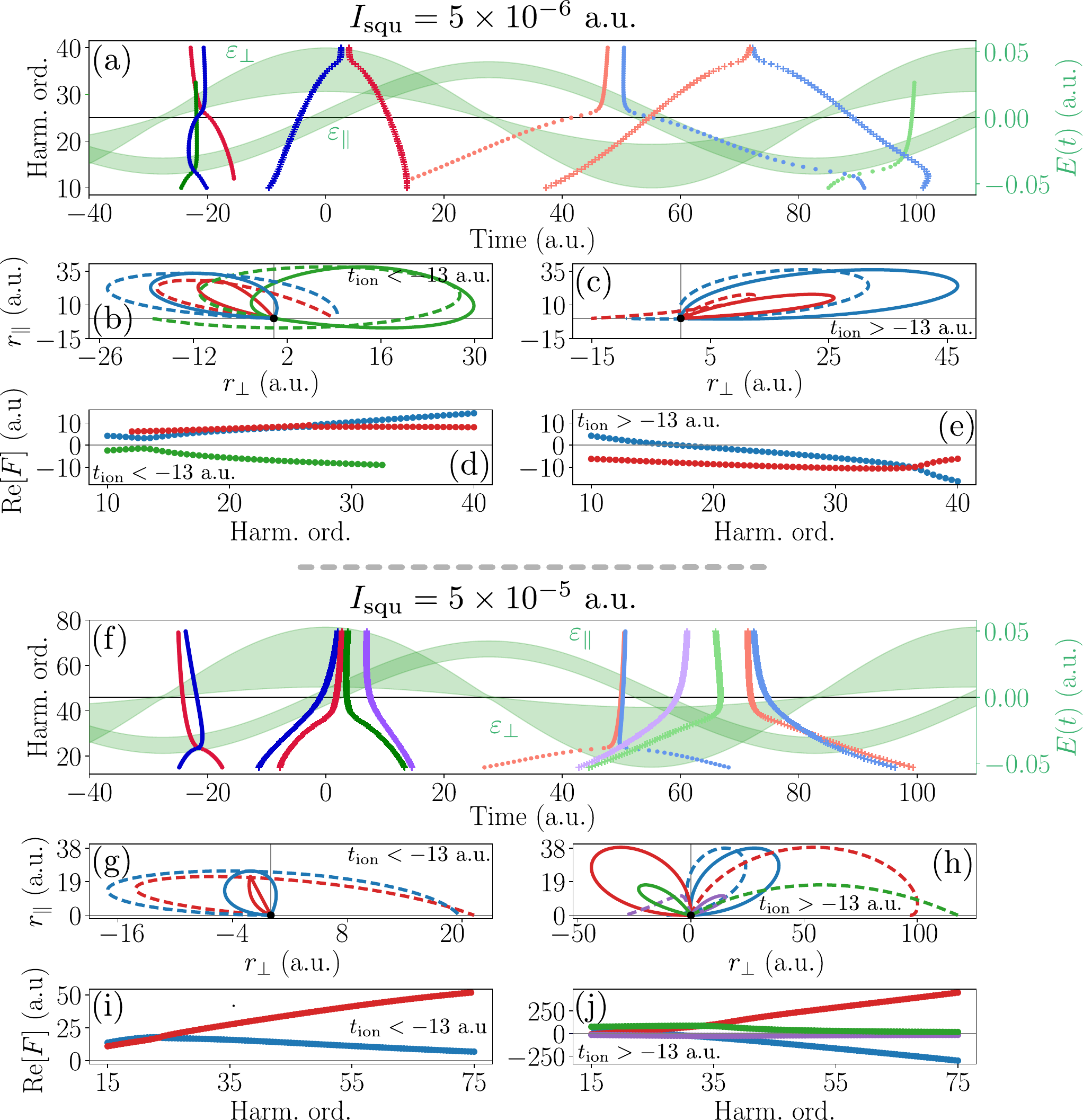}
	\caption{(a),(f) Real part of the ionization (bright colors) and recombination (soft colors) times, with the electric field represented for both polarizations in green; (b), (c), (g), (h) Electron trajectories in real space for the 31st harmonic order in the absence (dashed curves) and presence (solid curves) of field fluctuations. The colors match the trajectories in (a) and (f); (d), (e), (i), (j) Real part of the photon-statistical force.}
	\label{Fig:Traj:Field:yDSV}
\end{figure}

The dynamics change significantly when using phase squeezing, as shown in Fig.~\ref{Fig:Traj:Field:yDSV}~(a) and (f) for $I_{\text{squ}} = 5 \times 10^{-6}$ a.u. and $I_{\text{squ}} = 5 \times 10^{-5}$ a.u., respectively. In both cases, ionization and recombination events are observed, occurring at maxima of both $\lvert\varepsilon_{\parallel}\rvert$ and $\lvert\varepsilon_{\perp}\rvert$. Notably, ionization events at the $\lvert\varepsilon_{\perp}\rvert$ maxima, where intensity fluctuations are strongest, lead to a second, extended cutoff compared to ionization events at the $\lvert\varepsilon_{\parallel}\rvert$ maxima. These two cutoffs are located at $q_{c,1} \approx 27 $ and $q_{c,2} \approx 39 $ for $I_{\text{squ}} = 5 \times 10^{-6}$ a.u., and at $q_{c,1} \approx 25 $ and $q_{c,2} \approx 60$ for $I_{\text{squ}} = 5 \times 10^{-5}$ a.u., respectively. For low $I_{\text{squ}}$, ionization at $\lvert\varepsilon_{\perp}\rvert$ maxima produces well-defined short (in red) and long (in green) trajectories. As $I_{\text{squ}}$ increases, additional trajectories emerge, either longer (in blue) or shorter (in purple) than the standard ones, revealing a more complex pattern in electron dynamics induced by the enlarged intensity fluctuations.

The key factor enabling electron ionization and recombination are field fluctuations, represented by $\varepsilon^{(i)}_{\alpha,\perp}$. Without these, the electron cannot return to the parent ion for harmonic generation. We illustrate this by depicting the electron trajectories in real space in Fig.~\ref{Fig:Traj:Field:xDSV}~(b),(c) and (g),(h) for amplitude squeezing using $q= 33$, while in Fig.~\ref{Fig:Traj:Field:yDSV}~(a),(b) and (g),(h) for phase squeezing using $q= 31$. Across all cases, we observe that when $\varepsilon^{(i)}_{\alpha,\perp} = 0$ (dashed curves), recombination is not achieved, with the electron ending up the dynamics far away from the origin. However, when accounting for the saddle-point solution for $\varepsilon^{(i)}_{\alpha,\perp}$, recombination occurs, enabling harmonic generation.~From this trajectory perspective, the effect of $\varepsilon^{(i)}_{\alpha,\perp}$ can be effectively understood as a force that bends electron trajectories, facilitating recombination~\cite{even_tzur_photon-statistics_2023}. The results of this analysis are presented in Fig.~\ref{Fig:Traj:Field:xDSV}~(d),(e) and (i),(j) for amplitude squeezing, and in Fig.~\ref{Fig:Traj:Field:yDSV}~(d),(e) and (i),(j) for phase squeezing, illustrating how this photon-statistics-induced force acts on each electronic trajectory (indicated by matching colors to those in panels~(a) and (f)) as a function of the harmonic order. 

Comparing these forces with the real space electron trajectories, we find that the forces are positive for trajectories with $r_{\perp} < 0$ and negative for $r_{\perp} > 0$. The force magnitude generally increases with $I_{\text{squ}}$, although the specific behavior differs on the type of squeezing. For amplitude squeezing [Fig.~\ref{Fig:Traj:Field:xDSV}], ionization events occurring on either side of the $\lvert\varepsilon_{\perp}\rvert$ maxima, experience forces of similar magnitude. For phase squeezing [Fig.~\ref{Fig:Traj:Field:yDSV}] trajectories contributing to the highest cutoff experience notably stronger forces than those contributing to the first  cutoff, especially for $I_{\text{squ}} = 5 \times 10^{-5}$ a.u. at high harmonic orders.

\subsection{Numerical analysis}
The evaluation of saddle-point equations was entirely done in \texttt{Mathematica} using the \texttt{RB-SFA} package~\cite{RBSFA}. Specifically, we used the \texttt{FindComplexRoots} function which allows solving a set of complex-valued linear equations over a predefined grid in the complex space.

\section{The contribution of displaced thermal states}\label{Sec:Thermal:states:Spec}
In this section, we emphasize that the distinctive features presented in this manuscript emerge from the non-Poissonian nature of the driving light's photon statistics, irrespective of whether these statistics are sub- or super-Poissonian.~However, it is ultimately the photon statistics of the driver that shape the specific spectral features of the resulting HHG spectrum. While the main text focused on displaced amplitude- and phase-squeezed states, which either exhibit sub-Poissonian and super-Poissonian characteristics, here we compare these results with those obtained using thermal states of light.

\begin{figure}[h!]
	\centering
	\includegraphics[width=0.8\textwidth]{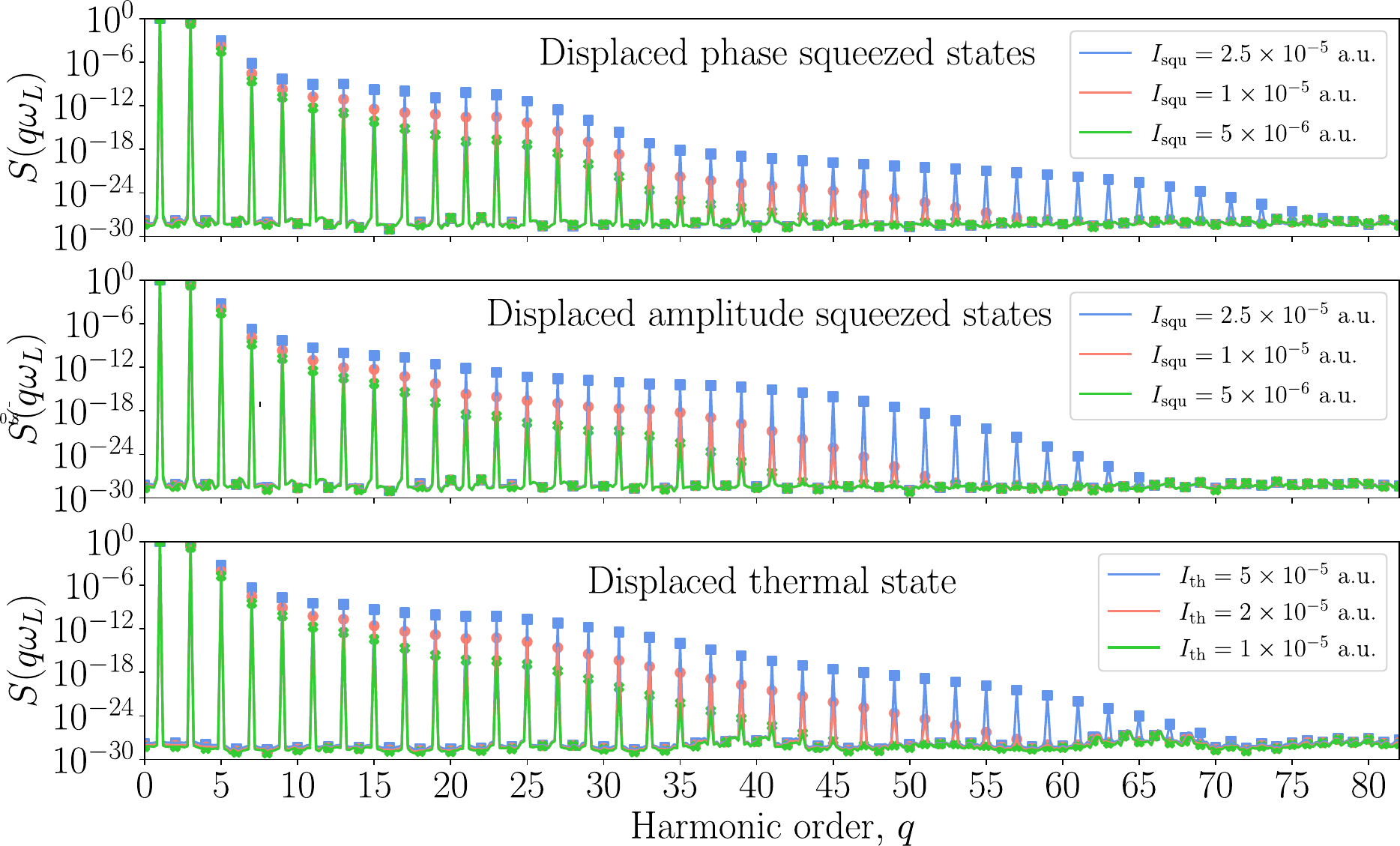}
	\caption{HHG spectrum obtained for the three states of light considered along the text. Each spectra has been normalized with respect to its maximum value. The field frequency has been set to $\omega = 0.057$ a.u., while the amplitudes to $\bar{\varepsilon}^{(x)}_{\parallel} = 0.053$ a.u., $\bar{\varepsilon}^{(y)}_{\parallel} = 0$ a.u., $\bar{\varepsilon}^{(x)}_{\perp} = 0$ a.u., $\bar{\varepsilon}^{(y)}_{\perp} = 0.053$ a.u.}
	\label{Fig:App:Spectrum:All}
\end{figure}

Using Eqs.~\eqref{Eq:App:Spectrum:x:DSV}, \eqref{Eq:App:Spectrum:y:DSV} and \eqref{Eq:App:Spectrum:thermal}, we present the results in Fig.~\ref{Fig:App:Spectrum:All}. Since the variance of the weighting probability distributions behave as $\sigma_i = 4 I_{\text{squ}}$ for squeezed states and $\sigma_{\text{th}} = 2 I_{\text{th}}$ for thermal states, we plot comparable cases by matching $I_{\text{th}} = 2 I_{\text{squ}}$. As reported in the main text, for the phase squeezed states, we observe a two-plateau structure, where plateau intensities vary with squeezing, while amplitude squeezed states produce a single plateau followed by a decay at higher harmonic orders. By contrast, displaced thermal states exhibit an extended cutoff, as noted in Ref.~\cite{gorlach_high-harmonic_2023}, which increases with thermal noise level but lacks the double-plateau pattern seen in the phase squeezed case.

Notably, the persistence of harmonic radiation in circularly polarized fields is not limited to squeezed states. Rather, any light state with non-Poissonian photon statistics and sufficient intensity to initiate HHG processes can produce harmonic radiation in circularly polarized configurations. Here, we illustrated this effect by comparing the spectra of phase squeezed states, amplitude squeezed states, and displaced thermal states.

\section{Evaluation of the time-zero second-order autocorrelation function}\label{App:g2}
To conclude with this Supplementary Material, we perform an analytical assessment to identify potential non-classical properties in the photon statistics of the emitted light by evaluating the time-zero second-order autocorrelation function, defined as
\begin{equation}
	g_{\vb{k},\mu}^{(2)}(0)
	= \dfrac{\langle(\hat{a}^\dagger_{\vb{k},\mu})^2\hat{a}^2_{\vb{k},\mu}\rangle}{\langle\hat{a}^\dagger_{\vb{k},\mu}\hat{a}_{\vb{k},\mu}\rangle^2},
\end{equation}
which allows us to distinguish between classical and non-classical light behavior based on photon statistics: for Poissonian and super-Poissonian statistics (classical light), $g_{\vb{k},\mu}^{(2)}(0)\geq 1$; for sub-Poissonian statistics (non-classical light), $0 \leq g_{\vb{k},\mu}^{(2)}(0) < 1$.

To evaluate $g_{\vb{k},\mu}^{(2)}(0)$, we start by computing $\langle(\hat{a}^\dagger_{\vb{k},\mu})^n\hat{a}^n_{\vb{k},\mu}\rangle$, which represents the numerator and denominator of our function when setting $n=2$ for the former and $n=1$ for the latter. For this general case, we obtain
\begin{equation}\label{Eq:App:n:order:correlator}
	\begin{aligned}
		\langle(\hat{a}^\dagger_{\vb{k},\mu})^n\hat{a}^n_{\vb{k},\mu}\rangle
		&= \epsilon^n
		\int \dd^2\alpha_{\parallel}
		\int \dd^2 \beta_{\parallel}
		\int \dd^2\alpha_{\perp}
		\int \dd^2 \beta_{\perp}
		\dfrac{P_{\parallel}(\alpha_{\parallel},\beta_{\parallel}^*)}{\braket{\beta_{\parallel}}{\alpha_{\parallel}}}
		\dfrac{P_{\perp}(\alpha_{\perp},\beta_{\perp}^*)}{\braket{\beta_{\perp}}{\alpha_{\perp}}}
		\big\{
		[d^*_{\boldsymbol{\beta},\mu}(\omega_{\vb{k}})]^n
		d^n_{\boldsymbol{\alpha},\mu}(\omega_{\vb{k}})
		\big\}
		\\&\hspace{6cm}\times
		\bra{\chi_{\boldsymbol{\beta},\vb{k_0}}}
		\hat{\vb{D}}^\dagger(\boldsymbol{\beta})
		\hat{\vb{D}}(\boldsymbol{\alpha})
		\ket{\chi_{\boldsymbol{\alpha},\vb{k_0}}}
		\prod_{\vb{k}\neq\vb{k}_0}
		\braket{\chi_{\boldsymbol{\beta},\vb{k}}}{\chi_{\boldsymbol{\alpha},\vb{k}}},
	\end{aligned}
\end{equation}
where we denote $\hat{\vb{D}}(\boldsymbol{\alpha}) = \hat{D}_{\parallel}(\alpha_{\parallel}) \otimes \hat{D}_{\perp}(\alpha_{\perp})$. In the following, we evaluate the matrix elements of this displacement operator. For the sake of clarity, we use a simplified notation
\begin{align}
	\bra{\chi_{\beta}}\hat{D}^\dagger(\beta)\hat{D}(\alpha)\ket{\chi_{\alpha}}
	&= e^{\frac12(\alpha \chi_{\alpha}^* - \alpha^*\chi_\alpha)}
	e^{\frac12(\beta^* \chi_{\beta} - \beta\chi_\beta^*)}
	\braket{\beta + \chi_\beta}{\alpha + \chi_\alpha}
	\\&
	=e^{\frac12(\alpha \chi_{\alpha}^* - \alpha^*\chi_\alpha)}
	e^{\frac12(\beta^* \chi_{\beta} - \beta\chi_\beta^*)}
	\exp{-\frac12
		\big[
		\abs{\beta + \chi_\beta}^2
		+ \abs{\alpha + \chi_\alpha}^2 
		- 2(\beta + \chi_\beta)^*(\alpha + \chi_\alpha)
		\big]}
	\\&= e^{\frac12(\alpha \chi_{\alpha}^* - \alpha^*\chi_\alpha)}
	e^{\frac12(\beta^* \chi_{\beta} - \beta\chi_\beta^*)}
	\exp{-\dfrac{1}{2}(\abs{\alpha}^2 + \abs{\beta}^2 -2\beta^*\alpha} \nonumber
	\\&\quad\times
	\exp{-\dfrac12
		\big[
		\abs{\chi_\beta}^2 + \abs{\chi_{\alpha}}^2 
		+ \beta^*\chi_\beta  + \beta\chi_\beta^*
		+ \alpha^*\chi_\alpha + \alpha\chi_\alpha
		-2 \beta^*\chi_\alpha - 2 \chi_\beta \alpha - 2 \chi_\beta^*\chi_\alpha
		\big]
	}
	\\& = \braket{\beta}{\alpha} f(d_{\alpha},d_{\beta},\epsilon_{\alpha},\epsilon_{\beta}),
\end{align}
where in the last equality we have identified a term accounting for the overlap $\braket{\beta}{\alpha}$, and defined
\begin{equation}
	\begin{aligned}
		f(d_{\alpha},d_{\beta},\epsilon_{\alpha},\epsilon_{\beta})
		&= e^{\frac14(\varepsilon_\alpha d_{\alpha}^* - \varepsilon_\alpha^*d_\alpha)}
		e^{\frac14(\varepsilon_\beta^* d_{\beta} - \varepsilon_\beta d_\beta^*)}
		\\&\quad \times
		\exp{-\dfrac14
			\big[
			2\epsilon^2\abs{d_\beta}^2 + 2\abs{d_{\alpha}}^2 
			+ \varepsilon_\beta^*d_\beta  + \varepsilon_\beta d_\beta^*
			+\varepsilon_ \alpha^*d_\alpha + \varepsilon_\alpha d^*_\alpha
			- \varepsilon_\beta^*d_\alpha - d_\beta^* \varepsilon_\alpha - 2\epsilon^2 d_\beta^*d_\alpha
			\big]
		},
	\end{aligned}
\end{equation}
where we have taken into account that $\chi_\alpha = \epsilon^2 d_{\alpha}$, with $d_{\alpha}$ a shorthand notation for the Fourier transform of the time-dependent dipole moment, and that $\varepsilon_{\alpha} = 2 \epsilon \alpha$. We observe than in the classical limit $\epsilon \to 0$, $f(d_{\alpha},d_{\beta},\epsilon_{\alpha},\epsilon_{\beta})$ remains finite. Since $\abs{\chi_{\alpha}} \ll 1$ at the single-atom level, we approximate $f(d_{\alpha},d_{\beta},\epsilon_{\alpha},\epsilon_{\beta}) \to 1$. This allows us to express Eq.~\eqref{Eq:App:n:order:correlator} as
\begin{equation}
	\begin{aligned}
		\langle(\hat{a}^\dagger_{\vb{k},\mu})^n\hat{a}^n_{\vb{k},\mu}\rangle
		&= \epsilon^n
		\int \dd^2\alpha_{\parallel}
		\int \dd^2 \beta_{\parallel}
		\int \dd^2\alpha_{\perp}
		\int \dd^2 \beta_{\perp}
		P_{\parallel}(\alpha_{\parallel},\beta_{\parallel}^*)
		P_{\perp}(\alpha_{\perp},\beta_{\perp}^*)
		\big\{
		[d^*_{\boldsymbol{\beta},\mu}(\omega_{\vb{k}})]^n
		d^n_{\boldsymbol{\alpha},\mu}(\omega_{\vb{k}})
		\big\},		
	\end{aligned}
\end{equation}
such that in the classical limit $\epsilon \to 0$---obtained as in Sec.~\ref{Sec:HHG:Spec}---$g^{(2)}_{\vb{k},\mu}(0)$ reads, for the case of coherent$_{\parallel}$ + squeezed$_{\perp}$ drivers, as follows
\begin{equation}
	g^{(2)}_{\vb{k},\mu}(0)
	= \dfrac{\smallint \dd \varepsilon_{\alpha,\perp}^{(i)} \big\{p(\varepsilon_{\alpha,\perp}^{(i)},\bar{\varepsilon}_{\alpha,\perp}^{(i)})\abs{d_{\boldsymbol{\alpha},\mu}(\omega_{\vb{k}})}^{4}\big\}}{\Big[\smallint \dd \varepsilon_{\alpha,\perp}^{(i)}\big\{ p(\varepsilon_{\alpha,\perp}^{(i)},\bar{\varepsilon}_{\alpha,\perp}^{(i)})\abs{d_{\boldsymbol{\alpha},\mu}(\omega_{\vb{k}})}^{2}\big\}\Big]^2},
\end{equation}
where we note that $p(\varepsilon_{\alpha,\perp}^{(i)},\bar{\varepsilon}_{\alpha,\perp}^{(i)}) \geq 0$ $\forall \varepsilon_{\alpha,\perp}^{(i)},\bar{\varepsilon}_{\alpha,\perp}^{(i)}$. This result matches with those derived in the Supplementary Material of Ref.~\cite{gorlach_high-harmonic_2023}.

We can further leverage the positive-semidefinite property of the weighting function to establish tighter bounds on the behavior of $g^{(2)}_{\vb{k},\mu}(0)$. Notably, both the numerator and denominator can be computed as the expected value of $(\hat{a}^\dagger)^n\hat{a}^n$ operators applied to the following effective state
\begin{equation}
	\hat{\rho}_{\text{eff}}
	= \int \dd \varepsilon_{\alpha,\perp}^{(i)}
	p(\varepsilon_{\alpha,\perp}^{(i)},\bar{\varepsilon}_{\alpha,\perp}^{(i)}) 
	\lvert\chi^{(\boldsymbol{\varepsilon_\alpha})}_{\vb{k},\mu}\rangle\!\langle\chi^{(\boldsymbol{\varepsilon_\alpha})}_{\vb{k},\mu}\rvert,
\end{equation}which represents a statistical mixture of various coherent states. States of this form are classical states of light and, as such, they yield to
\begin{equation}
	g^{(2)}_{\vb{k},\mu}(0) \geq 1,
\end{equation}
where equality holds when considering only coherent states of light. Importantly, these results are independent of the squeezing direction $i$, so we conclude that the harmonics will either exhibit Poissonian or super-Poissonian characteristics.

However, it is important to note that this is just one criterion for classifying states as classical or non-classical, and it is not unique. Other non-classicality measures could yield different results. Indeed, we expect entanglement to arise after the light-matter interaction when squeezed states are used as drivers. This expectation is based on the fact that our initial state is a pure state, which, due to the unitarity of the interaction, remains pure post-interaction. Yet, we have observed that evaluating a local observable on a specific subsystem---in this case, the optical mode $(\vb{k},\mu)$---can be done as if using a classical mixture of coherent states. This conversion from pure to mixed states is only possible if partial information has been traced out from the optical system, possibly indicating that the post-interaction state is entangled.

This argument is only valid for initial configurations of coherent$_{\parallel}$ + coherent$_{\perp}$ or coherent$_{\parallel}$ + squeezed$_{\perp}$ configurations. For a coherent$_{\parallel}$ + thermal$_{\perp}$ configuration, however, the initial state is already a mixture, so the entanglement argument is not applied and a more thorough analysis is therefore needed. Finally, for a coherent$_{\parallel}$ + coherent$_{\perp}$ driver, we find $g^{(2)}_{\vb{k},\mu}(0) = 1$, consistent with Refs.~\cite{lewenstein_theory_1994,rivera-dean_strong_2022,stammer_quantum_2023}.